\documentclass[12pt]{article}

\usepackage{algorithmic}
\usepackage{algorithm}
\usepackage{amsmath}

\usepackage{scicite}

\usepackage[backend=bibtex, giveninits=true, natbib=true, sorting=none, style=numeric-comp]{biblatex}
\makeatletter
\let\NAT@parse\undefined
\makeatother

\usepackage{url, amsmath, amssymb, tabularx, booktabs}
\usepackage{mathtools}
\usepackage{graphicx}
\usepackage{bm}
\usepackage{fmtcount}
\usepackage{cleveref}
\usepackage[normalem]{ulem}

\newcommand{\snr}{\mathrm{SNR}}
\newcommand{\pin}{p_{\rm in}}
\newcommand{\pout}{p_{\rm out}}
\newcommand{\cin}{c_{\rm in}}
\newcommand{\cout}{c_{\rm out}}
\newcommand{\din}{d_{\rm in}}
\newcommand{\dout}{d_{\rm out}}
\newcommand{\calE}{\mathcal{E}}
\newcommand{\calK}{\mathcal{K}}
\newcommand{\bmC}{\bm{C}}
\newcommand{\bmNB}{\bm{NB}}

\usepackage[usenames]{xcolor}
\definecolor{mpl0}{HTML}{1f77b4}
\definecolor{mpl1}{HTML}{ff7f0e}
\definecolor{mpl2}{HTML}{2ca02c}
\definecolor{mpl3}{HTML}{d62728}
\definecolor{mpl4}{HTML}{9467bd}
\definecolor{mpl5}{HTML}{8c564b}
\definecolor{mpl6}{HTML}{e377c2}
\definecolor{mpl7}{HTML}{7f7f7f}
\definecolor{mpl8}{HTML}{bcbd22}
\definecolor{mpl9}{HTML}{17becf}

\makeatletter
\def\mathcolor#1#{\@mathcolor{#1}}
\def\@mathcolor#1#2#3{
    \protect\leavevmode
    \begingroup
    \color#1{#2}#3
    \endgroup
}
\makeatother

\newenvironment{sciabstract}{
\begin{quote} \bf}
{\end{quote}}

\newcounter{lastnote}

\title{Higher order trade-offs in hypergraph community detection}

\author{
Jiaze Li$^{1}$, Michael T. Schaub$^{2}$, Leto Peel$^{1}$\\
\\
\parbox{\linewidth}{\centering
\normalsize{$^{1}$Department of Data Analytics and Digitalisation
,  Maastricht University, 6200 MD Maastricht, The Netherlands}\\
\normalsize{$^{2}$RWTH Aachen University, Aachen, Germany}\\}
}

\date{}

\begin{document}

\maketitle

\begin{sciabstract}
    Extending community detection from pairwise networks to hypergraphs introduces fundamental theoretical challenges.
    Hypergraphs exhibit structural heterogeneity with no direct graph analogue: hyperedges of varying orders can connect nodes across communities in diverse configurations. Crucially, this means that any hypergraph community detection algorithm must make a choice regarding which order or balance of hyperedges it prefers to maintain or split during partitioning. We address these challenges by developing a unified framework for community detection in non-uniform hypergraphs under the Hypergraph Stochastic Block Model.
    We introduce a general signal-to-noise ratio that enables a quantitative analysis of trade-offs unique to higher-order networks, such as which hypergedges we choose to split across communities and how we choose to split them.
    Building on this framework, we derive a Bethe Hessian operator for non-uniform hypergraphs that provides efficient spectral clustering with principled model selection.
    We characterize the resulting spectral detectability threshold and compare it to belief propagation limits, showing the methods coincide for uniform hypergraphs but diverge in non-uniform settings.
    Synthetic experiments confirm our analytical predictions and reveal systematic biases toward preserving higher-order and balanced-shape hyperedges.
    Application to empirical data demonstrates the practical relevance of these higher-order detectability trade-offs in real-world systems.
\end{sciabstract}

\section*{Introduction}
Community detection is central to interpreting mesoscale network structure, revealing groups of nodes with similar connectivity patterns that provide a coarse-grained summary often aligned with functional organization~\cite{fortunato2010community,porter2009communities}.
In the classical setting of dyadic, or \emph{pairwise}, networks, this intuition translates most commonly into identifying densely connected subgraphs, and a broad range of algorithmic and theoretical tools have been developed, including modularity maximization~\cite{blondel2008fast}, spectral clustering~\cite{von2007tutorial}, and probabilistic generative models such as the stochastic block model (SBM)~\cite{holland1983stochastic}. 
The SBM has enabled a comparatively mature theoretical understanding of community detection, particularly in the case of sparse graphs.
These theoretical contributions include sharp detectability thresholds, near-optimal inference algorithms and closely related efficient spectral methods based on the non-backtracking and the Bethe Hessian operators~\cite{decelle2011asymptotic,krzakala2013spectral,saade2014spectral}.

Yet, many complex systems cannot be faithfully represented by pairwise interactions alone.
In domains ranging from scientific collaboration and human contact patterns to microbial consortia and chemical reaction networks, interactions regularly involve more than two entities at a time, reflecting intrinsic multiway dependencies~\cite{bick2023higher,newman2018networks,sanchez2019high}.
Recognizing the importance of such higher-order interactions has led to growing interest in higher-order network representations and corresponding models and algorithms, with hypergraphs emerging as a natural and widely used framework; see, e.g., recent reviews~\cite{battiston2020networks,battiston2021physics,bick2023higher}.
Here we consider aspects of applying the established utility of community detection to the broadly applicable setting of hypergraphs.

\begin{figure}
    \centering
    \includegraphics[width=15cm]{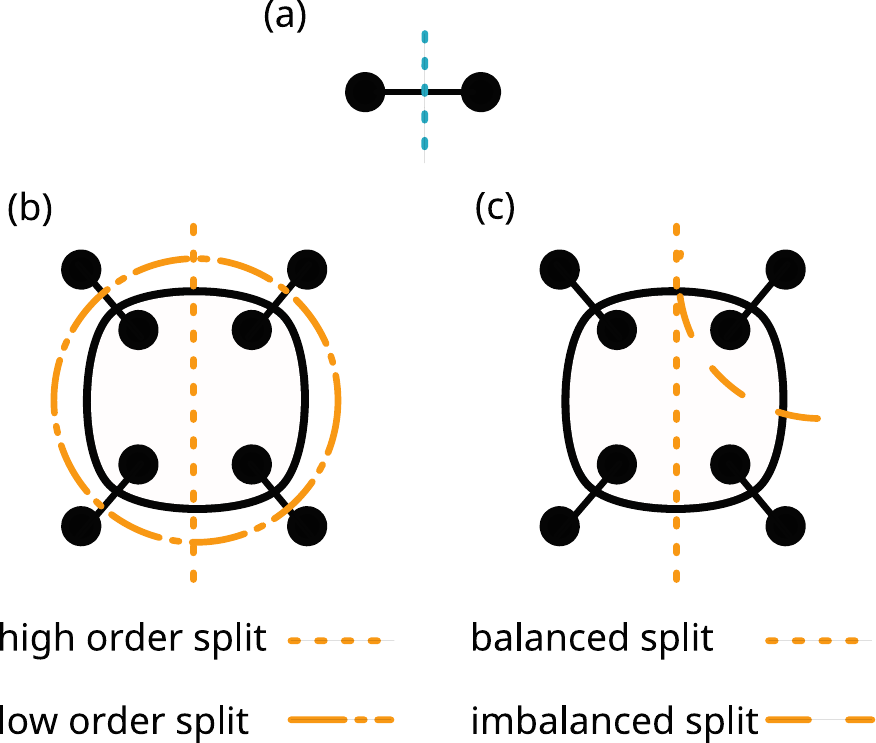}\\
	\caption{\small\it \textbf{Different configurations of community structure in graphs and hypergraphs.}
    (a) In dyadic networks, there is a single type of edge connecting pairs of nodes and edges can only be split across communities in one way.
    (b) In hypergraphs, hyperedges of different orders connect varying numbers of nodes and we must choose which order of hyperedge to split across communities, e.g., do we prefer to preserve the higher order 4-hyperedges or the lower order 2-hyperedges?
    (c) Hyperedges can connect more than two nodes and can be split across communities in multiple ways. Here, we consider two different shapes of a 4-hyperedge, a balanced shape and an imbalanced shape.
    \label{fig_1conceptualIllustration}
    }
\end{figure}

Transferring the notion of community structure from graphs to hypergraphs is far from straightforward.
In dyadic networks community structure can be largely described by edge density: the strength of community structure corresponds to the relative difference in the density of edges within communities compared to across communities.
Hyperedges connect varying numbers of nodes, and these multiway interactions can couple nodes from different communities in a variety of configurations. 
Figure~\ref{fig_1conceptualIllustration} illustrates some examples of these different configurations in community partitions in graphs and hypergraphs.
The resulting structural heterogeneity has no direct analogue in graphs. This implies that any community detection algorithm, regardless of its specific mechanism, must make a choice regarding which hyperedge orders and ``shapes'' (i.e., how hyperedges split across communities) to prioritize or preserve in the resulting partition. This leads to fundamental questions: how do these choices jointly influence the recoverability of planted structure, and how can we systematically characterize the resulting trade-offs?
Unlike the graph setting, where detectability and algorithmic optimality are now relatively well understood, the theoretical understanding of community detection in hypergraphs remains much less developed.

In particular, higher-order interactions introduce new trade-offs in describing strength of community structure that have not been characterized systematically so far.
Existing work has focused on extending graph-based community detection methods to hypergraphs without explicitly addressing these higher-order trade-offs. While different algorithms may make different choices, we argue that understanding the nature and impact of these choices is essential for any application of community detection to hypergraphs.
The study of hypergraph analogues of modularity maximization~\cite{chodrow2021generative}, spectral methods based on hypergraph Laplacians~\cite{ghoshdastidar2017consistency}, non-backtracking operators~\cite{chodrow2023nonbacktracking}, and approximate-inference schemes derived from belief propagation~\cite{angelini2015spectral,ruggeri2024message} provide useful building blocks, but a more unified picture of detectability in general non-uniform hypergraphs is still lacking.

In this work, we take an important step toward such a picture by studying the detectability of community structure in hypergraphs under the Hyper Stochastic Block Model (HSBM)~\cite{angelini2015spectral,chodrow2023nonbacktracking}.
In the HSBM nodes are assigned to latent communities and hyperedges of various orders are generated according to community-dependent probabilities.
This framework generalizes the SBM to higher-order interactions and provides a clear link between observed hypergraph structure and latent communities, enabling principled evaluation of community detection methods via their ability to recover planted structure.

Importantly, the HSBM also allows us to formulate and analyze detectability of communities in hypergraphs.
Previous work has derived BP-based detectability thresholds for uniform HSBMs, where all hyperedges have the same order~\cite{angelini2015spectral}, and for specific cases of non-uniform settings~\cite{chodrow2023nonbacktracking,ruggeri2024message}.
Here we generalize these results by introducing a unified signal-to-noise ratio (SNR) for hypergraphs that applies to arbitrary mixtures of hyperedge orders and any number of communities.
This SNR quantifies the relative strength of different hyperedge orders and shapes, and thereby provides an important tool to analyze the novel trade-offs that arise in hypergraph communities.

Our algorithmic approach builds on spectral clustering via the Bethe Hessian (BH)~\cite{saade2014spectral}.
We derive a BH operator for non-uniform hypergraphs that serves as a symmetric, $n\times n$ linearization of belief propagation for the HSBM and retains a spectral correspondence to the hypergraph non-backtracking operator~\cite{chodrow2023nonbacktracking,stephan2024sparse}.
This correspondence inherits principled model-order selection via negative BH eigenvalues and provides node-level embeddings via BH eigenvectors, while offering substantial computational advantages over non-backtracking operators.
Using our BH formulation, we characterize a spectral detectability threshold in terms of our SNR and compare it to a more general theoretical detectability limit derived for the same HSBM family using belief propagation.
We show that the two thresholds coincide in the uniform case, implying that BH-based spectral clustering reaches the theoretical limit, whereas for non-uniform HSBMs the BH spectral threshold is strictly weaker.

Synthetic experiments confirm these analytical predictions and further highlight systematic biases in BH-based clustering according to the order and shape of hyperedges, which we can explain in terms of our analytical SNR framework.
In particular, we identify situations where partitions favored by higher-order or more balanced shape hyperedges dominate the detected community structure, even when alternative partitions are equally or more strongly supported by lower-order interactions.
These phenomena exemplify detectability trade-offs that are unique to higher-order networks.
Finally, we apply our method to empirical hypergraphs, demonstrating its scalability, interpretability, and the practical relevance of these higher-order trade-offs in real-world data.

\section*{Community detection and detectability limits for hypergraphs}\label{sec:scbh-hyper}
Consider a hypergraph $\rm G_H$ with a set of nodes $\mathcal{V}=\{1,\ldots, n\}$, and a set of hyperedges $\mathcal{E} \subseteq \{e_j | e_j \subseteq \mathcal{V}, j=1,\ldots, m\}$ containing $|\mathcal{E}|=m$ hyperedges.
We can represent such a hypergraph using an incidence matrix $\bm{H} \in \left\{0, 1\right\}^{n\times m}$, whose rows are indexed by the nodes and columns are indexed by the hyperedges~\cite{ghoshdastidar2017consistency}.
Specifically, for a node $i\in \mathcal{V}$ and a hyperedge $e_j\in \mathcal{E}$, the value $\bm{H}_{ij}$ is
\begin{equation}
    \bm{H}_{ij} = \left\{
    \begin{matrix}
        1 & i\in e_j \\
        0 & \textrm{otherwise}
    \end{matrix}
    \right. \enspace .
\end{equation}

The degree $d_i = \sum_j \bm{H}_{ij}$ of a node $i$ is the total number of hyperedges that include node $i$ and the order of a hyperedge $|e_j| = \sum_i \bm{H}_{ij}$ is the number of nodes involved in a hyperedge $e_j$.
We refer to a hyperedge with order $|e|=\kappa$ as a $\kappa$-hyperedge or a hyperedge of order $\kappa$.
If every hyperedge in a hypergraph is a $\kappa$-hyperedge, the hypergraph is called a $\kappa$-uniform hypergraph. 
Any dyadic network (graph) can be viewed as 2-uniform hypergraph.
Hypergraphs with mixed order hyperedges are referred to as non-uniform hypergraphs. 
In non-uniform hypergraphs, we can distinguish the hyperedges by their order.
We denote the set of all $\kappa$-hyperedges as $\mathcal{E}^{(\kappa)}$, and the number of $\kappa$-hyperedges as $m^{(\kappa)}=|\mathcal{E}^{(\kappa)}|$.

\subsection*{The general Bethe Hessian for heterogeneous hypergraphs}
In this work, we consider a spectral approach for the detection of communities in hypergraphs based on the so-called Bethe-Hessian matrix for hypergraphs.

Specifically, the Bethe Hessian matrix for non-uniform hypergraphs, which can be derived from belief propagation and its linearization in terms of non-backtracking matrices, is defined as follows. 
\begin{equation}\label{nonuniformBH}
    \bm{B}^{(\mathcal{K})}_{\eta}=\bm{I}-\sum_{\kappa\in \mathcal{K}}
    \frac{(\kappa-1)}{(1-\eta)(\eta+\kappa-1)}\bm{D}^{(\kappa)} + \sum_{\kappa\in \calK}\frac{\eta}{(1-\eta)(\eta+\kappa-1)}\bm{A}^{(\kappa)}\enspace ,
\end{equation}
where $\mathcal{K}$ denotes the set of all hyperedge orders present in the hypergraph (formally defined in the next section), $\eta>0$ is a regularization parameter, $\bm{D}^{(\kappa)}$ is the diagonal degree matrix for $\kappa$-hyperedges, and $\bm{A}^{(\kappa)}$ is the one-mode projection matrix for $\kappa$-hyperedges, i.e., 
\begin{equation}
    \bm{A}^{(\kappa)}_{ij}=\left\{
    \begin{matrix}
        \left|\left\{e \mid i,j\in e \text{ and }|e|=\kappa\right\}\right | & i\neq j \\
        0 & i=j
    \end{matrix}
    \right. \enspace .
\end{equation}

Note that the matrix $\bm{A}^{(\kappa)}$ can also be expressed in terms of the $\kappa$-hyperedge incidence matrix $\bm{H}^{(\kappa)}$ as follows:
\begin{equation}
    \bm{A}^{(\kappa)}=\bm{H}^{(\kappa)}(\bm{H}^{(\kappa)})^T-\mathrm{diag}(\bm{H}^{(\kappa)}(\bm{H}^{(\kappa)})^T) \enspace .
\end{equation}

We remark that the Bethe Hessian has many desirable properties and deep connections to a variety of quantities and methods previously considered in the literature (see, e.g.,~\cite{saade2014spectral,angelini2015spectral,chodrow2023nonbacktracking,ruggeri2024message}).
We here consider the Bethe Hessian in its most general form in terms of heterogeneous hypergraphs, which generalizes prior works.
The derivation of the Bethe Hessian as well as its relations to the prior literature are discussed in the Appendix~\ref{app:nb2bh_hypergraph}.

\subsection*{The hyper stochastic block model}\label{hsbm}
The hyper stochastic block model (HSBM) is a generative model for uniform~\cite{ghoshdastidar2014consistency, angelini2015spectral} and non-uniform hypergraphs~\cite{ghoshdastidar2017consistency} with community structure, and is a generalization of the stochastic block model (SBM)~\cite{holland1983stochastic} for dyadic networks (see Appendix~\ref{app:sbm} for further details and definitions related to the SBM). 

We can use the HSBM to generate a hypergraph $\rm G_H$ with $q$ communities as follows.
We begin by assigning each node $i\in \mathcal{V}$ a community label $\psi_i\in \left\{1, 2, ..., q\right\}$.
We define the set $\mathcal{K}$ that collects all the possible hyperedge orders in our hypergraph.
If $\mathcal{K}$ only has one element, then the HSBM generates a uniform hypergraph, otherwise it generates a non-uniform hypergraph. 
For each $\kappa\in \mathcal{K}$, the probability of generating a $\kappa$-hyperedge is described by a $\kappa$-order tensor $\bm{\Omega}^{(\kappa)}\in [0, 1]^{q^\kappa}$,
i.e., for any set of $\kappa$ nodes $\{i, j, ..., \kappa\}$, the probability of generating a $\kappa$-hyperedge is given by $\bm{\Omega}^{(\kappa)}_{\psi_{i}, \psi_{j}, ..., \psi_{k}}$.

For a fixed number of nodes $n$ and $\kappa\in \mathcal{K}$, the number of possible $\kappa$-hyperedges is $\binom{n}{\kappa}$.
If we assume that hypergraphs are sparse, in the same way that many empirical dyadic networks are, then the number of $\kappa$-hyperedges should be $m^{(\kappa)}\sim \mathrm{O}(n)$ as $n\rightarrow \infty$. 
We can then reparameterize the hyperedge connectivity within and between communities, for hyperedges of order $\kappa$:
\begin{equation}\label{Cforsparse}
    \bm{\Omega}^{(\kappa)}=\frac{\bm{C}^{(\kappa)}}{n^{\kappa-1}} \enspace ,
\end{equation} 
where $\bm{C}^{(\kappa)}$ is a $\kappa$-order tensor that is constant with respect to $n$. 

For convenience, we define a symmetric HSBM for both uniform and nonuniform case in which all nodes have equal probability $1/q$ of being assigned to any of the $q$ communities and for any $\kappa\in \mathcal{K}$, the elements of tensor $\bm{C}^{(\kappa)}$ satisfy:
\begin{equation}\label{nonuniform_cincout}
   \bm{C}^{(\kappa)}_{\bm{z}} = \left\{
    \begin{matrix}
        c_{\rm in} & \text{if} \; \bm{z}_i=\bm{z}_j=...=\bm{z}_{\kappa} \\
        c_{\rm out} & \text{otherwise}
    \end{matrix}
   \right.\enspace ,
\end{equation}
where $\bm{z}$ represent the $\kappa$-dimensional vector of community indices. Note that in this symmetric HSBM, the parameters $\cin$ and $\cout$ are the same for all hyperedge orders $\kappa \in \mathcal{K}$, which simplifies analysis; the case of order-dependent parameters $c_{\rm in}^{(\kappa)}, c_{\rm out}^{(\kappa)}$ is a natural extension discussed in Appendix~\ref{app:sbm}.

\subsection*{Spectral clustering with the Bethe Hessian for hypergraphs}\label{sec:spectral_clustering_hyper}

Given a (possibly non-uniform) hypergraph $\mathrm{G_H}$ and its Bethe Hessian $\bm{B}^{(\mathcal{K})}_{\eta}$ from Eq.~\eqref{nonuniformBH}, we detect communities by spectral clustering on $\bm{B}^{(\mathcal{K})}_{\eta}$. 
As in the graph case~\cite{saade2014spectral}, informative eigenvalues of the non-backtracking operator correspond to negative eigenvalues of $\bm{B}^{(\mathcal{K})}_{\eta}$ when $\eta$ is chosen appropriately. 
Thus, the number of negative eigenvalues estimates the number of communities, and the associated eigenvectors provide an informative embedding of the nodes.

We now partition the hypergraph nodes according to the following procedure~\footnote{Code is available in \url{https://github.com/eggplantisme/HyperGraphBetheHessian}}:
\begin{enumerate}
    \item Choose a regularization parameter $\eta>0$. In practice, a simple degree-based rule works well: for uniform hypergraphs set $\eta \approx \sqrt{d^{(\kappa)}(\kappa-1)}$; for non-uniform hypergraphs set $\eta \approx \sum_{\kappa\in\mathcal{K}}\sqrt{d^{(\kappa)}(\kappa-1)}$ (the bulk radius of the non-backtracking spectrum; see Appendix~\ref{app:spectrum_hyperbh}).
    \item Let $\lambda_1(\eta)\leq \dots \leq \lambda_n(\eta)$ be the eigenvalues of $\bm{B}^{(\mathcal{K})}_{\eta}$. Set $\hat{q} = \big|\{i : \lambda_i(\eta) < 0\}\big|$ as the estimated number of communities. Note that $\hat{q}$ is a data-driven suggestion; one may also specify $q$ directly (e.g., if a target number of communities is known), in which case use the $q$ eigenvectors corresponding to the most negative eigenvalues.
    \item Form the matrix $\bm{U}\in\mathbb{R}^{n\times \hat{q}}$ whose columns are the eigenvectors corresponding to the $\hat{q}$ most negative eigenvalues, and apply $k$-means with $\hat{q}$ clusters to the rows of $\bm{U}$ to obtain community labels $\hat{\psi}_i$.
\end{enumerate}

This procedure generalizes the Bethe Hessian spectral clustering for graphs~\cite{saade2014spectral} and uniform hypergraphs~\cite{angelini2015spectral} to non-uniform hypergraphs.

\paragraph{Computational complexity.}
Computing $k$ eigenvectors of $\bm{B}^{(\mathcal{K})}_\eta$ via the Lanczos method costs $O(k(n + \mathrm{nnz}))$ time and $O(n + \mathrm{nnz})$ space, where $\mathrm{nnz} = \sum_\kappa m^{(\kappa)}$ is the total number of hyperedges across all orders. This is the same asymptotic cost as the hypergraph Laplacian, but substantially lower than the non-backtracking operator, which requires $O(k\sum_\kappa m^{(\kappa)}(\kappa-1)^2)$ time and space. Belief propagation, while achieving the theoretical detectability limit, additionally requires knowledge of the model parameters and costs $O(T \sum_\kappa m^{(\kappa)}\kappa^2)$ per iteration. A detailed complexity comparison is provided in Appendix~\ref{appendix-n&nnz_NB_BH}.

\section*{Generalized detectability limits for hypergraph community structure}\label{detectability_hsbm}
The detectability limit for community structure describes the boundary below which no algorithm can succeed in recovering the true community structure with accuracy better than random guessing.
Detectability limits of community structure in dyadic networks have been extensively studied in the literature~\cite{decelle2011asymptotic, mossel2015reconstruction, massoulie2014community, abbe2018community}.
In contrast, the theoretical understanding detectability in hypergraphs has received comparatively less attention.
Early work by Angelini et al.~\cite{angelini2015spectral} first analyzed the detectability limit of the uniform symmetric HSBM, employing a theoretical approach based on belief propagation (BP), similar to the analysis in dyadic networks~\cite{decelle2011asymptotic}.
Subsequently, Chodrow et al.~\cite{chodrow2023nonbacktracking} analyzed the spectral features of both the non-backtracking matrix (NB) and the Jacobian matrix of belief propagation (BP Jacobian). 
They established two limits related to the spectral clustering based on these two matrices for the non-uniform symmetric HSBM. However, their analysis was restricted to the case of only $q=2$ communities.
Chodrow et al. concluded that the limit derived from the NB is weaker than the limit from the BP Jacobian.
More recently, the work by Ruggeri et al. \cite{ruggeri2024message} derived the theoretical detectability limit using BP for a specialized variant called HySBM, a model structurally constrained by deriving all $\kappa$-order affinity tensors from a single underlying probability matrix, thereby limiting the independent modeling of different hyperedge order probabilities.

Here, we significantly extend these contributions.
First, we reparameterize and generalize the spectral detectability limit, build upon the work of Angelini et al.~\cite{angelini2015spectral} and Chodrow et al.~\cite{chodrow2023nonbacktracking} for the non-uniform symmetric HSBM with an arbitrary number of communities.
Second, we employ a methodology similar to that of Ruggeri et al.~\cite{ruggeri2024message} to derive the theoretical detectability limit with BP for the non-uniform symmetric HSBM. 
Finally, by comparing these two derived theoretical boundaries, we conclude that the spectral clustering limit is precisely equal to the theoretical limit in the uniform HSBM case, but is demonstrably weaker in the more complex non-uniform HSBM case.

\paragraph{Detectability limit for spectral methods on general hypergraphs}
We can describe the detectability limit for spectral clustering with the Bethe Hessian in terms of the average $\kappa$-in-degree $\din^{(\kappa)}$ and $\kappa$-out-degree $\dout^{(\kappa)}$.
To this end, we define the average $\kappa$-in-degree $\din^{(\kappa)}$ as the expected number of $\kappa$-hyperedges incident on a node in which all $\kappa$ nodes belong exclusively to the same community. 
Analogously, the average $\kappa$-out-degree $\dout^{(\kappa)}$ is the expected number of hyperedges that connect to at least one node belonging to a different community. 
Now, consider a network generated from a symmetric HSBM with $q$ equal-sized communities and symmetric tensor $\bm{C}$ satisfying~\eqref{nonuniform_cincout}.
Then, $\din^{(\kappa)}$ and $\dout^{(\kappa)}$ can be computed as (See formulas~\eqref{din^k} and~\eqref{dout^k} in Appendix~\ref{append:detectability_literature} for details)
\begin{equation}\label{dindout_cincout}
        \din^{(\kappa)}=\frac{\cin}{q^{\kappa-1}(\kappa-1)!} \qquad
        \dout^{(\kappa)}=\frac{\cout}{q^{\kappa-1}(\kappa-1)!} \enspace,
\end{equation}
and the average $\kappa$-degree for the node is
\begin{equation}\label{d_kappa}
    \begin{aligned}
        d^{(\kappa)}
        &=\din^{(\kappa)}+(q^{\kappa-1}-1)\dout^{(\kappa)}\enspace.
    \end{aligned}
\end{equation}

The detectability limits for spectral methods previously derived in the literature~\cite{angelini2015spectral, chodrow2023nonbacktracking} can now be generalized to the non-uniform symmetric HSBM with any $q$ as follows (for details see Appendix~\ref{append:detectability_literature}):
\begin{equation}\label{phi_by_dindout}
    \begin{aligned}
        \snr_{\rm BH}:=\frac{\left(\sum_{\kappa\in\mathcal{K}}(\kappa-1)(\din^{(\kappa)}- \dout^{(\kappa)})\right)^2}{\sum_{\kappa\in\mathcal{K}}(\kappa-1)d^{(\kappa)}}=1
    \end{aligned}\enspace .
\end{equation}

\paragraph{Detectability limit in general hypergraphs}
We also derive the theoretical detectability limit with BP for non-uniform symmetric HSBM, employing a similar methodology as in the work by Ruggeri et al.~\cite{ruggeri2024message}.
To establish the general detectability limit, we first determine two essential characteristics of the network structure: the average degree $d$ of a node and the average order $\hat{\kappa}$ of a hyperedge. 
Both  the average degree $d$ and the average order $\hat{\kappa}$ can be expressed based
on the order-$\kappa$ average degree $d^{(\kappa)}$ given in~\eqref{d_kappa}.
The average degree $d$ is directly calculated by summing the average degrees across all hyperedge orders:
\begin{equation}\label{averaged}
    d=\sum_{\kappa\in \mathcal{K}}d^{(\kappa)} \enspace ,
\end{equation}
where the average order $\hat{\kappa}$ can be expressed as (See Appendix~\ref{appendix-stabilityBPfixedpoint} for details):
\begin{equation}
    \hat{\kappa}=\frac{d}{\sum_{\kappa\in \calK}\frac{d^{(\kappa)}}{\kappa}}\enspace .
\end{equation}

Now the detectability for belief propagation on the non-uniform symmetric HSBM can be explicitly expressed as:
\begin{equation}\label{detectability_bp}
     \snr_{\rm BP}:=d(\hat{\kappa}-1)\prod_{\kappa\in \mathcal{K}}\left(\frac{\din^{(\kappa)}-\dout^{(\kappa)}}{d^{(\kappa)}}\right)^{2\frac{\hat{\kappa}d^{(\kappa)}}{d\kappa}}=1 \enspace .
\end{equation}

Analogously, to prior works~\cite{decelle2011asymptotic}, we posit that this represents the theoretical detectability limit for
non-uniform hypergraphs with $q$ communities. 
For simplicity, the detailed derivation of~\eqref{detectability_bp} is deferred to the Appendix~\ref{appendix-bp&detectability}.

In the uniform HSBM, The $\snr_{\rm BP}$~\eqref{detectability_bp} will degenerate to $\snr_{\rm BH}$~\eqref{phi_by_dindout}.
That means that in the uniform case, spectral clustering (e.g., using the Bethe Hessian or the non-backtracking matrix) can detect community structure down to the same threshold as belief propagation. 
In the non-uniform case, however, the spectral detectability threshold is \emph{weaker} than the BP threshold. Here, ``weaker detectability'' means that the spectral method requires a higher signal-to-noise ratio to succeed: at the BP threshold ($\snr_{\rm BP}=1$), the BH has $\snr_{\rm BH}<1$ and therefore cannot detect communities better than random guessing, while BP can still succeed. Equivalently, the BH threshold $\epsilon_{\rm BH}^*$ is strictly smaller than $\epsilon_{\rm BP}^*$, meaning there is a regime in which BP detects structure but the BH spectral method does not. 
It is important to note that being above this threshold does not guarantee the detection of all community structures; instead, it ensures that the recovered partition is correlated with the true structure, which can include, for example, a coarser hierarchical partition or the detection of dominant (majority) communities~\cite{peel2024detectability, li2026detectability}.

We validate the thresholds $\snr_{\rm BH}=1$ and $\snr_{\rm BP}=1$ empirically in the following experiment on synthetic hypergraphs.
We fix the average degree $d$ of a hypergraph and vary the ratio $\epsilon=\frac{\cout}{\cin}$ to control the community strength. 
When $\epsilon$ is small, the community structure is strong, and when $\epsilon$ is large, the community structure is weak. 
We denote the critical $\epsilon$ for which $\snr_{\rm BH}=1$ as $\epsilon_{\rm BH}^*$, and the critical $\epsilon$ that make $\snr_{\rm BP}=1$ as $\epsilon_{\rm BP}^*$.
We use the adjusted mutual information $\rm AMI$~\cite{vinh2009information} to evaluate the community detection result. 
The $\rm AMI$ score lies in range $[0, 1]$, where $1$ indicates perfect agreement between the detected communities and the true communities, while $0$ indicates that the detected communities are no better than random guessing.

\begin{figure}
    \centering
    \includegraphics[width=15cm]{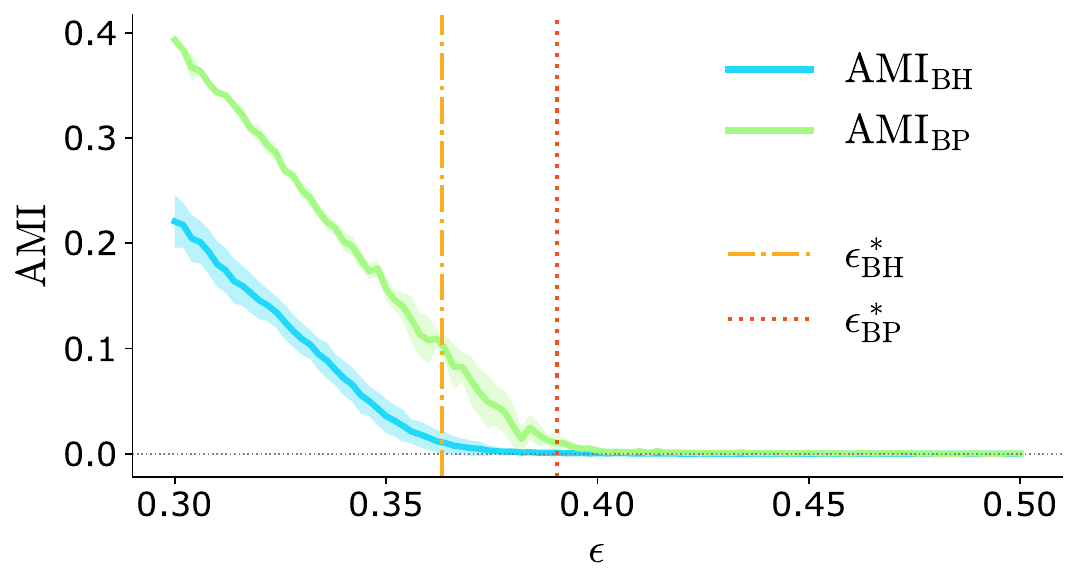}\\
	\caption{\small\it \textbf{Spectral methods exhibit a weaker detectability limit in non-uniform hypergraphs.}
    An experiment with nonuniform hypergraphs ($n=30,000$ nodes, $q=3$ communities, $\calK=\{2,3\}$, mean degree $d=10$). For each $\epsilon$, we generate independent HSBM samples and run BH spectral clustering 100 times (with random initialisation) and BP 5 times; solid lines show means and shaded bands show 95\% confidence intervals. BH runs use random Lanczos initialisation; BP is deterministic given the graph, so trials correspond to independent graph samples. The yellow and red vertical dashed lines show $\epsilon_{\rm BH}^*$ and $\epsilon_{\rm BP}^*$ respectively, where $\snr_{\rm BH}=1$ and $\snr_{\rm BP}=1$. \label{fig_detectability_exp}
    }
\end{figure}

In Figure~\ref{fig_detectability_exp} we see the results of experiments conducted on non-uniform hypergraphs with $q=3$ communities.
We see that $\epsilon_{\rm BH}^*$ indicates well the point that $\rm AMI_{BH}$ transitions from zero to nonzero. Similarly, $\epsilon_{\rm BP}^*$ identifies the transition of $\rm AMI_{BP}$ from zero to nonzero.

\paragraph{Using the SNR in practice.}
On synthetic HSBM data, $\snr_{\rm BH}$ and $\snr_{\rm BP}$ are computed directly from the model parameters. On real data, one may obtain an empirical SNR estimate post-hoc: first obtain community assignments via spectral clustering, then compute $\din^{(\kappa)}$ and $\dout^{(\kappa)}$ from the partition (counting within- and across-community hyperedges per order). The resulting empirical SNR serves as a diagnostic heuristic: $\snr \gg 1$ indicates robust structure; $\snr \approx 1$ signals proximity to detectability boundary; $\snr \ll 1$ suggests cautious interpretation. This heuristic does not require knowledge of ground-truth communities and complements the qualitative analysis of real-data partitions.

\section*{Additional trade-offs for detecting hypergraph community structure}

In dyadic networks, community structure is most commonly characterized by \emph{assortativity}: a higher density of edges within communities compared to the edge density between communities. (Disassortative structure, where edges are more frequent between communities, is also possible but less common in practice.) Detecting communities in this assortative setting involves identifying a partition that maximizes within-community edge density while minimizing cross-community edges.

However, for higher-order networks, even this relatively simple density-based notion of community structure is less straightforward to conceptualize, due to the fact that hyperedges can connect more than two nodes and that hyperedges within the same hypergraph can vary in size.
For instance, consider the case of detecting a partition of a network into two communities. 
An edge connecting two nodes that crosses a community boundary will always have one node in each community. 
For a hyperedge connecting three nodes that crosses a community boundary, there will be one node in one community and two nodes in another community. 
However, for a hyperedge connecting four nodes that crosses a community boundary, there are two possible configurations: it can connect one node in one community with three nodes in another community, or it can connect two nodes in each community.
This observation raises the question of the relationship between these different configurations and the strength of community structure in hypergraphs.
To probe this relationship further, we conduct experiments on hypergraphs in which we plant two ``competing'' community structures and see which one is easier to detect. 

\subsection*{Shape preference in hypergraph community structure}
\begin{figure}
    \centering
    \includegraphics[width=15cm]{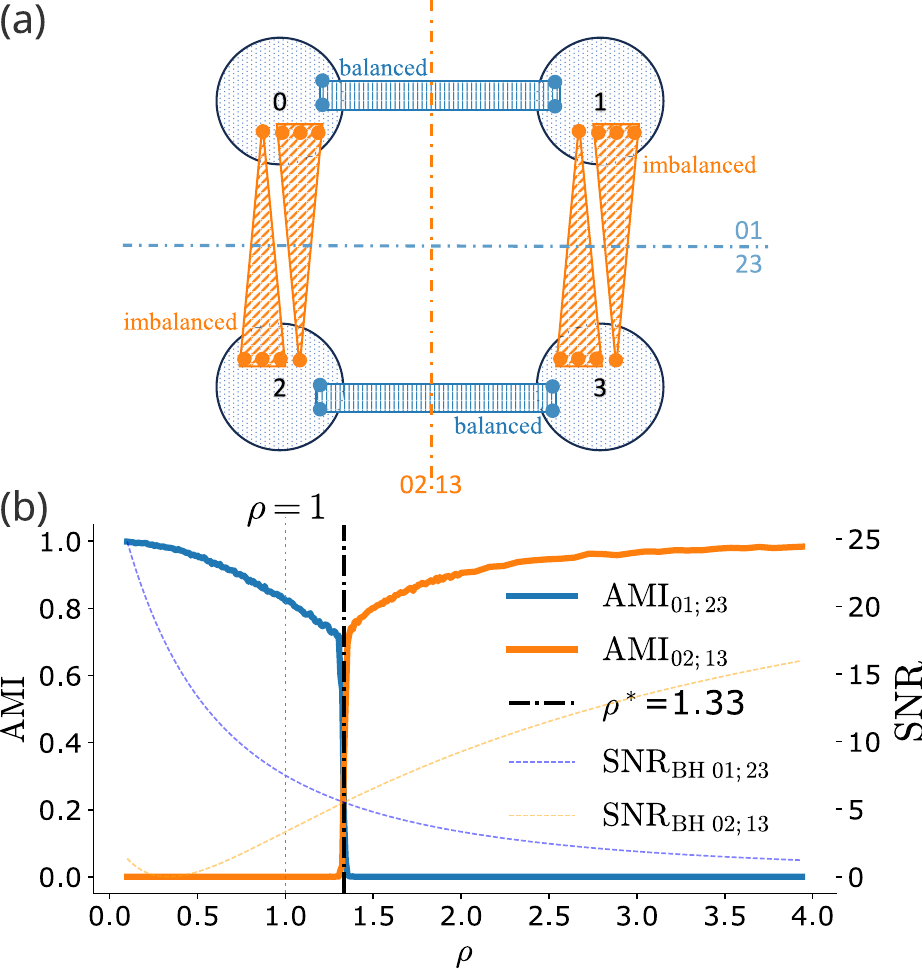}\\
	\caption{\small\it \textbf{Hyperedge shape preferences in community detection.}
    (a) An illustration of the planted community structure in hypergraphs used in experiments to explore the effect of hyperedge shape. All the hyperedges are 4-order and connect 2 of the 4 communities. The \emph{balanced} hyperedges are represented by rectangle shapes between communities $\left\{0, 1\right\}$ or $\left\{2, 3\right\}$. The  \emph{imbalanced} hyperedges are represented by triangle shapes between communities $\left\{0, 2\right\}$ or $\left\{1, 3\right\}$. (b) Experimental results with $n=8000$ nodes, $d=10$ mean node degree, and uniform 4-order hyperedges. We plot $\mathrm{AMI}_{01;23}$ and $\mathrm{AMI}_{02;13}$ as we vary $\rho$, the ratio of imbalanced to balanced hyperedges. The vertical dashed line at $\rho=1$ marks equal numbers of each type. The plot shows that at $\rho=1$ the partition $01;23$ (balanced split) is preferred, i.e., the algorithm tends to split balanced hyperedges and preserve imbalanced ones (3 nodes in one community, 1 in the other).  \label{fig_shapeeffect4order}}
\end{figure}

We first explore how different configurations of hyperedges crossing community boundaries affect the strength of community structure in hypergraphs.
We refer to these different configurations as the \emph{shape} of a hyperedge crossing community boundaries.
We consider uniform 4-order hypergraphs with two planted community structures that differ only in the shape of the hyperedges crossing community boundaries: \emph{balanced} hyperedges that connect two nodes in each community, and \emph{imbalanced} hyperedges that connect one node in one community with three nodes in another community.
Figure~\ref{fig_shapeeffect4order}(a) illustrates the experimental setup. 
Each hypergraph has 4 planted communities, where hyperedges connecting communities $\{0, 1\}$ and $\{2, 3\}$ are balanced, while hyperedges connecting communities $\{0, 2\}$ and $\{1, 3\}$ are imbalanced. We then apply spectral clustering with $q=2$ (deliberately below the true $q=4$) to force the algorithm to choose between the two competing partitions: $\{0,1\}|\{2,3\}$ vs.\ $\{0,2\}|\{1,3\}$. Note that the BH has 4 negative eigenvalues in this setting, but we manually override $q$ to 2 in order to probe relative detectability. We vary the ratio $\rho = \frac{m^{\rm imbalanced}}{m^{\rm balanced}}$ of imbalanced to balanced hyperedges.

Following the rationale of~\cite{peel2024detectability}, we predict the relative detectability of the two configurations by calculating which configuration yields a higher value for $\snr_{\rm BH}$~\eqref{phi_by_dindout}. Figure~\ref{fig_shapeeffect4order}(b) shows the results of the experiment, where we plot the $\rm AMI$ of the detected two-community partition with respect to the planted balanced and imbalanced configurations as a function of $\rho$.
We find that the balanced configuration is easier to detect than the imbalanced one, i.e., there is a preference to split the hyperedges such that we have 3 nodes in one community and 1 node in another community, relative to splitting the hyperedges such that we have 2 nodes in each community. 
This result indicates that preserving more nodes of a hyperedge within the same community creates a stronger community structure.
The transition point where the two configurations are equally detectable aligns well with the prediction based on $\snr_{\rm BH}$ and in this case occurs at $\rho^*=1.33$, indicating that we require $33\%$ more imbalanced hyperedges to achieve the same signal-to-noise ratio as balanced hyperedges.

\subsection*{Order preference in hypergraph community structure}
Now we investigate how hyperedges of different orders crossing community boundaries affect the strength of community structure in hypergraphs.
Similar to the previous experiment, we consider non-uniform hypergraphs with planted community structures that differ only in the order of hyperedges crossing community boundaries: low-order hyperedges of order $\kappa$ and high-order hyperedges of order $\kappa^*$.
Each hypergraph has 4 planted communities, where hyperedges connecting communities $\{0, 1\}$ and $\{2, 3\}$ are of order $\kappa^*$, while hyperedges connecting communities $\{0, 2\}$ and $\{1, 3\}$ are of order $\kappa < \kappa^*$.
As before, we apply spectral clustering with $q=2$ to probe which of the two competing partitions is preferred. We vary the ratio $\rho=\frac{m^{(\kappa)}}{m^{(\kappa^*)}}$ of low-order to high-order hyperedges. 
Following the same rationale as before, we predict the relative signal-to-noise of the two configurations by calculating which configuration yields a higher value for $\snr_{\rm BH}$~\eqref{phi_by_dindout}.
This time, however, the difference in hyperedge orders means that we need to adjust $\snr_{\rm BH}$ accordingly.
In the case of non-uniform hypergraphs we find the signal-to-noise ratio for each configuration is equal when
\begin{equation}
    \frac{\snr_{\rm BH \; 02;13}}{\kappa+1}=\frac{\snr_{\rm BH \; 01;23}}{\kappa^*+1} \enspace .
\end{equation}

\begin{figure}
    \centering
    \includegraphics{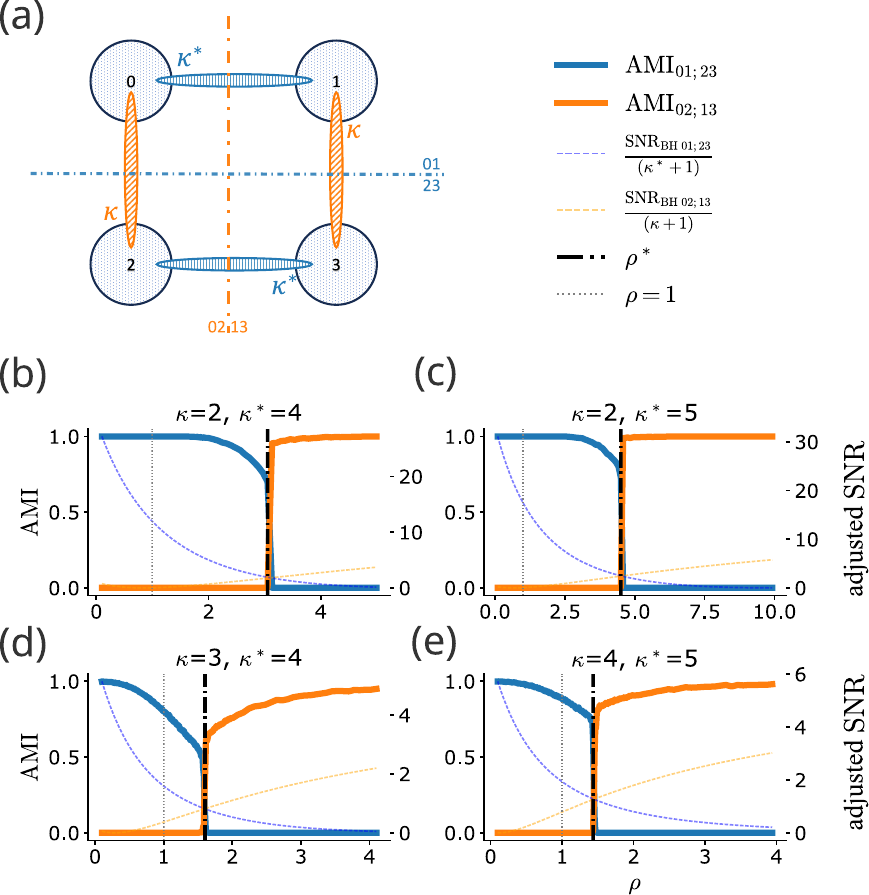}
    \caption{\small\it \textbf{Hyperedge order preferences in community detection.}
    (a) Synthetic hypergraphs with 4 planted communities and all the hyperedges between communities $\left\{0, 1\right\}$ or $\left\{2, 3\right\}$ are of order $\kappa^*$, the hyperedges between communities $\left\{0, 2\right\}$ or $\left\{1, 3\right\}$ are of order $\kappa$. 
    All hypergraphs have $n=8000$ nodes. 
    The average degree is either $d=50$ (b-c) or $d=10$ (d-e). 
    We see the $\mathrm{AMI}_{01;23}$ and $\mathrm{AMI}_{02;13}$ as we vary $\rho$ the ratio of high-order to low-order hyperedges.
    We observe a preference for  preserving higher-order $\kappa^*$-hyperedges  and splitting lower-order $\kappa$-hyperedges.  \label{fig_ordereffect}
    }
\end{figure}

Figure~\ref{fig_ordereffect} shows the results of the experiment, where we plot the $\rm AMI$ of the detected two-community partition with respect to the planted low-order and high-order configurations as a function of $\rho$.
We find that the high-order configuration is easier to detect than the low-order one, i.e., there is a preference to split the lower order hyperedges across community boundaries over splitting higher order ones.
This result indicates that hyperedges of lower order create a stronger community structure.

\section*{Experiments on empirical data}
We test the Bethe Hessian on some hypergraphs constructed from empirical data: two human contact interaction datasets and one user review dataset. 
The two human contact interaction datasets are from face-to-face interactions using wearable RFID tags in a primary school~\cite{stehle2011high} and a high school~\cite{Mastrandrea-2015-contact}.
Students and teachers are represented as nodes, and a hyperedge is constructed for each group of individuals who all have a recorded interaction with each other at a given time. 
The nodes have labels that indicate the school class in which they belong. 
The user review dataset is from the Yelp website for business reviews in the United States and Canada~\cite{yelp2025}.
The yelp dataset contains reviews for numerous businesses across 13 states in the United States and one province in Canada. 
We use nodes to represent the businesses and hyperedges to represent groups of businesses that are reviewed by the same user. 
We summarize the hypergraphs constructed from these empirical datasets in Table~\ref{empiricaltable}.
\begin{table}
\centering
\begin{tabular}{ccccc}
 Hyper Graph & Nodes & Hyperedges & Max Order &  \\ \hline
 primary school & 242 & 12704 & 5 &  \\
 high school & 327 & 7818 & 5 &  \\
 yelp & 150346 & 851921 & 3048 & 
\end{tabular}
\caption{Empirical Data}
\label{empiricaltable}
\end{table}

\subsection*{Human contact interaction}
\begin{figure}
    \centering
    \includegraphics[width=15cm]{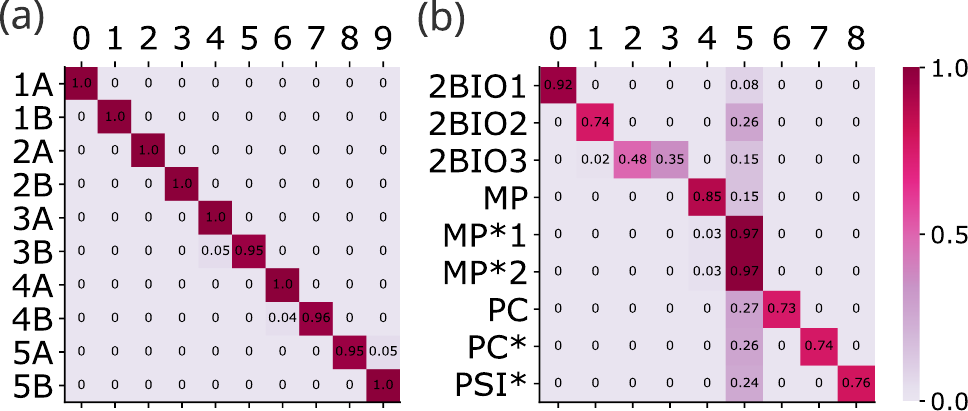}\\
	\caption{\small\it \textbf{Comparison between detected communities and class structure in human contact hypergraphs.}
    Confusion matrices of communities detected and school classes in hypergraphs of human contact interactions in (a) the primary school and (b) the high school. 
    The rows are normalized such that each entry represents the proportion of individuals in the class assigned to each community.
    \label{figPrimary}}
\end{figure}

First, we detect communities of the primary school~\footnote{\url{https://www.cs.cornell.edu/~arb/data/contact-primary-school-labeled/}} detected using spectral clustering with the Bethe Hessian and compare  with the school classes. 
To facilitate comparison, we manually set the number of communities to match the number of classes in the school.
Figure~\ref{figPrimary}(a) shows the confusion matrix between the detected communities and the classes.
We observe a high correspondence, with the vast majority of students ($\ge 95\%$) from each class assigned to a single corresponding community.
This result is similar to community detection results on dyadic networks constructed from the same primary school data~\cite{stehle2011high, tremblay2014graph}. 

Next we perform a similar comparison on the high school data~\footnote{https://www.cs.cornell.edu/~arb/data/contact-high-school-labeled/}.
Figure~\ref{figPrimary}(b) shows a more fragmented correspondence compared to the primary school data.
This result deviates from the results of community detection on dyadic networks constructed from the same high school data (e.g., using the Bethe Hessian~\cite{schaub2023hierarchical}), where the dense within-class connections dominate the mesoscale structure.
Instead, we observe that class 2BIO3 is split into two main communities (communities 2 and 3, with $48\%$ and $35\%$ of its members respectively), while classes MP*1 and MP*2 are largely concentrated in community~5~($97\%$).

\begin{figure}
    \centering
    \includegraphics[width=15cm]{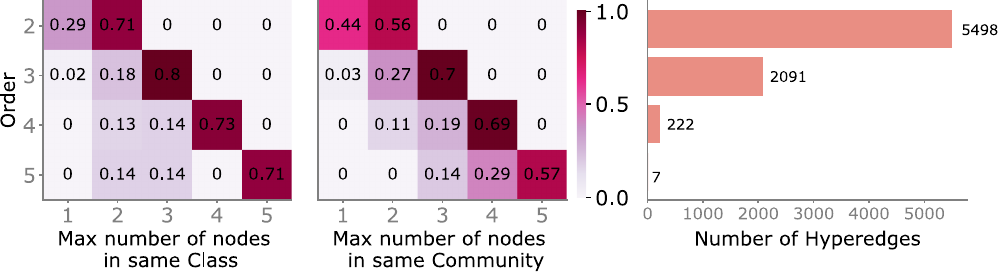}\\
	\caption{\small\it \textbf{Hyperedge distributions in the high school data.}
    The distribution of hyperedges by the maximum number of nodes in the same school class (left panel) and maximum number of nodes in the same detected community (middle panel). The frequency of hyperedges of each order (right panel). 
    \label{figHighDis}}
\end{figure}

One might be tempted to attribute this deviation from class structure and previous results on dyadic networks as a failure of our method.
However, it is important to consider that there may be multiple valid ways to partition the nodes in empirical data~\cite{peel2017ground}, and that we should expect a different community structure to emerge when we consider higher-order interactions.
To better understand the composition of these hypergraph communities, we analyze the distribution of hyperedges across communities in terms of their order and shape.
Figure~\ref{figHighDis} reveals that the detected communities are not simply composed of homogeneous hyperedges where all nodes belong to the same community.
Instead, we observe a preference for hyperedge shapes that preserve a higher number of nodes within a single community.
For example, for order-3 hyperedges, the fraction of imbalanced hyperedges that connect 2 nodes within a community and 1 outside increases from 18\% in the school classes to 27\% in the detected communities.
This quantifies the tendency of the algorithm to favor these near-homogeneous splitting shapes.
The fact that higher-order hyperedges are not always favored within communities is explained by the much larger proportion of lower-order hyperedges in the data (see right panel of Figure~\ref{figHighDis}).
These results align with our earlier findings on the preference for imbalanced and higher-order hyperedges in planted community structures.

\subsection*{User reviews of businesses}
\begin{figure}
    \centering
    \includegraphics[width=15cm]{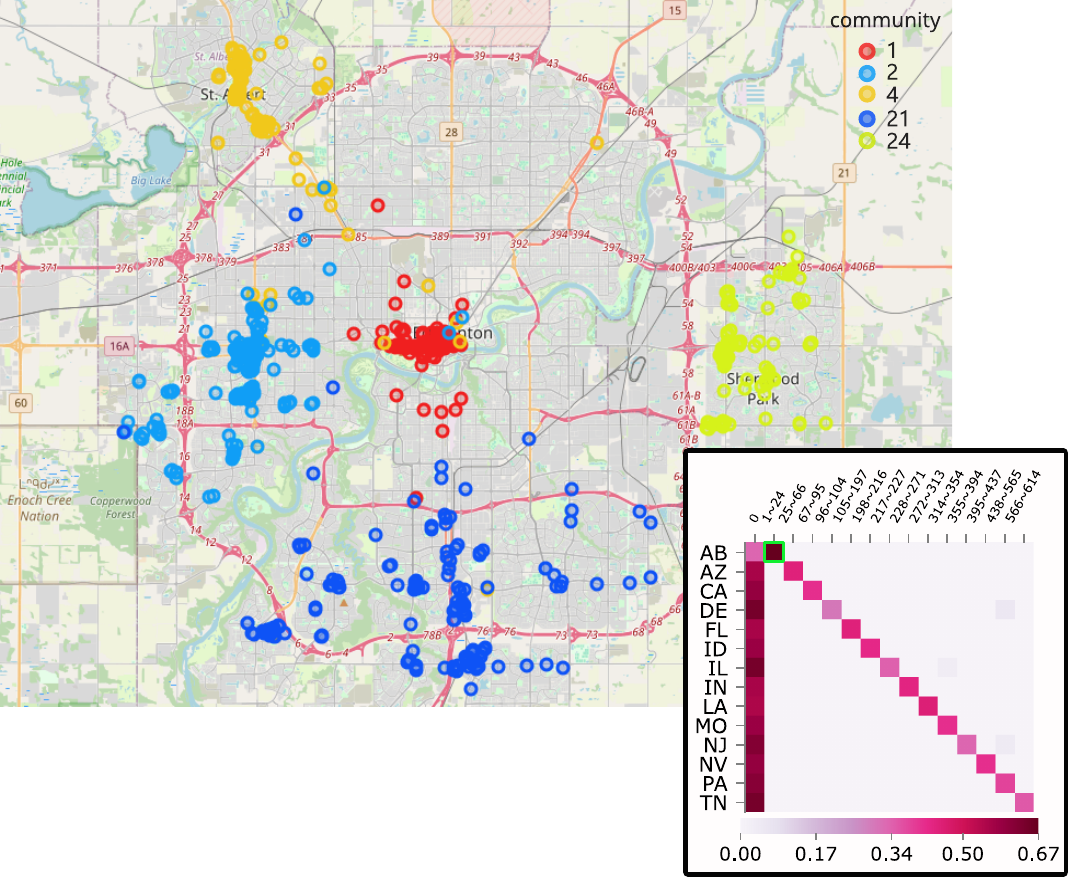}\\
	\caption{\small\it 
    \textbf{Geographical distribution of detected communities in the Yelp data.}
    The bottom right corner is the confusion matrix of communities detected by Bethe Hessian-based spectral clustering with state metadata for yelp data. We highlight some communities composed of business in the Alberta province (AB) and show their geographical distribution.\label{figYelpCM}}
\end{figure}

Finally, we apply our method for spectral hypergraph clustering with the Bethe Hessian to the Yelp dataset~\footnote{\url{https://business.yelp.com/data/resources/open-dataset/}}.
Here, we determine the number of communities according to the number of negative eigenvalues of $\mathrm{BH_{\eta}}$, resulting in the identification of 615 communities. 
Figure~\ref{figYelpCM}(inset) presents a confusion matrix between the detected communities and the states in which the businesses are located; for clarity, we merge communities primarily composed of businesses from the same state. 
We see that most of these communities predominantly consist of businesses located within a single state, except community 0 that includes businesses across all states because these businesses are reviewed by users who review businesses across multiple states. 
We also find geographical organization within states, indicating a hierarchical community structure~\cite{schaub2023hierarchical} that corresponds to different resolutions of geographical regions. 
Figure~\ref{figYelpCM} shows the geographical distribution of businesses in some of the detected communities in the Alberta province in Canada.
We see that community 1 is mainly located in the central region of Edmonton, community 2 is in the western region, community 21 is in the southern region, community 24 is in Sherwood Park, and community 4 is in St. Albert.

\section*{Discussion}
Community detection in hypergraphs presents fundamental challenges that have no direct analogue in dyadic networks.
While pairwise networks exhibit a relatively simple trade-off between within- and across-community edge densities, hypergraphs introduce structural heterogeneity through hyperedges of varying orders that can split across communities in multiple configurations.
Existing approaches to hypergraph community detection have largely overlooked a critical question: when hyperedges of different orders coexist and can be partitioned across communities in different ways, which structures are preferentially preserved by inference algorithms, and what are the implications for interpreting detected communities?
This work provides the first systematic characterization of these higher-order trade-offs, introducing a unified framework for understanding how hyperedge order and splitting shape fundamentally determine what community structure can be detected.

Our main theoretical contributions are a generalized signal-to-noise ratio that quantifies the detectability of communities in hypergraphs for arbitrary mixtures of hyperedge orders and shapes, and a corresponding Bethe Hessian operator for non-uniform hypergraphs that enables efficient spectral clustering.
Critically, this SNR framework reveals that the relative detectability of different community partitions depends not only on which hyperedge orders we choose to split across communities, but also on how we split them---a trade-off unique to higher-order networks that has not been considered before.
We show that spectral methods exhibit systematic biases toward preserving higher-order hyperedges and those with more balanced shapes within the same community, even when alternative partitions are equally or more strongly supported by lower-order interactions.
This finding has direct implications for interpreting communities in real-world hypergraphs: detected partitions reflect not just edge density, but also the order and shape preferences inherent in the detection method.
Through comparison with general theoretical limits, we establish that spectral methods based on the Bethe Hessian achieve optimal detectability for uniform hypergraphs but exhibit a weaker threshold in non-uniform settings.
Our synthetic experiments confirm these analytical predictions, while empirical applications demonstrate the practical relevance of these order and shape preferences: we recover known structure in human contact networks while revealing alternative organizational patterns that align with preferences for specific hyperedge configurations, and successfully identify hierarchical geographical communities in large-scale data with over 150,000 nodes and hyperedge orders exceeding 3,000.

While we frame these results in terms of hypergraphs, they apply equally to community detection in bipartite graphs, where hypergraphs can be represented via their incidence structure and hyperedge orders correspond to the degrees of one mode.
However, the hypergraph perspective provides a more natural lens for understanding these trade-offs, as the notions of order and splitting shape emerge directly from the multiway interaction structure rather than indirectly through degree heterogeneity in one mode of a bipartite network.
The connection between hyperedge orders and node degrees in bipartite graphs suggests that our framework may open new directions for defining detectability limits in degree-heterogeneous networks, a problem that has only been considered in graphs for the restricted cases of two communities~\cite{radicchi2013detectability} or when community structure correlates with degree~\cite{zhang2016community}.

Several directions warrant further investigation.
First, our current framework assumes homogeneous degree and order distributions within communities.
Extending the Bethe Hessian to incorporate degree and order heterogeneity, analogous to degree-corrected models for dyadic networks~\cite{dall2019revisiting}, would broaden applicability to real-world systems.
However, beyond developing such extensions, it remains an open empirical question whether degree and order correction are as relevant for hypergraph community detection as degree correction has proven to be for dyadic networks.
Second, while our analysis focuses on assortative community structure, disassortative structures in hypergraphs are less straightforward to characterize than in graphs.
How should we define disassortativity when hyperedges can split across communities in different shapes?
The notion of hyperedge shape introduced here may provide a natural lens for addressing this question, as different splitting patterns could encode various forms of disassortative organization.
Finally, rigorous proofs of our conjectured detectability thresholds and signal-to-noise ratio characterizations remain an important theoretical challenge, as does developing spectral methods that close the gap between the spectral and overal theoretical limits in non-uniform settings.

\section*{Acknowledgments}
JL acknowledges the financial support from the China Scholarship Council under grant number 202106380033. MTS acknowledges support from the European Union (ERC, HIGH-HOPeS, 101039827). Views and opinions expressed are however those of the author(s) only and do not necessarily reflect those of the European Union or the European Research Council Executive Agency. Neither the European Union nor the granting authority can be held responsible for them. LP was supported in part by the Dutch Research Council (NWO) Talent Programme ENW-Vidi 2021 under grant number VI.Vidi.213.163.

\printbibliography

\clearpage

\appendix
\setcounter{figure}{0}
\renewcommand{\thefigure}{S\arabic{figure}}

\section{Detectability limit and the Bethe Hessian for dyadic networks}\label{app:scbh_network}
In this section, we first introduce the Stochastic Block Model, a generative model for networks with community structure. Then we review the non-backtracking matrix and Bethe Hessian matrix used for spectral clustering. Lastly, we briefly review the detectability limit of SBM.

\subsection{The stochastic block model}\label{app:sbm}
We consider an unweighted, undirected dyadic network $G_A$ with node set $\mathcal{V}$ and edge set $\mathcal{E}$, where the number of nodes $|\mathcal{V}|=n$, number of edges $|\mathcal{E}|=m$, such that $G_A$ can be represented as an adjacency matrix $\bm{A}\in\left\{0, 1\right\}^{n\times n}$. For any two nodes $i, j\in \mathcal{V}$, $\bm A$ is defined as
\begin{equation}
    \bm{A}=\left\{
    \begin{matrix}
        1 & \text{if $(i, j)\in \mathcal{E}$} \\
        0 & \text{otherwise}
    \end{matrix}
    \right. \enspace .
\end{equation}

The degree of any node is defined by the number of edges connected to that node. For a node $i$, its degree $d_i$ is
\begin{equation}
d_i = \sum_{j\in \mathcal{V}}\bm{A}_{ij} \enspace ,
\end{equation}
and the average degree of whole network as $d$, then 
\begin{equation}
    d=\frac{\sum_{i\in \mathcal{V}}d_i}{n}=\frac{2m}{n} \enspace .
\end{equation}

We can use the SBM to generate network $G_A$ with $q$ communities. We assign each node $i\in \mathcal{V}$ a community label $\psi_i\in\left\{1, 2, ..., q\right\}$. Let matrix $\bm{\Omega}$ be the edge probability matrix, $\bm{\Omega}\in[0, 1]^{q\times q}$. For each pair of nodes $i, j\in \mathcal{V}$, we generate an edge $(i, j)$ with probability $\bm{\Omega}_{\psi_i\psi_j}$.
We assume the network $G_A$ is sparse based on the observation that many empirical networks tend to have constant average degree relative to the network size. This sparsity implies that $m\sim \mathrm{O}(n)$ and we can assume the matrix $\bm{\Omega}$ has the form
\begin{equation}
    \begin{aligned}
        \bm{\Omega}=\frac{\bm{C}}{n} \enspace, 
    \end{aligned}
\end{equation}
where $\bm{C}$ is a matrix that is constant as we vary $n$.

A common variant of the SBM is the symmetric stochastic block model. The symmetric SBM assumes that all $q$ communities have the same number of nodes and the matrix $\bm{\Omega}$ is composed of two distinct probabilities: intra-community edge probability $\pin$ and inter-community edge probability $\pout$. If nodes $i$ and $j$ belong to same community, $\psi_i=\psi_j$, then $\bm{\Omega}_{\psi_i\psi_j}=\pin$, otherwise $\bm{\Omega}_{\psi_i\psi_j}=\pout$. Similarly, the rescaled edge probability matrix $\bm{C}=n\bm{\Omega}$ also has two distinct values: $\cin=n\pin$, $\cout=n\pout$.

\subsection{Detectability limit of community structure in networks}
The goal of community detection is to infer the community labels $\bm{\psi}$ based solely on observing the network $G_A$. 
A community structure is said to be detectable if there exists an algorithm that can infer the community labels better than random guessing. 
The detectability of community structure is related to the parameters of the SBM used to generate the network, specifically the differences in edge densities within and between communities. 
For instance, if $\cin = \cout$, the network is essentially an Erd\H{o}s--R\'{e}nyi random graph~\cite{erdds1959random} without any community structure, making it impossible to detect communities. 
Conversely, when $\cin$ is significantly larger than $\cout$, the communities are more densely connected internally than externally, making them easier to detect.
If a community detection algorithm can detect $\bm{\psi}$ better than random guessing within polynomial time complexity, then the community structure is said to be detectable~\cite{abbe2018community}. 
This detectability threshold coincides with the Kesten-Stigum threshold~\cite{decelle2011asymptotic, moore2017computer}.
Below the detectability threshold, no algorithm can perform better than random guessing, and the community structure is said to be undetectable. 

For the symmetric SBM, the theoretical detectability threshold occurs when
\begin{equation}
    \begin{aligned}
        |\cin-\cout|>q\sqrt{d} \enspace ,
    \end{aligned}
\end{equation}
where the average degree $d=\frac{\cin + (q-1)\cout}{q}$. With a little transformation, we can obtain the threshold: 
\begin{equation}\label{ssbmSNR}
    \mathrm{SNR}:=\frac{(\cin-\cout)^2}{q^2d}>1 \enspace,
\end{equation}
where $\rm SNR$ is the signal-to-noise ratio. 
So for any network $\rm G_A$ generated by the symmetric SBM, if $\mathrm{SNR}>1$, the community structure in $\rm G_A$ is  considered detectable. 

To see the $\rm SNR$ for general SBM~\cite{abbe2018community}, we first define the matrix $\bm N$ which is a $(q\times q)$-dimension diagonal matrix, with diagonal entries equal to the number of nodes in each community.
Ordering the non-decreasing magnitude eigenvalues of matrix $\bm{N} \bm{\Omega}$ as $\lambda_1, \lambda_2, ...$,the general $\rm SNR$ is defined as
\begin{equation}\label{generalSNR}
    \mathrm{SNR}=\frac{\lambda_2^2}{\lambda_1} \enspace ,
\end{equation}
which is equivalent to the $\rm SNR$~\eqref{ssbmSNR} above for the symmetric SBM.

\subsection{The non-backtracking matrix and Bethe Hessian for dyadic networks}\label{chapterBH}
Spectral clustering using the non-backtracking matrix is an effective community detection method that is consistent with inference on the SBM.
The non-backtracking matrix is a linearization of the belief propagation algorithm (BP), an approximate inference algorithm for the SBM. The BP can detect community structure with asymptotically optimal accuracy~\cite{decelle2011asymptotic}. 
The Bethe Hessian matrix is a compact formulation of the non-backtracking matrix that shares the same properties of detectability of the non-backtracking matrix~\cite{saade2014spectral}.

The non-backtracking matrix $\bmNB$ of a network is a matrix of dimension $2m\times 2m$ such that each row and column corresponds to a directed edge in the network. Note that the non-backtracking matrix is defined on directed edges, so in the case of an undirected network, each edge $(i, j)$ is represented as two directed edges $i\to j$ and $j\to i$. Thus, for a network with $m$ undirected edges, there are $2m$ directed edges in total.
For the directed edges $i\to j$ and $k\to l$ in the network, the entry of the non-backtracking matrix $\bm{NB}_{i\to j, k\to l}$ is defined as
\begin{equation}
    \bm{NB}_{i\to j, k\to l}=\left\{\begin{matrix}
1 & j\neq k\ \textrm{and}\ i=l  \\ 
0 & \rm{otherwise}
\end{matrix}\right. \enspace .
\end{equation}

Following the standard approach to spectral clustering, we compute the first $q$ eigenvectors $\bm{V}_q = [\bm{\nu}_{\cdot,1}, \bm{\nu}_{\cdot,2}, \ldots, \bm{\nu}_{\cdot,q}]$ of the non-backtracking matrix $\bmNB$ corresponding to the leading $q$ eigenvalues. These eigenvectors are then used to cluster the nodes in the network into $q$ communities. Since we are interested in clustering the $n$ nodes of the network, rather than the $2m$ directed edges, the clustering is performed by first pooling the $2m$-length eigenvectors of $\bmNB$ into vectors of length $n$. This operation is performed using a pooling matrix which aggregate relevant components of the eigenvectors for each node.
The pooling matrix $\bm{P}$ is defined as follows:
\begin{equation}
    \bm{P}_{i, k\to l}=\left\{\begin{matrix}
1 & i=l\ \textrm{and}\ k\in\partial i  \\
0 & \rm{otherwise}
\end{matrix}\right. \enspace ,
\end{equation}
where $\partial i$ is the set of directed edges that point to node $i$. The pooling matrix $\bm{P}$ effectively selects the components of the eigenvectors that correspond to the directed edges pointing to each node.
The spectral clustering of the network is then performed by applying a clustering algorithm, such as k-means, to the rows of the pooled matrix $\bm{U}_q = \bm{P} \bm{V}_q$. The clustering algorithm groups the nodes based on the pooled eigenvectors, resulting in a partition of the nodes into $q$ communities.

The construction and eigen-decomposition of the non-backtracking matrix can be expensive in terms of time and memory, especially for large networks. A better alternative is to construct the Bethe Hessian that has eigenvectors that correspond to the pooled  eigenvectors $\bm{U} = [\bm{\mu}_{\cdot,1}, \bm{\mu}_{\cdot,2}, \ldots, \bm{\mu}_{\cdot,n}]$. The Bethe Hessian is defined as
\begin{equation}\label{bethehessian}
    \bm{B}_\eta = (\eta^2-1)\bm{I}-\eta \bm{A} + \bm{D}\enspace ,
\end{equation}
where $\bm I$ is identity matrix, $\bm D$ is diagonal matrix of node degrees, $\eta$ is a regularization parameter, which typically works well with the setting $\eta=\sqrt{d}$.

The Bethe Hessian can be derived from non-backtracking matrix~\cite{krzakala2013spectral, saade2016spectral}. To see how, we start with the eigen decomposition of the non-backtracking matrix $\bmNB\cdot\bm{V} = \bm{\Lambda \cdot V}$ (where $\bm{\Lambda}$ is the diagonal matrix of eigenvalues) and consider in terms of a specific directed edge $i\to j$:
\begin{equation}\label{networkBH_expandNB}
    \bm{\lambda} \bm{\nu}_{i\to j} = \sum_{k\in \partial i/j}\bm{\nu}_{k\to i} \enspace, 
\end{equation}
and notice that the pooling operation $\bm{\mu P} = \bm{\nu}$ can be expressed in terms of equation \eqref{networkBH_expandNB} and an additional term $\bm{\nu}_{j\to i}$, 
\begin{align}\label{networkBH_expandPooling}
    \bm{\mu}_{i} &= \sum_{k \in \partial i} \bm{\nu}_{k\to i}\\
    &= \sum_{k\in \partial i/j}\bm{\nu}_{k\to i} + \bm{\nu}_{j\to i} \\
    &= \lambda \bm{\nu}_{i\to j} + \bm{\nu}_{j\to i} \enspace.
\end{align}

Similarly for node $j$, we have:
\begin{equation}
\bm{\mu}_{j}=\lambda \bm{\nu}_{j\to i}+\bm{\nu}_{i\to j}\enspace ,
\end{equation}
then we can subtract $\bm{\mu}_{i}-\lambda \bm{\mu}_{j}$ to eliminate $\bm{\nu}_{i\to j}$:
\begin{equation}
\bm{\mu}_{i}-\lambda\bm{\mu}_{j}=\bm{\nu}_{j\to i} - \lambda^2 \bm{\nu}_{j\to i}\enspace ,
\end{equation}
and rearrange to get
\begin{equation}\label{eq_nu_itoj}
\bm{\nu}_{j\to i} = \frac{\bm{\mu}_{i}-\lambda \bm{\mu}_{j}}{1-\lambda^2} \enspace .
\end{equation}

Substitute $\bm{\nu}_{j\to i}$ from equation~\eqref{eq_nu_itoj} above into equation~\eqref{networkBH_expandPooling} and we get:
\begin{align}
\bm{\mu}_{i}&=\sum_{k \in \partial i}\frac{\bm{\mu}_{i}-\lambda \bm{\mu}_{k}}{1-\lambda^2} \\
\bm{\mu}_{i}&=d_i\frac{\bm{\mu}_{i}}{1-\lambda^2}-\frac{\lambda}{1-\lambda^2}\sum_{k \in \partial i}\bm{\mu}_{k}\\
(1-\frac{d_i}{1-\lambda^2})\bm{\mu}_{i}+\frac{\lambda}{1-\lambda^2}\sum_{k \in \partial i}\bm{\mu}_{k}&=0 \\
\frac{1}{\lambda^2-1}\left[(\lambda^2-1+d_i)\bm{\mu}_{i}-\lambda\sum_{k \in \partial i}\bm{\mu}_{k}\right]&=0 \enspace .
\end{align}

Omitting the term $1/(\lambda^2-1)$, we have the matrix form:
\begin{align}\label{bh*mu=0}
\left[(\lambda^2-1)\bm{I}+\bm{D}-\lambda \bm{A}\right]\bm{\mu} &= 0\\
\bm{B}_\lambda\bm{\mu} &= 0 \enspace ,
\end{align}
where $\bm{B}_\lambda$ is the Bethe Hessian matrix with regularization parameter $\eta=\lambda$.

From equation~\eqref{bh*mu=0} we can conclude that $\bm{\mu}$ is related with null space of $\bm{B}_{\lambda}$. In other words, if we set the regularization parameter $\eta=\lambda$ to an eigenvalue of $\bmNB$, then $\bm{\mu}$ will be an eigenvector of $\bm{B}_{\lambda}$ corresponding with eigenvalue 0. Note that although the above relationship does not hold when the parameter $\eta$ is not an eigenvalue of $\bmNB$, the eigenvector of $\bm{B}_{\sqrt{d}}$ is still found to be a good approximation of $\bm{\mu}$ in practice~\cite{saade2016spectral}.

\begin{figure}
    \centering
    \includegraphics[width=15cm]{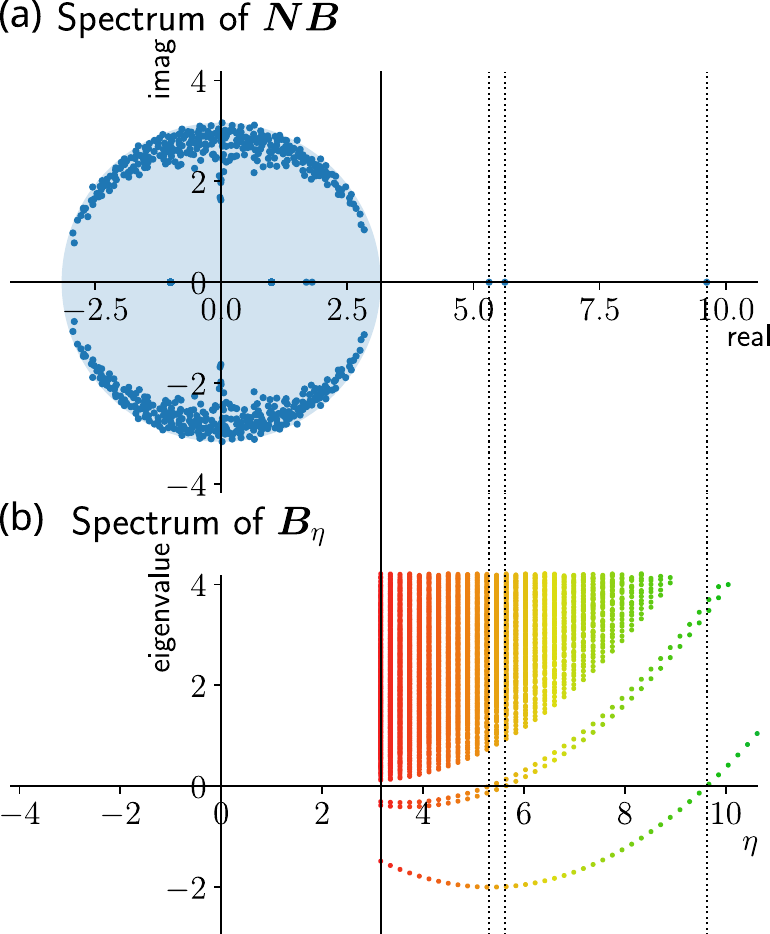}\\
	\caption{\small\it The spectrum of non-backtracking matrix (a) and Bethe Hessian matrix (b) for a network generate by symmetric SBM. 
    The number of nodes $n=300$, number of communities $q=3$, average degree $d=10$, edge probability ratio $\pout/\pin=0.2$. 
    The light blue circle is the bulk with radius $\sqrt{d}$. Each column of red to green points in (b) is the spectrum of $\bm{B}_{\eta}$ where $\eta$ is x-axis value. Here the spectrum of $\bm{B}_{\eta}$ is scaled for better presenting. \label{figNBBHspectrum}}
\end{figure}

Figure~\ref{figNBBHspectrum} gives an example spectrum of $\bmNB$ and $\bm{B}_{\eta}$ for a network generated by the symmetric SBM. The blue points in the plot above represent the eigenvalues of the non-backtracking matrix $\bmNB$ and the light blue circle with radius $\sqrt{d}$ indicates approximately the uninformative bulk within which most of the eigenvalues lie. The red to green points in the plot below represent the eigenvalues of the Bethe Hessian matrix $\bm{B}_{\eta}$ for corresponding values of $\eta$. When $\eta$ is set to one of the real informative eigenvalues of $\bmNB$, the corresponding eigenvector $\bm{\mu}$ can be used to optimally cluster the nodes in the network~\cite{dall2019revisiting}.

When the communities are detectable, there are a few informative real eigenvalues outside the bulk, which correspond to the number of communities in the network. We see that when $\eta$ is sufficiently large, all eigenvalues of $\bm{B}_{\eta}$ are positive. As $\eta$ decreases, new negative eigenvalues appear each time $\eta$ crosses a real informative eigenvalue of $\bmNB$. Consequently, when $\eta$ is equal to the radius of the uninformative bulk, the number of negative eigenvalues of $\bm{B}_{\eta}$ can also be used to determine the number of communities in the network. The eigenvectors corresponding with these negative eigenvalues can be used as a good approximation of $\bm{\mu}$ corresponding with the informative eigenvalues of $\bmNB$. The exact et of eigenvectors $\bm{U}_q$ can be obtained by identifying the leading eigenvector of the Bethe Hessian matrix $\bm{B}_{\eta}$ with $\eta$ set to the each of the $q$ informative eigenvalues of $\bmNB$.  

\section{Spectral clustering for hypergraphs}\label{app:nb2bh_hypergraph}
In this section, we first introduce the non-backtracking matrix for hypergraphs. Then we use this non-backtracking matrix to derive the Bethe Hessian for uniform hypergraphs and non-uniform hypergraphs.

\subsection{The non-backtracking matrix for hypergraphs}
The uniform hypergraph non-backtracking matrix~\cite{angelini2015spectral} is a generalization of the non-backtracking matrix for dyadic networks~\cite{krzakala2013spectral}. 
Similar to the non-backtracking matrix for dyadic networks, the non-backtracking matrix for hypergraphs is defined on directed hyperedges.
These directed hyperedges are simply defined from undirected hyperedges such that each node $i$ in hyperedge $e$ induces a directed hyperedge $i\to e$.
Consequently, each undirected hyperedge of size $\kappa$ induces $\kappa$ directed hyperedges.
The non-backtracking matrix, $\bmNB^{(\kappa)}$, of a $\kappa$-uniform hypergraph is defined as follows. 
For two $\kappa$-hyperedges $e_1$ and $e_2$, if $i\in e_1$ and $j\in e_2$, the non-backtracking matrix is given by
\begin{equation}
    \bmNB^{(\kappa)}_{i\to e_1, j\to e_2}=\left\{\begin{matrix}
1 & j\in e_1/i \ \textrm{and}\ e_2\neq e_1  \\   
0 & \textrm{otherwise}
\end{matrix}\right. \enspace ,
\end{equation}
where $i\to e_1$ is a directed hyperedge~\cite{chodrow2023nonbacktracking}
if node $i$ is in hyperedge $e_1$. 
Since each node in a hyperedge induces a directed hyperedge in the non-backtracking matrix, the dimension of the non-backtracking matrix is equal to the total sum of node-degrees, $\sum_i d_i$, which is equal to $\kappa m$ in the case of a $\kappa$-uniform hypergraph.

The non-backtracking matrix for a non-uniform hypergraph~\cite{chodrow2023nonbacktracking} is constructed in the same way, however, we can order the rows and columns by the order of the hyperedges,
\begin{equation}\label{NB_nonuniform}
    \bmNB^{(\mathcal{K})}=\left[\begin{matrix}
\bmNB^{(2\to 2)} & \bmNB^{(2\to 3)} & ... & \bmNB^{(2\to \kappa)} \\
\bmNB^{(3\to 2)} & \bmNB^{(3\to 3)} & ... & \bmNB^{(3\to \kappa)} \\
... & ... & ... & ... \\
\bmNB^{(\kappa\to 2)} & \bmNB^{(\kappa\to 3)} & ... & \bmNB^{(\kappa\to \kappa)}
\end{matrix}\right] \enspace ,
\end{equation}
where $\bmNB^{(\kappa'\to \kappa)}$ is the non-backtracking matrix from the $\kappa'$-hyperedges, $\mathcal{E}^{(\kappa')}$ to the $\kappa$-hyperedges, $\mathcal{E}^{(\kappa)}$. 
The dimension of each $\bmNB^{(\kappa'\to \kappa)}$ is $\kappa' m^{(\kappa')}\times \kappa m^{(\kappa)}$.

The leading $q$ eigenvectors of the non-backtracking matrix can be used to perform spectral clustering. Since we are interested in clustering the nodes in the network, we can ``pool'' the eigenvectors to an $n$-dimensional vector $\bm \mu$,
\begin{equation}\label{nonuniformBH_expandPooling}
    \bm{\mu}_i=\sum_{\kappa\in \calK}\sum_{e\in\partial_{\kappa}i}\bm{\nu}_{i\to e} \enspace ,
\end{equation}
and use the pooled vectors $\mu$ as the input to a clustering algorithm, e.g., k-means.

\subsection{The Bethe Hessian for hypergraphs}

The construction and eigen decomposition of the non-backtracking matrix is computationally intensive in terms of time and space resources, especially for large hypergraphs. 
For dyadic networks, the Bethe Hessian~\cite{saade2014spectral} is a lower-dimensional symmetric matrix that captures the same desirable spectral properties of the non-backtracking matrix that makes it effective for spectral clustering.
The Bethe Hessian has previously been extended for hypergraphs~\cite{stephan2024sparse}, but is only applicable if the hypergraph is uniform.  
Here we present a new version of the Bethe Hessian that is suitable for non-uniform hypergraphs, following a similar derivation as in the dyadic case (see Appendix~\ref{app:scbh_network}).

We consider an informative eigenvector $\bm{\nu}$ and corresponding eigenvalue $\lambda$ of the non-backtracking matrix $\bmNB^{(\mathcal{K})}$, such that $\bmNB^{(\mathcal{K})}\bm{\nu}=\lambda\bm{\nu}$. For a given directed hyperedge $i\to e_1$, we expand the eigenvalue equation,
\begin{equation}\label{nonuniformBH_expandNB}
    \lambda\bm{\nu}_{i\to e_1}=\sum_{\kappa\in \calK}\sum_{j\in e_1/i}\sum_{e_2\in \partial_\kappa j/e_1}\bm{\nu}_{j\to e_2} \enspace ,
\end{equation}
where $\partial_\kappa j$ is all $\kappa$-hyperedges including $j$.
We can reorder the sum in equation~\eqref{nonuniformBH_expandNB} and substitute in the pooled vector $\bm{\mu}$ from~\eqref{nonuniformBH_expandPooling}:
\begin{align}
    \lambda\bm{\nu}_{i\to e_1}
    =&\sum_{j\in e_1/i}\sum_{\kappa\in \calK}\sum_{e_2\in \partial_\kappa j/e_1}\bm{\nu}_{j\to e_2}\\
    =&\sum_{j\in e_1/i}\left(\sum_{\kappa\in \calK}\sum_{e_2\in \partial_\kappa j}\bm{\nu}_{j\to e_2}-\bm{\nu}_{j\to e_1}\right)\\
    =&\sum_{j\in e_1/i}\left(\bm{\mu}_j-\bm{\nu}_{j\to e_1}\right) \\
    \label{nonuniformBH_relationmunu}
    \lambda\bm{\nu}_{i\to e_1}+\sum_{j\in e_1/i}\bm{\nu}_{j\to e_1}=&\sum_{j\in e_1/i}\bm{\mu}_j \enspace . 
\end{align}

Note that equation~\eqref{nonuniformBH_relationmunu} above repeats $\kappa$ times, once for each node $i\in e_1$ if $e_1$ is a $\kappa$-hyperedge. 
Summing over all of nodes in $e_1$, we get:
\begin{equation}\label{nonuniformBH_sumoveri}
    (\lambda+\kappa-1)\sum_{i\in e_1}\bm{\nu}_{i\to e_1}
    =(\kappa-1)\sum_{j\in e_1}\bm{\mu}_j \enspace ,
\end{equation}
then dividing by $(\lambda + \kappa - 1)$  and subtracting~\eqref{nonuniformBH_relationmunu}, 
\begin{align}
    (1-\lambda)\bm{\nu}_{i\to e_1}=&\frac{\kappa-1}{\lambda+\kappa-1}\bm{\mu}_i-\frac{\lambda}{\lambda+\kappa-1}\sum_{j\in e_1/i}\bm{\mu}_j \\
    \bm{\nu}_{i\to e_1}=&\frac{\kappa-1}{(1-\lambda)(\lambda+\kappa-1)}\bm{\mu}_i-\frac{\lambda}{(1-\lambda)(\lambda+\kappa-1)}\sum_{j\in e_1/i}\bm{\mu}_j
    \enspace .
\end{align}

Now we substitute the above definition of $\bm{\nu}_{i\to e_1}$ into pooling equation~\eqref{nonuniformBH_expandPooling},
\begin{align}
    \bm{\mu}_i=&\sum_{\kappa\in \calK}\sum_{e\in\partial_{\kappa}i}\left[\frac{\kappa-1}{(1-\lambda)(\lambda+\kappa-1)}\bm{\mu}_i-\frac{\lambda}{(1-\lambda)(\lambda+\kappa-1)}\sum_{j\in e/i}\bm{\mu}_j\right]\\
    =&\sum_{\kappa\in \calK}\frac{(\kappa-1)d_i^{(\kappa)}}{(1-\lambda)(\lambda+\kappa-1)}\bm{\mu}_i-\lambda\sum_{\kappa\in \calK}\sum_{e\in\partial_{\kappa}i}\sum_{j\in e/i}\frac{1}{(1-\lambda)(\lambda+\kappa-1)}\bm{\mu}_j
    \enspace ,
\end{align}
where $d_i^{(\kappa)}=|\partial_{\kappa}i|$ is the $\kappa$-degree of $i$, i.e., the number of $\kappa$-hyperedges that $i$ is included in. We can rearrange the above result to set the right-hand side to zero,
\begin{equation}\label{nonuniformBH_detail}
    (1-\sum_{\kappa\in \calK}\frac{(\kappa-1)d_i^{(\kappa)}}{(1-\lambda)(\lambda+\kappa-1)})\bm{\mu}_i+\lambda\sum_{\kappa\in \calK}\sum_{e\in\partial_{\kappa}i}\sum_{j\in e/i}\frac{1}{(1-\lambda)(\lambda+\kappa-1)}\bm{\mu}_j=0 \enspace .
\end{equation}

Finally, we can rewrite the above equation in matrix form, which is the Bethe Hessian matrix for non-uniform hypergraphs:
\begin{equation}
    \bm{B}^{(\mathcal{K})}_{\eta}=\bm{I}-\sum_{\kappa\in \mathcal{K}}
    \frac{(\kappa-1)}{(1-\eta)(\eta+\kappa-1)}\bm{D}^{(\kappa)} + \sum_{\kappa\in \calK}\frac{\eta}{(1-\eta)(\eta+\kappa-1)}\bm{A}^{(\kappa)}\enspace ,
\end{equation}
where $\eta$ is a regularization parameter, $\bm{D}^{(\kappa)}$ is the diagonal degree matrix for $\kappa$-hyperedges, and $\bm{A}^{(\kappa)}$ is the one-mode projection matrix for $\kappa$-hyperedges, i.e., 
\begin{equation}
    \bm{A}^{(\kappa)}_{ij}=\left\{
    \begin{matrix}
        |\left\{e|i,j\in e, |e|=\kappa\right\}| & i\neq j \\
        0 & i=j
    \end{matrix}
    \right. \enspace .
\end{equation}

Note that the matrix $\bm{A}^{(\kappa)}$ can also be expressed in terms of the $\kappa$-hyperedge incidence matrix $\bm{H}^{(\kappa)}$ as follows:
\begin{equation}
    \bm{A}^{(\kappa)}=\bm{H}^{(\kappa)}(\bm{H}^{(\kappa)})^T-\mathrm{diag}(\bm{H}^{(\kappa)}(\bm{H}^{(\kappa)})^T) \enspace .
\end{equation}

\subsection{The spectrum of Bethe Hessian matrix}~\label{app:spectrum_hyperbh}
The spectrum of Bethe Hessian relates to the spectrum of non-backtracking matrix such that the negative eigenvalues of Bethe Hessian correspond to the informative eigenvalues of non-backtracking matrix. 
Of particular importance to the task of detecting community structure is that we preserve the relationship between the eigenvectors.
Specifically, when we take an informative eigenvalue $\lambda$ and corresponding eigenvector $\bm{\nu}$ and set the regularization parameter $\eta=\lambda$, then the eigenvector associated with the eigenvalue $0$ of the Bethe Hessian is the pooled vector $\bm{\mu}$.

\begin{figure}
    \centering
    \includegraphics[width=15cm]{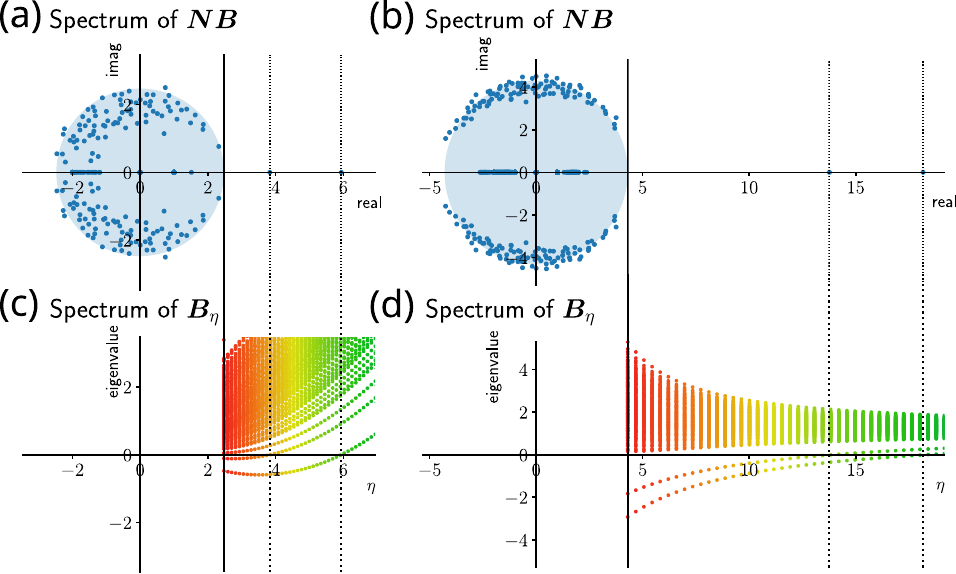}\\
	\caption{\small\it The spectrum of non-backtracking matrix (a, b) and Bethe Hessian matrix (c, d) for hypergraphs generated by $\kappa=3$ uniform HSBM (a, c) and $\mathcal{K}=\left\{2, 3, 4\right\}$ nonuniform HSBM (b, d). In both settings, $n=100$, $q=2$ and degree parameters $\din=10$ and $\dout=1$. The light blue circle is the uninformative bulk of $\bmNB$ with radius $\sum_{\kappa\in \mathcal{K}}\sqrt{d^{(\kappa)}(\kappa-1)}$.\label{figNBBHspectrumHSBM}}
\end{figure}

Figure~\ref{figNBBHspectrumHSBM} provides an empirical demonstration of the relationship between the spectra of the non-backtracking matrix and the Bethe Hessian for two hypergraphs with $q=2$ communities, generated using the HSBM described earlier.
Figure~\ref{figNBBHspectrumHSBM}(a, c) shows the $\kappa$-uniform case with $\kappa=3$.
The radius of the bulk of uninformative eigenvalues is $\sqrt{d(\kappa-1)}$, where $d$ is the average degree of the hypergraph.
For the non-uniform case [Fig.~\ref{figNBBHspectrumHSBM}(b, d)], we set $\mathcal{K}=\left\{2, 3, 4\right\}$. 
This time the radius of the bulk is $\sum_{\kappa\in \mathcal{K}}\sqrt{d^{(\kappa)}(\kappa-1)}$, where $d^{(\kappa)}$ is the average degree of the $\kappa$-order hyperedges, $d^{(\kappa)}=\mathrm{Tr}(\bm{D}^{(\kappa)})/n$.

Figure~\ref{figNBBHspectrumHSBM} shows the same relationship between the spectra of the non-backtracking matrix and Bethe Hessian that we see for dyadic networks in Figure~\ref{figNBBHspectrum}. 
When parameter $\eta$ is equal to the radius of uninformative bulk, then negative eigenvalues of $\bm{B}_{\eta}$ correspond one-to-one with the real informative eigenvalues of $\bmNB$. We can also use the eigenvectors corresponding with these negative eigenvalues of $\bm{B}_{\eta}$ to perform spectral clustering, as described in Algorithm~\ref{scwithbh}. 
\begin{algorithm}
    \caption{Spectral Clustering with Bethe Hessian}
    \label{scwithbh}
    \renewcommand{\algorithmicrequire}{\textbf{Input:}}
    \renewcommand{\algorithmicensure}{\textbf{Output:}}
    
    \begin{algorithmic}[1]
        \REQUIRE incidence matrix $\bm{H}$, all possible hyperedge order $\calK$
        \ENSURE communities labels $\overrightarrow{\psi}$
        \STATE  Compute $\eta=\sum_{\kappa\in \calK}\sqrt{d^{(\kappa)}(\kappa-1)}$
        \STATE  Compute $\bm{B}_{\eta}$ by~\eqref{nonuniformBH}
        \STATE  Compute spectral decomposition $\bm{B}_{\eta}=\bm{V}\bm{\Lambda} \bm{V}^T$.
        \STATE  Compute $q=|\left\{\bm{\Lambda}_{i, i}: \bm{\Lambda}_{i, i}<0\right\}|$.
        \STATE  Select first $q$ eigenvectors of $\bm V$ based on the increasing order of corresponding eigenvalues as $\bm{W}_q$.
        \STATE Run k-means clustering on the rows of $\bm{W}_q$: $\overrightarrow{\psi}= k\textrm{-means}(\bm{W}_q^T, q)$
        \RETURN $\overrightarrow{\psi}$
    \end{algorithmic}
\end{algorithm}

\section{Time complexity comparison of spectral clustering operators}\label{appendix-n&nnz_NB_BH}
The computational cost of spectral clustering is primarily dominated by the eigen decomposition of the associated operator.
For instance, when employing the Lanczos algorithm for eigen-decomposition~\cite{golub2013matrix}, the time complexity is approximately $O(k(n+nnz))$, where $k\ll n$ is the number of Lanczos iterations, $n$ is the size of the matrix and $nnz$ is the number of non-zero entries. Crucially, the Lanczos algorithm operates more efficiently on symmetric matrices, giving the Bethe Hessian matrix a distinct advantage over the $\bmNB$.

Let us first consider the dyadic network case. 
For a network with $n$ nodes and $m$ edges, the dimension of its non-backtracking matrix $\bmNB$ is $n^*_{\bmNB}=2m$ and the number  
of non-zero entries is $nnz_{\bmNB}=\sum_{i\in \mathcal{V}}d_i(d_i-1)$.
For the Bethe Hessian matrix $\bm{B}$, the dimension is $n^*_{\bm{B}} = n$, and the number of non-zero entries is $nnz_{\bm{B}}=m+n$.
Therefore, to compare the time complexity of $\bmNB$ based and $\bm{B}$ based spectral clustering, we compute
\begin{align}\label{appendix_compareNBBH_network}
        n^*_{\bmNB}+nnz_{\bmNB}-n^*_{\bm{B}}-nnz_{\bm{B}}&=2m+\sum_id_i(d_i-1)-n-n-m \notag\\
        &=\sum_id_i(d_i-1)+m-2n \notag\\
        &=\sum_id_i^2-2m+m-2n \notag\\
        &=\sum_id_i^2-m-2n  \notag\\
        &\geq nd^2-nd/2-2n \notag\\
        &=n(d^2-d/2-2) \enspace ,
\end{align} 
where $d$ is the average degree. In a large connected network, $d>2$, which makes~\eqref{appendix_compareNBBH_network} strictly greater than zero. Thus, the Bethe Hessian based spectral clustering has lower time complexity than the one based on non-backtracking matrix.

Now, consider a hypergraph containing hyperedges of all possible orders $\kappa\in \calK$. 
Let $d_i^{(\kappa)}$ denote the order $\kappa$ degree of node $i$. 
From the definition of the non-backtracking matrix~\eqref{NB_nonuniform}, its dimension is equal to the total number of directed hyperedges: 
\begin{equation}
    n^*_{\bmNB}=\sum_i\sum_{\kappa}d_i^{(\kappa)} \enspace ,
\end{equation}
and the number of non-zero entries is
\begin{equation}
    nnz_{\bmNB}=\sum_i\sum_{\kappa}d_i^{(\kappa)}(\kappa-1)(d_i-1) \enspace ,
\end{equation}
where the total degree of node $i$ is $d_i=\sum_{\kappa\in \calK}d_i^{(\kappa)}$.
Based on the Hypergraph Bethe Hessian matrix~\eqref{nonuniformBH}, the dimension is $n^*_{\bm{B}}=n$. The number of non-zero entries in $\bm{B}$ has the form $nnz_{\bm{B}}=n+x$, where $x$ is the number of non-zero entries in its off-diagonal part. 
Each order $\kappa$ hyperedge contributes $\binom{\kappa}{2}$ non-zero entries in $\bm{B}$, but different hyperedges may overlap in their contributions. Therefore we can upper bound $x$ as,
\begin{equation}
        x<\sum_\kappa m^{(\kappa)}\binom{\kappa}{2}=\sum_\kappa\sum_i\frac{d_i^{(\kappa)}}{\kappa}\frac{\kappa(\kappa-1)}{2}=\sum_i\sum_\kappa d_{i}^{(\kappa)}\frac{(\kappa-1)}{2} \enspace .
\end{equation}

We then have:
\begin{align}
    &n^*_{\bmNB}+nnz_{\bmNB}-n^*_{\bm{B}}-nnz_{\bm{B}} \notag\\
    >&\sum_i\sum_\kappa d_i^{(\kappa)}+\sum_i\sum_\kappa d_i^{(\kappa)}(\kappa-1)(d_i-1)-n-\sum_i\sum_\kappa d_i^{(\kappa)}\frac{(\kappa-1)}{2} \notag\\
    =&\sum_i\sum_\kappa\left[d_i^{(\kappa)}(1+(\kappa-1)(d_i-\frac{3}{2}))-\frac{1}{|\calK|}\right] \enspace .
\end{align}

\begin{figure}
    \centering
    \includegraphics[width=15cm]{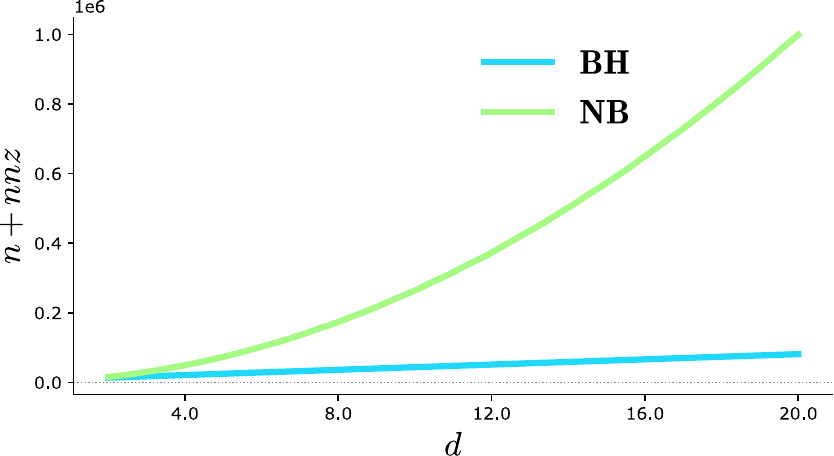}\\
	\caption{\small\it The $n+nnz$ for operators $\bm{B}$ and $\bmNB$ which is a signal for the time complexity of spectral clustering. With different average degree, we sample the hypergraphs with $n=3000$, $q=3$, $\kappa=(2, 3)$ and $\epsilon=0.3$  in detectable phase.\label{figNBBHcomplexity}}
\end{figure}

For a large sparse hypergraph, $d_i^{(\kappa)}\sim O(n)$. Thus, as $n\to \infty$, the above expression remains positive, leading to the same conclusion: for both dyadic networks and hypergraphs, the eigen decomposition of the Bethe Hessian has lower time complexity than the non-backtracking operator. This is also illustrated in Figure~\ref{figNBBHcomplexity}.

\section{Generalizing the spectral detectability limit of symmetric HSBM}\label{append:detectability_literature}
To introduce the detectability limit of symmetric HSBM in literature~\cite{angelini2015spectral, chodrow2023nonbacktracking}, we need to introduce the definition appeared in both of them:
\begin{equation}\label{hatcak}
    \begin{aligned}
        \widehat{c}_{a}^{(\kappa)} = \frac{1}{(\kappa-1)!}\sum_{\left\{\overrightarrow{z}|\overrightarrow{z}_0=a, \overrightarrow{z}_{1:}\in[q]^{\kappa-1}\right\}}\bm{C}_{\overrightarrow{z}}^{(\kappa)}\prod_{l=1}^{\kappa-1}p_{\overrightarrow{z}_l}
    \end{aligned}\enspace ,
\end{equation}
and
\begin{equation}\label{hatcabk}
    \begin{aligned}
        \widehat{c}_{ab}^{(\kappa)} = \frac{1}{(\kappa-2)!}\sum_{\left\{\overrightarrow{z}|\overrightarrow{z}_0=a, \overrightarrow{z}_1=b, \overrightarrow{z}_{2:}\in[q]^{\kappa-2}\right\}}\bm{C}_{\overrightarrow{z}}^{(\kappa)}\prod_{l=2}^{\kappa-1}p_{\overrightarrow{z}_l}
    \end{aligned}\enspace ,
\end{equation}
where $[q]^{\kappa}$ represent the cartesian product for $\kappa$ sets $[q] = \left\{1, 2, ..., q\right\}$, $p_r$ is the normalized community size of community $r$.
In the case that all communities have the same size, $p_r=\frac{1}{q}$ and we can replace the product with the factor $\frac{1}{q^{\kappa-1}}$ in \eqref{hatcak} and with  $\frac{1}{q^{\kappa-2}}$ in \eqref{hatcabk}.
The relationship between $\widehat{c}_{a}^{(\kappa)}$ and $\widehat{c}_{ab}^{(\kappa)}$ is
\begin{equation}\label{relationcacab}
\begin{aligned}
    \widehat{c}_{a}^{(\kappa )} = \frac{1}{\kappa -1}\sum_{b\in [q]}\widehat{c}_{ab}^{(\kappa )}n_b
\end{aligned}
\end{equation}

Under the symmetric HSBM case, $p_r=\frac{1}{q}$ and $\bm{C}$ satisfy~\eqref{nonuniform_cincout} with parameters $\cin$ and $\cout$, a node can be incident on a maximum of $\binom{n/q-1}{\kappa-1}$ possible $\kappa$-hyperedges within the same community. 
Then we define the average $\kappa$-in-degree $\din^{(\kappa)}$ can be calculated as:
\begin{equation}\label{din^k}
    \begin{aligned}
        \din^{(\kappa)}
        &=\lim_{n\to \infty}\binom{n/q-1}{\kappa-1}\frac{\cin}{n^{\kappa-1}}\\
        &=\lim_{n\to \infty}\frac{\cin}{n^{\kappa-1}}\frac{(n/q-1)!}{(\kappa-1)!(n/q-\kappa)!}\\
        &=\lim_{n\to \infty}\frac{\cin}{n^{\kappa-1}}\frac{(n/q-1)...(n/q-\kappa+1)}{(\kappa-1)!}\\
        &=\frac{\cin}{n^{\kappa-1}}\frac{(n/q)^{\kappa-1}}{(\kappa-1)!}\\
        &=\frac{\cin}{q^{\kappa-1}(\kappa-1)!} \enspace .
    \end{aligned}
\end{equation}
A node can be connected to a maximum of $\binom{n-1}{\kappa-1}-\binom{n/q-1}{\kappa-1}$ possible $\kappa$-hyperedges that connect to at least one node in a different community. For each of the $q^{\kappa-1}-1$ ordered community combination of other $\kappa-1$ nodes, the out-community $\kappa$-degree of the node is
\begin{equation}\label{dout^k}
    \begin{aligned}
        \dout^{(\kappa)}
        &=\lim_{n\to \infty}\frac{\binom{n-1}{\kappa-1}-\binom{n/q-1}{\kappa-1}}{q^{\kappa-1}-1}\frac{\cout}{n^{\kappa-1}}\\
        &=\frac{\cout}{n^{\kappa-1}}\frac{n^{\kappa-1}-(n/q)^{\kappa-1}}{(\kappa-1)!(q^{\kappa-1}-1)}\\
        &=\frac{\cout}{q^{\kappa-1}(\kappa-1)!}
    \end{aligned} \enspace .
\end{equation}
The average $\kappa$-degree for the node is
\begin{equation}
    \begin{aligned}
        d^{(\kappa)}
        &=\din^{(\kappa)}+(q^{\kappa-1}-1)\dout^{(\kappa)}
    \end{aligned}\enspace .
\end{equation}

With our definition of $\din^{(\kappa)}$ and $\dout^{(\kappa)}$ above, we can represent the $\widehat{c}_{ab}^{(\kappa)}$ and $\widehat{c}_{a}^{(\kappa)}$ with notation $\din^{(\kappa)}$ and $\dout^{(\kappa)}$:
\begin{equation}\label{hatcincout}
    \begin{aligned}
        \widehat{c}_{ab}^{(\kappa )}=\left\{\begin{matrix}
        \begin{aligned}
            &\widehat{c}_{\rm in}^{(\kappa )}=\frac{c_{\rm in}+(q^{\kappa -2}-1)c_{\rm out}}{q^{\kappa -2}(\kappa -2)!}=q(\kappa-1)(\din^{(\kappa)}+(q^{\kappa-2}-1)\dout^{(\kappa)})& ,a=b\\
    	  &\widehat{c}_{\rm out}^{(\kappa )}=\frac{q^{\kappa -2}c_{\rm out}}{q^{\kappa -2}(\kappa -2)!}=q(\kappa-1)q^{\kappa-2}\dout^{(\kappa)}& ,a \neq b
        \end{aligned}
    	\end{matrix}\right.
    \end{aligned} \enspace ,
\end{equation}
and with equation~\eqref{relationcacab}, we have
\begin{equation}\label{symuniform_ck}
    \begin{aligned}
    \widehat{c}_{a}^{(\kappa )} &= \frac{\widehat{c}_{\rm in}^{(\kappa )}+(q-1)\widehat{c}_{\rm out}^{(\kappa )}}{q(\kappa -1)} \\
    &=\frac{\frac{c_{\rm in}+(q^{\kappa -2}-1)c_{\rm out}}{q^{\kappa -2}(\kappa -2)!} + (q-1)\frac{q^{\kappa -2}c_{\rm out}}{q^{\kappa -2}(\kappa -2)!}}{q(\kappa -1)} \\
        &=\frac{c_{\rm in}+(q^{\kappa -1}-1)c_{\rm out}}{q^{\kappa -1}(\kappa -1)!}\\
    &=\din^{(\kappa)}+(q^{\kappa-1}-1)\dout^{(\kappa)}\\
    &=d^{(\kappa)}
    \end{aligned}\enspace .
\end{equation}

For the $\kappa$-uniform symmetric HSBM introduced in~\cite{angelini2015spectral}, the detectability threshold is
\begin{equation}\label{symuniform_conjectureddetectability}
    \begin{aligned}
    \frac{(\widehat{c}_{\rm in}^{(\kappa )}-\widehat{c}_{\rm out}^{(\kappa )})^2}{q^2d^{(\kappa )}(\kappa -1)}&=1\\
    \frac{\left((\kappa-1)(\din^{(\kappa)}-\dout^{(\kappa)})\right)^2}{(\kappa-1)d^{(\kappa)}}&=1
    \end{aligned} \enspace .
\end{equation}
which can degenerated from our format of detectability limit in~\eqref{phi_by_dindout} by restricting $\mathcal{K}=\left\{\kappa\right\}$. 

For non-uniform symmetric HSBM with $q=2$ considered in~\cite{chodrow2023nonbacktracking}, they define:
\begin{equation}\label{alpha_beta}
    \begin{aligned}
        \alpha_\kappa &=d^{(\kappa )}(\kappa -1) \\
        \beta_\kappa &=\frac{\widehat{c}_{\rm in}^{(\kappa )}-\widehat{c}_{\rm out}^{(\kappa )}}{2}
    \end{aligned} \enspace .
\end{equation}
Then their conjectured detectability threshold of spectral clustering with non-backtracking matrix~\cite{chodrow2023nonbacktracking} (also for Bethe Hessian matrix since their relationship as shown in Appendix~\ref{app:nb2bh_hypergraph}) is $\snr_{\rm NB} =1$ where
\begin{equation}\label{detectabilityforq=2}
    \begin{aligned}
        \snr_{\rm NB} &:= \frac{(\sum_{\kappa \in \mathcal{K}}\beta_\kappa)^2}{\sum_{\kappa \in \mathcal{K}}\alpha_\kappa}\\
        &=\frac{\left(\sum_{\kappa \in \mathcal{K}}\widehat{c}_{\rm in}^{(\kappa )}-\widehat{c}_{\rm out}^{(\kappa )}\right)^2}{2^2\sum_{\kappa \in \mathcal{K}}d^{(\kappa )}(\kappa -1)} \\
        &=\frac{\left(\sum_{\kappa \in \mathcal{K}}(\kappa-1)(\din^{(\kappa)}-\dout^{(\kappa)})\right)^2}{\sum_{\kappa \in \mathcal{K}}(\kappa -1)d^{(\kappa )}}
    \end{aligned} \enspace .
\end{equation}
By simple generalizing the $\beta_\kappa$ in~\eqref{alpha_beta} into
\begin{equation}
    \beta_{\kappa, q}=\frac{\widehat{c}_{\rm in}^{(\kappa)}-\widehat{c}_{\rm out}^{(\kappa )}}{q} \enspace ,
\end{equation}
we can get the spectral detectability limit of non-uniform symmetric HSBM for any $q$:
\begin{equation}\label{append:detectabilityforanyq}
    \begin{aligned}
        \snr_{\rm BH} &:= \frac{(\sum_{\kappa \in \mathcal{K}}\beta_\kappa)^2}{\sum_{\kappa \in \mathcal{K}}\alpha_\kappa}\\
        &=\frac{\left(\sum_{\kappa \in \calK}\widehat{c}_{\rm in}^{(\kappa )}-\widehat{c}_{\rm out}^{(\kappa )}\right)^2}{q^2\sum_{\kappa \in \calK}(\kappa -1)d^{(\kappa )}}\\
        &=\frac{\left(\sum_{\kappa \in \mathcal{K}}(\kappa-1)(\din^{(\kappa)}-\dout^{(\kappa)})\right)^2}{\sum_{\kappa \in \mathcal{K}}(\kappa -1)d^{(\kappa )}}
    \end{aligned} \enspace .
\end{equation}

\section{Belief propagation and detectability limit of symmetric HSBM}\label{appendix-bp&detectability}
\begin{figure}[H]
    \centering
    \includegraphics[width=15cm]{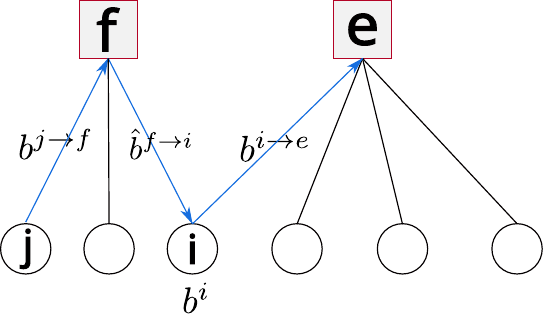}\\
	\caption{\small\it The factor graph representation of hypergraph. The red square represent the hyperedge, the circle represent the node. The message and marginal probability on directed hyperedge and 
    node are also shown as $b^{j\to f}$, $b^{i\to e}$, $\hat{b}^{f\to i}$ and $b^{i}$. The blue directed hyperedge form a non-backtracking path $b^{j\to f}\to \hat{b}^{f\to i} \to b^{i\to e}$.\label{figBPillustration}}
\end{figure}

Similar with belief propagation in graph~\cite{decelle2011asymptotic}, the main process of belief propagation in hyper graph is to update the message $b^{i\to e}_{\psi_i}$ in all directed hyperedges, and then update the marginal probability $b^{i}_{\psi_i}$ that node $i$ belong to community $\psi_i$ for all nodes. Here $b^{i\to e}_{\psi_i}$ represent the marginal probability that node $i$ belong to community $\psi_i$ without hyperedge $e$. However, unlike graph, we need another intermediate hyperedge to node message $\hat{b}^{e\to i}_{\psi_i}$ when we update the node to hyperedge message $b^{i\to e}_{\psi_i}$. To make a distinction with all exist hyperedges $\calE$, we use $\boldsymbol{E}$ represent all possible hyperedges. Then the belief propagation can be formalized as
\begin{equation}\label{bp_form}
    \begin{aligned}
    \text{node to hyperedge message}:b^{i\to e}_{\psi_i}&\propto n_{\psi_i}\prod_{f\in \partial i/e\cap \boldsymbol{E}}\hat{b}^{f\to i}_{\psi_i} \\
    \text{hyperedge to node message}:\hat{b}^{e\to i}_{\psi_i}&\propto\sum_{\overrightarrow{\psi_{e/i}}\in[q]^{|e|-1}}\left(\bm{\Omega}_{\overrightarrow{\psi_e}}\right)^{A_e}\left(1-\bm{\Omega}_{\overrightarrow{\psi_e}}\right)^{1-A_e}\prod_{j\in e/i}b^{j\to e}_{\psi_j}\\
    \text{node marginal}:b^{i}_{\psi_i}&\propto n_{\psi_i}\prod_{f\in \partial i\cap \boldsymbol{E}}\hat{b}^{f\to i}_{\psi_i}
    \end{aligned} \enspace ,
\end{equation}
where $A_e=1$ only if $e\in \calE$, otherwise $A_e=0$. $\overrightarrow{\psi_{e/i}}$ means the community assignment for the nodes in hyperedge $e$ except node $i$, $n_{\psi_i}$ means number of nodes in community $\psi_i$ by percentage. We illustrate $b^{i\to e}$, $\hat{b}^{e\to i}$ and $b^{i}$ with factor graph~\cite{mezard2009information} in Figure~\ref{figBPillustration}.
Next we discuss the $\hat{b}^{e\to i}_{\psi_i}$ based on existence of hyperedge $e$.
\subsection{Discussion on hyperedge to node message}\label{appendix-discuss_bp_h2n}
When hyperedge $e$ exist, $A_e=1$, we have
\begin{equation}\label{betoi_einE}
    \begin{aligned}
    \hat{b}^{e\to i}_{\psi_i}&\propto \sum_{\overrightarrow{\psi_{e/i}}\in[q]^{|e|-1}}\bm{\Omega}_{\overrightarrow{\psi_e}}\prod_{j\in e/i}b^{j\to e}_{\psi_j}\\
    &\propto \sum_{\overrightarrow{\psi_{e/i}}\in[q]^{|e|-1}}\bm{C}_{\overrightarrow{\psi_e}}\prod_{j\in e/i}b^{j\to e}_{\psi_j}
    \end{aligned} \enspace .
\end{equation}
However if $A_e=0$, then
\begin{equation}\label{betoi_1}
    \begin{aligned}
    \hat{b}^{e\to i}_{\psi_i}&\propto \sum_{\overrightarrow{\psi_{e/i}}\in[q]^{|e|-1}}\left(1-\bm{\Omega}_{\overrightarrow{\psi_e}}\right)\prod_{j\in e/i}b^{j\to e}_{\psi_j}\\
    &= \sum_{\overrightarrow{\psi_{e/i}}\in[q]^{|e|-1}}\left(1-\frac{\bm{C}_{\overrightarrow{\psi_e}}}{n^{|e|-1}}\right)\prod_{j\in e/i}b^{j\to e}_{\psi_j}\\
    &=\sum_{\overrightarrow{\psi_{e/i}}\in[q]^{|e|-1}}\prod_{j\in e/i}b^{j\to e}_{\psi_j}-\sum_{\overrightarrow{\psi_{e/i}}\in[q]^{|e|-1}}\frac{\bm{C}_{\overrightarrow{\psi_e}}}{n^{|e|-1}}\prod_{j\in e/i}b^{j\to e}_{\psi_j}\\
    &=\prod_{j\in e/i}\sum_{\psi_{j}\in[q]}b^{j\to e}_{\psi_j}-\sum_{\overrightarrow{\psi_{e/i}}\in[q]^{|e|-1}}\frac{\bm{C}_{\overrightarrow{\psi_e}}}{n^{|e|-1}}\prod_{j\in e/i}b^{j\to e}_{\psi_j}\\
    &=1-\sum_{\overrightarrow{\psi_{e/i}}\in[q]^{|e|-1}}\frac{\bm{C}_{\overrightarrow{\psi_e}}}{n^{|e|-1}}\prod_{j\in e/i}b^{j\to e}_{\psi_j}
    \end{aligned}\enspace ,
\end{equation}
where the last line use $\sum_{\psi_{j}\in[q]}b^{j\to e}_{\psi_j}=1$. According to sparse assumption, $\bm{C}_{\overrightarrow{\psi_e}}$ is a constant compared to $n$, so
\begin{equation}\label{limbetoi}
    \begin{aligned}
        \lim_{n\to\infty}\hat{b}^{e\to i}_{\psi_i}=\frac{1}{q}
    \end{aligned}\enspace .
\end{equation}
According to~\eqref{bp_form}, we have
\begin{equation}\label{relation_bjtoe_bj}
    \begin{aligned}
    b^{j\to e}_{\psi_j}&\propto n_{\psi_j}\prod_{f\in \partial j/e}\hat{b}^{f\to j}_{\psi_j} \\
    &=\frac{n_{\psi_j}}{\hat{b}^{e\to j}_{\psi_j}}\prod_{f\in \partial j}\hat{b}^{f\to j}_{\psi_j}\\
    &=\frac{b^{j}_{\psi_j}}{\hat{b}^{e\to j}_{\psi_j}}
    \end{aligned}\enspace .
\end{equation}
Based on~\eqref{limbetoi} and~\eqref{relation_bjtoe_bj}, we have
\begin{equation}\label{limbjtoe}
    \lim_{n\to \infty}b^{j\to e}_{\psi_j}=b^{j}_{\psi_j} \enspace .
\end{equation}
We substitute~\eqref{limbjtoe} into~\eqref{betoi_1} and consider $1+x\approx \mathrm{exp}(x)$ when $x\to 0$, then when $n\to \infty$ 
\begin{equation}\label{betoi_2}
    \hat{b}^{e\to i}_{\psi_i}=\mathrm{exp}\left({-\frac{1}{n^{|e|-1}}\sum_{\overrightarrow{\psi_{e/i}}\in[q]^{|e|-1}}\bm{C}_{\overrightarrow{\psi_e}}\prod_{j\in e/i}b^{j}_{\psi_j}}\right)\enspace .
\end{equation}
\subsection{Update of node to hyperedge message and external field}\label{section_externalfield}
Based on $\hat{b}^{e\to i}_{\psi_i}$ in different case~\eqref{betoi_einE} and~\eqref{betoi_2}, we can update node to hyperedge message as
\begin{equation}
    \begin{aligned}
    b^{i\to e}_{\psi_i}&\propto n_{\psi_i}\prod_{f\in \partial i/e\cap \boldsymbol{E}}\hat{b}^{f\to i}_{\psi_i}\\
    &=n_{\psi_i}\prod_{f\in \partial i/e \cap \calE}\hat{b}^{f\to i}_{\psi_i}\prod_{f\in \partial i/e \cap \boldsymbol{E}/\calE}\hat{b}^{f\to i}_{\psi_i}\\
    &=n_{\psi_i}\prod_{f\in \partial i/e \cap \calE}\hat{b}^{f\to i}_{\psi_i}\prod_{f\in \partial i/e \cap \boldsymbol{E}/\calE}\mathrm{exp}\left({-\frac{1}{n^{|f|-1}}\sum_{\overrightarrow{\psi_{f/i}}\in[q]^{|f|-1}}\bm{C}_{\overrightarrow{\psi_f}}\prod_{j\in f/i}b^{j}_{\psi_j}}\right)\\
    &=n_{\psi_i}\prod_{f\in \partial i/e \cap \calE}\hat{b}^{f\to i}_{\psi_i}\mathrm{exp}\left({-\sum_{f\in \partial i/e \cap \boldsymbol{E}/\calE}\frac{1}{n^{|f|-1}}\sum_{\overrightarrow{\psi_{f/i}}\in[q]^{|f|-1}}\bm{C}_{\overrightarrow{\psi_f}}\prod_{j\in f/i}b^{j}_{\psi_j}}\right)\\
    &=n_{\psi_i}\prod_{f\in \partial i/e \cap \calE}\hat{b}^{f\to i}_{\psi_i}\mathrm{exp}\left(-h^{i\to e}_{\psi_i}\right)
    \end{aligned}\enspace .
\end{equation}
Because of $e\in \calE$ and the existed hyperedge set $\calE$ is far smaller than all possible hyperedge set $\boldsymbol{E}$ in sparse regime, so 
\begin{equation}\label{general_external}
    h^{i\to e}_{\psi_i}=h^{i}_{\psi_i}={\sum_{f\in \partial i \cap \boldsymbol{E}}\frac{1}{n^{|f|-1}}\sum_{\overrightarrow{\psi_{f/i}}\in[q]^{|f|-1}}\bm{C}_{\overrightarrow{\psi_f}}\prod_{j\in f/i}b^{j}_{\psi_j}} \enspace .
\end{equation}
We call the $h^{i}_{\psi_i}$ as external field. The external field can not be reduced anymore for general tensor $\bm{C}$. However, considering the specific form of $\bm{C}$~\eqref{nonuniform_cincout} in symmetric HSBM, we can reduce $h^{i}_{\psi_i}$ to $h_{\psi_i}$
\begin{equation}\label{reduce_external}
    \begin{aligned}
        h^{i}_{\psi_i}&=\sum_{f\in \partial i \cap \boldsymbol{E}}\frac{1}{n^{|f|-1}}\left(c_{\rm in}\prod_{j\in f/i}b^{j}_{\psi_i}+c_{\rm out}\sum_{\overrightarrow{\psi_{f/i}}\in[q]^{|f|-1}/\underbracket{[\psi_i]}_{|f|-1}}\prod_{j\in f/i}b^{j}_{\psi_j}\right)\\
        &=\sum_{\kappa\in \calK}\frac{1}{n^{\kappa-1}}\sum_{f\in \boldsymbol{E}^{(\kappa-1)}}\left(c_{\rm in}\prod_{j\in f}b^{j}_{\psi_i}+c_{\rm out}\sum_{\overrightarrow{\psi_{f}}\in[q]^{\kappa-1}/\underbracket{[\psi_i]}_{\kappa-1}}\prod_{j\in f}b^{j}_{\psi_j}\right)\\
        &=\sum_{\kappa\in \calK}\frac{1}{n^{\kappa-1}}\sum_{f\in \boldsymbol{E}^{(\kappa-1)}}\left(c_{\rm out}+(c_{\rm in}-c_{\rm out})\prod_{j\in f}b^{j}_{\psi_i}\right)\\
        &\approx\sum_{\kappa\in \calK}\left[c_{\rm out}+\frac{c_{\rm in}-c_{\rm out}}{n^{\kappa-1}(\kappa-1)!}(\sum_{j\in[n]}b^j_{\psi_i})^{\kappa-1}\right]=h_{\psi_i}
    \end{aligned}\enspace ,
\end{equation}
where the second last line is derived based on $\sum_{\overrightarrow{\psi_{f}}\in[q]^{\kappa-1}}\prod_{j\in f}b^{j}_{\psi_j}=1$. The last line is derived with similar idea in paper~\cite{angelini2015spectral}.

\subsection{Update of node marginal and summary}
The node marginal $b^{i}_{\psi_i}$ is updated similarly to $b^{i\to e}_{\psi_i}$, 
\begin{equation}
    \begin{aligned}
    b^{i}_{\psi_i}&\propto n_{\psi_i}\prod_{f\in \partial i}\hat{b}^{f\to i}_{\psi_i}\\
    &=n_{\psi_i}\prod_{f\in \partial i \cap \calE}\hat{b}^{f\to i}_{\psi_i}\prod_{f\in \partial i \cap \boldsymbol{E}/\calE}\hat{b}^{f\to i}_{\psi_i}\\
    &=n_{\psi_i}\mathrm{exp}\left({-h^i_{\psi_i}}\right)\prod_{f\in \partial i \cap \calE}\hat{b}^{f\to i}_{\psi_i}
    \end{aligned}\enspace .
\end{equation}
Based on above derivation, we can summary the belief propagation as
\begin{equation}\label{reduced_bp}
    \begin{aligned}
    \text{node to hyperedge message}:b^{i\to e}_{\psi_i}&\propto n_{\psi_i}\mathrm{exp}\left({-h^i_{\psi_i}}\right)\prod_{f\in \partial i/e \cap \calE}\hat{b}^{f\to i}_{\psi_i} \\
    \text{hyperedge to node message}:\hat{b}^{e\to i}_{\psi_i}&\propto\sum_{\overrightarrow{\psi_{e/i}}\in[q]^{|e|-1}}\bm{C}_{\overrightarrow{\psi_e}}\prod_{j\in e/i}b^{j\to e}_{\psi_j}\\
    \text{external field}: h^i_{\psi_i}&=\sum_{f\in \partial i \cap \boldsymbol{E}}\frac{1}{n^{|f|-1}}\sum_{\overrightarrow{\psi_{f/i}}\in[q]^{|f|-1}}\bm{C}_{\overrightarrow{\psi_f}}\prod_{j\in f/i}b^{j}_{\psi_j}\\
    \text{node marginal}:b^{i}_{\psi_i}&\propto n_{\psi_i}\mathrm{exp}\left({-h^i_{\psi_i}}\right)\prod_{f\in \partial i\cap \calE}\hat{b}^{f\to i}_{\psi_i}
    \end{aligned}
\end{equation}
In practice, we apply dropout on the messages with same setting in Ruggeri et al.~\cite{ruggeri2024message}.

\subsection{The detectability of belief propagation}
Similarly with paper~\cite{decelle2011asymptotic} and~\cite{ruggeri2024message}, we get the  detectability by deriving the stability of trivial fixed point in belief propagation. We will focus on symmetric HSBM case~\eqref{nonuniform_cincout}. In this case, $n_{\psi_i}=\frac{1}{q}$ and we can verify that there is a trivial fixed point of belief propagation based on~\eqref{reduced_bp} and external field~\eqref{reduce_external}
\begin{equation}
    \begin{aligned}
    b^{i\to e}_{\psi_i}&=\frac{1}{q}\\
    \hat{b}^{e\to i}_{\psi_i}&=\frac{1}{q}\\
    b^{i}_{\psi_i}&=\frac{1}{q}
    \end{aligned} \enspace .
\end{equation}
This trivial fixed point means nothing but any node belong to any community with same probability, so we need this trivial fixed point should be unstable. One way to discuss the stability of the fixed point is to get the largest modular eigenvalue of Jacobian Matrix of belief propagation~\cite{moore2017computer}. Here we use another approach which is inherited from~\cite{decelle2011asymptotic} and~\cite{ruggeri2024message}.

\begin{figure}[H]
    \centering
    \includegraphics[width=8cm]{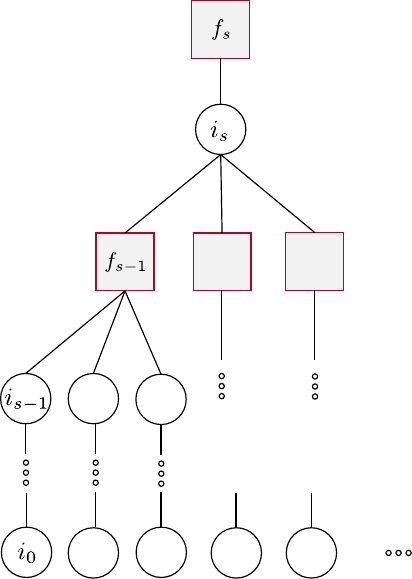}\\
	\caption{\small\it The approximate message passing tree in belief propagation.\label{figBPtree}}
\end{figure}

We assume the message passing as a tree with depth $s$, as shown in Figure~\ref{figBPtree}. We will check if the node to hyperedge message perturbation around trivial fixed point will be amplified or reduced during its propagation from all the leaves to root. First, we need to know how the perturbation propagate between adjacent node to hyperedge message, such as $b^{i_s\to f_s}$ and $b^{i_{s-1}\to f_{s-1}}$ in Figure~\ref{figBPtree}. This propagation progress can be formalized as transition matrix $\hat{\rm T}$ with dimension $q\times q$, defined as
\begin{equation}
    \hat{\rm T}^{i\to e;j\to f}_{x;y}=\frac{\partial b^{i\to e}_x}{\partial b^{j\to f}_y}\enspace ,
\end{equation}
where $e\neq f$ and $i\in f/j$, as this short message passing path is $b^{j\to f}\to \hat{b}^{f\to i}\to b^{i\to e}$. Next we will derive this transition matrix at trivial fixed point.

\subsubsection{Transition matrix at trivial fixed point}\label{transitionmatrix}
Based on belief propagation~\eqref{reduced_bp}, we can write the full update function of $b^{i\to e}_x$ without hyperedge to node message as
\begin{equation}\label{full_nodetohyper_message}
    b^{i\to e}_{x}=\frac{1}{Z^{i\to e}} n_{x}\mathrm{exp}({-h^i_{x}})\prod_{g\in \partial i/e \cap \calE}\sum_{\overrightarrow{\psi_{g/i}}\in[q]^{|g|-1},\overrightarrow{\psi_i}=x}\bm{C}_{\overrightarrow{\psi_g}}\prod_{l\in g/i}b^{j\to g}_{\psi_l} \enspace ,
\end{equation}
where $Z^{i\to e}$ is normalization term over community label $x$. We define the numerator part of~\eqref{full_nodetohyper_message} as $\beta^{i\to e}_{x}$, then the transition matrix can be transformed as
\begin{equation}\label{hatT}
    \begin{aligned}
        \hat{\rm T}^{i\to e;j\to f}_{x;y}&=\frac{\partial b^{i\to e}_x}{\partial b^{j\to f}_y}\\
        &=\frac{1}{Z^{i\to e}}\frac{\partial \beta^{i\to e}_x}{\partial b^{j\to f}_y}-\frac{1}{(Z^{i\to e})^2}\frac{\partial Z^{i\to e}}{\partial b^{j\to f}_y}\beta^{i\to e}_x\\
    \end{aligned}\enspace .
\end{equation}
In symmetric HSBM case, the external field $h^i_{\psi_i}$ is a constant at trivial fixed point based on~\eqref{reduce_external}
\begin{equation}
    \begin{aligned}
        h^i_{\psi_i}=h_{\psi_i}&=\sum_{\kappa\in \calK}\left[c_{\rm out}+\frac{c_{\rm in}-c_{\rm out}}{n^{\kappa-1}(\kappa-1)!}(\sum_{j\in[n]}\frac{1}{q})^{\kappa-1}\right]\\
        &=\sum_{\kappa\in \calK}\left[c_{\rm out}+\frac{c_{\rm in}-c_{\rm out}}{q^{\kappa-1}(\kappa-1)!}\right]=h
    \end{aligned}\enspace .
\end{equation}
The average $\kappa$-order degree~\eqref{symuniform_ck} is also a constant. Based on~\eqref{hatcak}, we have
\begin{equation}
    \sum_{\overrightarrow{\psi_{g/i}}\in[q]^{|g|-1},\overrightarrow{\psi_i}=x}\bm{C}_{\overrightarrow{\psi_g}}\prod_{l\in g/i}\frac{1}{q}=d^{(|g|)}(|g|-1)! \enspace .
\end{equation}
Then at trivial fixed point, $\beta^{i\to e}_{x}$ is
\begin{equation}\label{beta}
\begin{aligned}
    \beta^{i\to e}_{x}&=n_{x}\mathrm{exp}({-h^i_{x}})\prod_{g\in \partial i/e \cap \calE}\sum_{\overrightarrow{\psi_{g/i}}\in[q]^{|g|-1},\overrightarrow{\psi_i}=x}\bm{C}_{\overrightarrow{\psi_g}}\prod_{l\in g/i}\frac{1}{q}\\
    &=\frac{\mathrm{exp}(-h)}{q}\prod_{g\in \partial i/e \cap \calE}d^{(|g|)}(|g|-1)!
\end{aligned}\enspace ,
\end{equation}
and $Z^{i\to e}$ is
\begin{equation}\label{Z}
    \begin{aligned}
        Z^{i\to e}&=\sum_{x\in [q]}\beta^{i\to e}_{x}\\
        &=\sum_{x\in [q]}\frac{\mathrm{exp}(-h)}{q}\prod_{g\in \partial i/e \cap \calE}d^{(|g|)}(|g|-1)!\\
        &=\mathrm{exp}(-h)\prod_{g\in \partial i/e \cap \calE}d^{(|g|)}(|g|-1)!
    \end{aligned}\enspace .
\end{equation}
Next the partial derivative $\frac{\partial \beta^{i\to e}_x}{\partial b^{j\to f}_y}$ is
\begin{equation}
    \begin{aligned}
        \frac{\partial \beta^{i\to e}_x}{\partial b^{j\to f}_y}=& \frac{\partial\left\{ n_{x}\mathrm{exp}({-h^i_{x}})\prod_{g\in \partial i/e \cap \calE}\sum_{\overrightarrow{\psi_{g/i}}\in[q]^{|g|-1},\overrightarrow{\psi_i}=x}\bm{C}_{\overrightarrow{\psi_g}}\prod_{l\in g/i}b^{l\to g}_{\psi_l} \right\} }{\partial b^{j\to f}_y} \\
        =& \frac{\mathrm{exp}({-h})}{q}\left(\prod_{g\in \partial i/e,f \cap \calE}\sum_{\overrightarrow{\psi_{g/i}}\in[q]^{|g|-1},\overrightarrow{\psi_i}=x}\bm{C}_{\overrightarrow{\psi_g}}\prod_{l\in g/i}b^{l\to g}_{\psi_l}\right)\\ &\times \sum_{\overrightarrow{\psi_{f/i, j}}\in[q]^{|f|-2},\overrightarrow{\psi_i}=x, \overrightarrow{\psi_j}=y}\bm{C}_{\overrightarrow{\psi_f}}\prod_{l\in f/i,j}b^{l\to f}_{\psi_l}\\
        =&\frac{\mathrm{exp}({-h})}{q}\left(\prod_{g\in \partial i/e,f \cap \calE} d^{(|g|)}(|g|-1)! \right) \sum_{\overrightarrow{\psi_{f/i, j}}\in[q]^{|f|-2},\overrightarrow{\psi_i}=x, \overrightarrow{\psi_j}=y}\bm{C}_{\overrightarrow{\psi_f}}\prod_{l\in f/i,j}\frac{1}{q}\\
        =&\frac{\mathrm{exp}({-h})}{q}\left(\prod_{g\in \partial i/e,f \cap \calE} d^{(|g|)}(|g|-1)! \right)\hat{c}_{xy}^{(|f|)}(|f|-2)!
    \end{aligned}\enspace ,
\end{equation}
where the last line is based on definition~\eqref{hatcabk}. In symmetric HSBM case, $\hat{c}_{xy}^{(|f|)}$ is follow~\eqref{hatcincout}, so we have
\begin{equation}\label{partialbetapartialb}
    \begin{aligned}
        \frac{\partial \beta^{i\to e}_x}{\partial b^{j\to f}_y}=& \frac{\mathrm{exp}({-h})}{q}\left(\prod_{g\in \partial i/e,f \cap \calE} d^{(|g|)}(|g|-1)! \right)\frac{c_{\rm in}\delta_{x,y}+c_{\rm out}(1-\delta_{x, y})+(q^{|f|-2}-1)c_{\rm out}}{q^{|f|-2}}
    \end{aligned}\enspace ,
\end{equation}
where $\delta_{x,y}=1$ if $x=y$. Then we have
\begin{equation}\label{partialZpartialb}
    \begin{aligned}
        \frac{\partial Z^{i\to e}}{\partial b^{j\to f}_y}&=\sum_{x\in [q]}\frac{\partial\beta^{i\to e}_x}{\partial b^{j\to f}_y}\\
        &=\mathrm{exp}({-h})\left(\prod_{g\in \partial i/e,f \cap \calE} d^{(|g|)}(|g|-1)! \right)\left(\frac{c_{\rm in}+(q^{|f|-2}-1)c_{\rm out}}{q^{|f|-1}} + \frac{c_{\rm out}}{q}(q-1)\right)\\
        &=\mathrm{exp}({-h})\left(\prod_{g\in \partial i/e,f \cap \calE} d^{(|g|)}(|g|-1)! \right)\left(\frac{c_{\rm in}+(q^{|f|-1}-1)c_{\rm out}}{q^{|f|-1}}  \right)
    \end{aligned}
\end{equation}
Finally substitute~\eqref{beta},~\eqref{Z},~\eqref{partialbetapartialb},~\eqref{partialZpartialb} into~\eqref{hatT}, we get the transition matrix at trivial fixed point
\begin{equation}
    \begin{aligned}
        \hat{\rm T}^{i\to e;j\to f}_{x;y}=&\frac{1}{Z^{i\to e}}\frac{\partial \beta^{i\to e}_x}{\partial b^{j\to f}_y}-\frac{1}{(Z^{i\to e})^2}\frac{\partial Z^{i\to e}}{\partial b^{j\to f}_y}\beta^{i\to e}_x\\
        =&\frac{\frac{\mathrm{exp}({-h})}{q}\left(\prod_{g\in \partial i/e,f \cap \calE} d^{(|g|)}(|g|-1)! \right)\frac{c_{\rm in}\delta_{x,y}+c_{\rm out}(1-\delta_{x, y})+(q^{|f|-2}-1)c_{\rm out}}{q^{|f|-2}}}{\mathrm{exp}(-h)\prod_{g\in \partial i/e \cap \calE}d^{(|g|)}(|g|-1)!}\\
        &-\frac{\mathrm{exp}({-h})\left(\prod_{g\in \partial i/e,f \cap \calE} d^{(|g|)}(|g|-1)! \right)\left(\frac{c_{\rm in}+(q^{|f|-1}-1)c_{\rm out}}{q^{|f|-1}}  \right)}{q\mathrm{exp}(-h)\prod_{g\in \partial i/e \cap \calE}d^{(|g|)}(|g|-1)!}\\
        =&\frac{1 }{qd^{(|f|)}(|f|-1)!}\frac{c_{\rm in}\delta_{x,y}+c_{\rm out}(1-\delta_{x, y})+(q^{|f|-2}-1)c_{\rm out}}{q^{|f|-2}}\\
        &-\frac{1 }{qd^{(|f|)}(|f|-1)!}\frac{c_{\rm in}+(q^{|f|-1}-1)c_{\rm out}}{q^{|f|-1}}\\
        =&\frac{1}{d^{(|f|)}(|f|-1)!}\frac{qc_{\rm in}\delta_{x,y}+qc_{\rm out}(1-\delta_{x, y})+q(q^{|f|-2}-1)c_{\rm out} - c_{\rm in}-(q^{|f|-1}-1)c_{\rm out}}{q^{|f|}}\\
        =&\frac{1}{d^{(|f|)}(|f|-1)!}\frac{c_{\rm in}(q\delta_{x,y}-1)+c_{\rm out}(1-q\delta_{x, y})}{q^{|f|}}\\
        =&\frac{1}{d^{(|f|)}(|f|-1)!}\frac{(q\delta_{x,y}-1)(c_{\rm in}-c_{\rm out})}{q^{|f|}}
    \end{aligned} \enspace .
\end{equation}
We define $\rm J$ is all ones matrix, then the transition matrix $\hat{\rm T}$ is
\begin{equation}
\begin{aligned}
    \hat{\rm T}^{i\to e;j\to f} &= \frac{1}{d^{(|f|)}(|f|-1)!q^{|f|}}\left[q(c_{\rm in}-c_{\rm out})\mathrm{I}+(c_{\rm out}-c_{\rm in})\mathrm{J}\right] \\
    &=\frac{1}{d^{(|f|)}(|f|-1)!q^{|f|}}\mathrm{T}
\end{aligned}\enspace .
\end{equation}

\subsubsection{Stability of trivial fixed point}\label{appendix-stabilityBPfixedpoint}
We first focus on the message passing path in Figure~\ref{figBPtree} from $i_0$ to $f_s$. We define $\varepsilon^r$ as the perturbation of message $b^{i_r\to f_r}$ around trivial fixed point, as $b^{i_r\to f_r}=\frac{1}{q}+\varepsilon^r$. Then we have
\begin{equation}
    \begin{aligned}
        \varepsilon^{r+1} &= \hat{\rm T}^{i_{r+1}\to f_{r+1};i_{r}\to f_{r}}\varepsilon^{r}\\
        &=\frac{1}{d^{(|f_r|)}(|f_r|-1)!q^{|f_r|}}\mathrm{T}\varepsilon^{r}
    \end{aligned}\enspace .
\end{equation}
The perturbation $\varepsilon^0$ in the leaf will propagate $s$ times to the $\varepsilon^s$, so
\begin{equation}\label{ep0toeps}
\begin{aligned}
    \varepsilon^s &= \left(\prod_{r=0}^{s-1}\frac{1}{d^{(|f_r|)}(|f_r|-1)!q^{|f_r|}}\right)\mathrm{T}^s\varepsilon^0 \\
\end{aligned}\enspace .
\end{equation}
When $s\to \infty$, we define
\begin{equation}
    \begin{aligned}
        \mu&=\lim_{s\to\infty}\prod_{r=0}^{s-1}\frac{1}{d^{(|f_r|)}(|f_r|-1)!q^{|f_r|}}\\
        &=\mathrm{exp}\left(\lim_{s\to\infty}\sum_{r=0}^{s-1}\mathrm{log}\frac{1}{d^{(|f_r|)}(|f_r|-1)!q^{|f_r|}}\right)\\
        &\approx\mathrm{exp}\left(s\mathbb{E}_{f}\left(\mathrm{log}\frac{1}{d^{(|f|)}(|f|-1)!q^{|f|}}\right)\right)
    \end{aligned}\enspace ,
\end{equation}
where $\mathbb{E}_{f}$ is the expectation over randomly drawn hyperedges $f\in \calE$. On the other hand, the leading eigenvalue of $\rm T$ is $\lambda=q(c_{\rm in}-c_{\rm out})$. So the equation~\eqref{ep0toeps} is approximated as
\begin{equation}\label{epsilontransition}
    \varepsilon^s = \mu\lambda^s\varepsilon^0 \enspace .
\end{equation}

Before proceeding with the following derivation, we observe that the perturbation of node marginal $b^{i}$ follow the same transition as perturbation of $b^{i\to e}$. That is equal to say the $\hat{\rm T}^{i;j\to f}=\hat{\rm T}^{i\to e;j\to f}$, where
\begin{equation}
    \hat{\rm T}^{i;j\to f}_{x;y} = \frac{\partial b^{i}_x}{\partial b^{j\to f}_y} \enspace.
\end{equation}
This can be derived with similar way in section~\ref{transitionmatrix}, just remove $e$ in each equation. 
So now, we can define $\varepsilon^r$ also to be the perturbation of $b^{i_r}$ except $r=0$, and the equation~\eqref{epsilontransition} still holds. However in this case, each node can average has $d$ hyperedges as its son nodes in message passing tree, each hyperedge still average has $\hat{\kappa}-1$ nodes as its son nodes in message passing tree. Then there are totally $\left[d(\hat{\kappa}-1)\right]^s$ leave nodes whose perturbation will finally affect $\varepsilon^s$ in root node $i_s$. Here, the $\hat{\kappa}$ is the average order of hyperedges. To derive $\hat{\kappa}$, we first estimate the expected number of order-$\kappa$ hyperedges $m^{(\kappa)}$ under the symmetric HSBM. 
As $n\to \infty$, this estimation yields:
\begin{equation}\label{Ek}
    \begin{aligned}
        m^{(\kappa)}&=\lim_{n\to \infty}q\binom{\frac{n}{q}}{\kappa } \frac{\cin}{n^{\kappa -1}}+(\binom{n}{\kappa }-q\binom{\frac{n}{q}}{\kappa })\frac{\cout}{n^{\kappa -1}}\\
        &=\frac{n}{\kappa!q^{\kappa-1}}\cin+\frac{n(q^{\kappa-1}-1)}{\kappa!q^{\kappa-1}}\cout\\
        &=\frac{n}{\kappa}d^{(\kappa)}
    \end{aligned}\enspace ,
\end{equation}
where $d^{(\kappa)}$ is from~\eqref{d_kappa} and the asymptotic limits of the combinatorial parts are given by:
\begin{equation}
    \begin{aligned}
        \lim_{n\to \infty}q\binom{\frac{n}{q}}{\kappa }\frac{1 }{n^{\kappa-1 }}&=\lim_{n\to \infty}\frac{(n/q)!}{\kappa !(n/q-\kappa )!}\frac{q}{n^{\kappa-1} }=\frac{1}{(\kappa)!}\frac{(n/q)^\kappa q}{n^{\kappa-1} }=\frac{n}{\kappa!q^{\kappa -1}} \\
        \lim_{n\to \infty}\left(\binom{n}{\kappa }-q\binom{\frac{n}{q}}{\kappa }\right)\frac{1 }{n^{\kappa-1 }}&=\frac{n}{(\kappa)!}-\frac{n}{(\kappa)!q^{\kappa -1}}=\frac{n(q^{\kappa -1}-1)}{\kappa !q^{\kappa -1}}
    \end{aligned} \enspace .
\end{equation}
Then the average order $\hat{\kappa}$ is then calculated as the ratio of the total order of hyperedges to the total number of hyperedges:
\begin{equation}\label{hatk}
    \begin{aligned}
        \hat{\kappa}&=\frac{\sum_{\kappa\in \mathcal{K}}\kappa m^{(\kappa)}}{\sum_{\kappa\in \mathcal{K}}m^{(\kappa)}}\\
        &=\frac{\sum_{\kappa\in \mathcal{K}}nd^{(\kappa)}}{\sum_{\kappa\in \mathcal{K}}\frac{n}{\kappa}d^{(\kappa)}}\\
        &=\frac{d}{\sum_{\kappa\in \calK}\frac{d^{(\kappa)}}{\kappa}}
    \end{aligned}\enspace .
\end{equation}

Next we aggregate perturbations in all leaves in the tree. Because the expectation of any perturbation $\varepsilon$ is 0, we consider its variance
\begin{equation}
    \begin{aligned}
        \mathbb{E}\left((\varepsilon^s)^2\right) &= \mathbb{E}\left(\left(\sum_{i=1}^{[d(\hat{\kappa}-1)]^s}\mu\lambda^s\varepsilon^{0;i}\right)^2\right)\\
        &\approx \mu^2\lambda^{2s}\left[d(\hat{\kappa}-1)\right]^s\mathbb{E}\left((\varepsilon^0)^2\right)
    \end{aligned}\enspace .
\end{equation}
The stability of trivial fixed point is decided by
\begin{equation}\label{stability_fixedpoint}
    \begin{aligned}
    \sqrt[s]{\mu^2\lambda^{2s}[d(\hat{\kappa}-1)]^s}&<1\\
    \mathrm{exp}\left({2\mathbb{E}\left(\log(\frac{1}{d^{|f|}(|f|-1)!q^{|f|}})\right)}\right)q^2(c_{\rm in}-c_{\rm out})^2[d(\hat{\kappa}-1)]&<1\\
    \end{aligned}\enspace ,
\end{equation}
where 
\begin{equation}\label{expect_log}
    \begin{aligned}
        \mathbb{E}\left(\log(\frac{1}{d^{|f|}(|f|-1)!q^{|f|}})\right)&=\sum_{\kappa\in \calK}\frac{m^{(\kappa)}}{\sum_{\kappa'\in \calK}m^{(\kappa')}}log(\frac{1}{d^{(\kappa)}(\kappa-1)!q^{\kappa}})\\
        &=\sum_{\kappa\in \calK}\frac{\frac{n}{\kappa}d^{(\kappa)}}{\sum_{\kappa'\in \calK}\frac{n}{\kappa'}d^{(\kappa')}}log(\frac{1}{d^{(\kappa)}(\kappa-1)!q^{\kappa}})\\
        &=\sum_{\kappa\in \calK}\frac{d^{(\kappa)}/\kappa}{\sum_{\kappa'\in \calK}\frac{d^{(\kappa')}}{\kappa'}}\log(\frac{1}{d^{(\kappa)}(\kappa-1)!q^{\kappa}})\\
        &=\sum_{\kappa\in \calK}\frac{\hat{\kappa}d^{(\kappa)}}{d\kappa}\log(\frac{1}{d^{(\kappa)}(\kappa-1)!q^{\kappa}})
        \end{aligned}\enspace .
\end{equation}
Here number of order-$\kappa$ hyperedge $m^{(\kappa)}$ is from equation~\eqref{Ek}, average order~\eqref{hatk} $\hat{\kappa}=\frac{d}{\sum_{\kappa\in \calK}\frac{d^{(\kappa)}}{\kappa}}$ and average degree~\eqref{averaged} $d=\sum_{\kappa\in \calK}d^{(\kappa)}$. Because the detectability is equal to instability of trivial fixed point, so based on equation~\eqref{stability_fixedpoint} and~\eqref{expect_log}, the detectability threshold of belief propagation is
\begin{equation}\label{appendix-bp-detectabilitythreshold}
\begin{aligned}
    q^2(c_{\rm in}-c_{\rm out})^2[d(\hat{\kappa}-1)]\prod_{\kappa\in \calK}\left(\frac{1}{d^{(\kappa)}(\kappa-1)!q^{\kappa}}\right)^{2\frac{\hat{\kappa}d^{(\kappa)}}{d\kappa}}&>1 \\
    d(\hat{\kappa}-1)\prod_{\kappa\in \calK}\left(\frac{q(\cin-\cout)}{d^{(\kappa)}(\kappa-1)!q^{\kappa}}\right)^{2\frac{\hat{\kappa}d^{(\kappa)}}{d\kappa}}&>1\\
    d(\hat{\kappa}-1)\prod_{\kappa\in \calK}\left(\frac{(\cin-\cout)}{d^{(\kappa)}(\kappa-1)!q^{\kappa-1}}\right)^{2\frac{\hat{\kappa}d^{(\kappa)}}{d\kappa}}&>1\\
    d(\hat{\kappa}-1)\prod_{\kappa\in \calK}\left(\frac{\din^{(\kappa)}-\dout^{(\kappa)})}{d^{(\kappa)}}\right)^{2\frac{\hat{\kappa}d^{(\kappa)}}{d\kappa}}&>1\\
\end{aligned}\enspace ,
\end{equation}
where the simplification leading to the second line relies on:
\begin{equation}
    \sum_{\kappa \in \calK}\frac{\hat{\kappa}d^{(\kappa)}}{d\kappa}=1 \enspace, 
\end{equation}
while the last line is a direct result of substituting the explicit expressions for $\din^{(\kappa)}$ and $\dout^{(\kappa)}$ as defined in~\eqref{dindout_cincout}. We denote the left formula of inequality~\eqref{appendix-bp-detectabilitythreshold} as
\begin{equation}
    \snr_{\rm BP}:=d(\hat{\kappa}-1)\prod_{\kappa\in \calK}\left(\frac{\din^{(\kappa)}-\dout^{(\kappa)})}{d^{(\kappa)}}\right)^{2\frac{\hat{\kappa}d^{(\kappa)}}{d\kappa}} \enspace .
\end{equation}

\section{Supplementary experiment: detectability limits of symmetric HSBM} 
We use numeric experiment to verify the detectability limit of Bethe Hessian-based spectral clustering~\eqref{phi_by_dindout} and belief propagation~\eqref{detectability_bp}.
First, we run experiment on $\kappa $-uniform hyper graph to verify the detectability limit~\eqref{symuniform_conjectureddetectability} in literature~\cite{angelini2015spectral}. Then we run experiment on nonuniform hyper graph with $q=2$ to verify the $\snr_{\rm BH}$~\eqref{phi_by_dindout} in literature~\cite{chodrow2023nonbacktracking} and $\snr_{\rm BP}$~\eqref{detectability_bp} in this paper. 

We use $\rm AMI$~\cite{vinh2009information} to evaluate the community detection result. Assume $\bm \psi^*$ is the real community label vector and $\bm \psi$ is detected community label vector. Then $\rm AMI$ is defined by
\begin{equation}
    \mathrm{AMI}(\bm \psi, \bm \psi^*)=\frac{\mathrm{MI}(\bm \psi, \bm \psi^*)-\mathbb{E}(\mathrm{MI}(\bm \psi, \bm \psi^*))}{\frac{1}{2}(H(\bm \psi) + H(\bm \psi^*))-\mathbb{E}(\mathrm{MI}(\bm \psi, \bm \psi^*))}\enspace ,
\end{equation}
where $\mathrm{MI}$ represent mutual information, $\mathbb{E}$ means the expectation. The $\rm AMI$ score lies in range $[0, 1]$, with 0 means the detected communities have no difference to randomly detection, 1 means the detected communities is perfect fit the true communities.

For each of our experiments in this chapter, we fix the average degree $d$ of hyper graph. 
Assume a hyper graph generated by nonuniform symmetric HSBM has average degree $d$, then 
\begin{equation}
\begin{aligned}
    d&=\frac{\sum_{\kappa \in \calK}q\binom{\frac{n}{q}}{\kappa }\kappa \frac{c_{\rm in}}{n^{\kappa -1}}+(\binom{n}{\kappa }-q\binom{\frac{n}{q}}{\kappa })\kappa \frac{c_{\rm out}}{n^{\kappa -1}}}{n}\\
    &=\sum_{\kappa \in \calK}q\binom{\frac{n}{q}}{\kappa }\frac{\kappa }{n^{\kappa }}c_{\rm in}+\left(\binom{n}{\kappa }-q\binom{\frac{n}{q}}{\kappa }\right)\frac{\kappa }{n^{\kappa }}c_{\rm out}
\end{aligned}\enspace .
\end{equation}
After given $d$ and setting $\epsilon=\frac{c_{\rm out}}{c_{\rm in}}$, we have
\begin{equation}\label{cincoutby_d_epsilon}
    \begin{aligned}
        c_{\rm in}&= \frac{d}{\sum_{\kappa \in \calK}q\binom{\frac{n}{q}}{\kappa }\frac{\kappa }{n^{\kappa }}+\epsilon\left(\binom{n}{\kappa }-q\binom{\frac{n}{q}}{\kappa }\right)\frac{\kappa }{n^{\kappa }}}\\
        c_{\rm out}&= \frac{d\epsilon}{\sum_{\kappa \in \calK}q\binom{\frac{n}{q}}{\kappa }\frac{\kappa }{n^{\kappa }}+\epsilon\left(\binom{n}{\kappa }-q\binom{\frac{n}{q}}{\kappa }\right)\frac{\kappa }{n^{\kappa }}}
    \end{aligned}\enspace .
\end{equation}
We can generate a hyper graph by sample from Bernoulli distribution with probability $c_{\rm in}/n^{\kappa-1}$ or $c_{\rm out}/n^{\kappa-1}$ for each of $\binom{n}{\kappa}$ possible hyperedges and $\kappa\in \calK$. When $n$ is large enough, the high time complexity $O(n^{\mathrm{max}(\calK)})$ of this method makes it difficult to apply. So we use another more efficient approach to generate hyper graph with HSBM. The details of the approach is presented in Appendix~\ref{efficientgeneration}.

Beyond the Bethe Hessian spectral clustering Algorithm~\ref{scwithbh}, we also implement the belief propagation with the assumption of symmetric HSBM based on the code in paper~\cite{ruggeri2024message}. The main difference is the calculation of external field~\eqref{reduce_external} shown in Appendix~\ref{section_externalfield} and the update of hyperedge to node message shown in Appendix~\ref{dp_hyperedgetonode}.

\subsection{Experiment on uniform hyper graph}
For a $\kappa $-uniform hyper graph generated by $\kappa $-uniform symmetric HSBM, we can get the critical $\epsilon^*$ at detectability threshold $\snr_{\rm BH}=1$~\eqref{phi_by_dindout} when $\mathcal{K}=\left\{\kappa\right\}$.
Considering assortative case $\din^{(\kappa)} > \dout^{(\kappa)}$, then at the detectability threshold, we have
\begin{equation}
    \begin{aligned}
        (\kappa-1)(\din^{(\kappa)}-\dout^{(\kappa)}) &= \sqrt{(\kappa -1)d^{(\kappa)}} \\
    \end{aligned} \enspace .
\end{equation}
Combined with average $\kappa$-degree~\eqref{d_kappa}, we can get the $\din^{(\kappa)}, \dout^{(\kappa)}$ at the detectability threshold:
\begin{equation}
    \begin{aligned}
        \din^{(\kappa)} &= \frac{1}{q^{\kappa-1}}\left(d^{(\kappa )}+(q^{\kappa -1}-1)\sqrt{\frac{d^{(\kappa )}}{\kappa -1}}\right)\\
        \dout^{(\kappa)} &= \frac{1}{q^{\kappa-1}}\left(d^{(\kappa )}-\sqrt{\frac{d^{(\kappa )}}{\kappa -1}}\right)\\
    \end{aligned} \enspace .
\end{equation}
With setting $\epsilon=\frac{c_{\rm out}}{c_{\rm in}}$ and relation between $\din^{(\kappa)}, \dout^{(\kappa)}$ and $\cin, \cout$~\eqref{dindout_cincout}, the critical $\epsilon^*$ is
\begin{equation}
    \epsilon^*=\frac{\dout^{(\kappa)}}{\din^{(\kappa)}}=\frac{\sqrt{d^{(\kappa )}(\kappa -1)}-1}{\sqrt{d^{(\kappa )}(\kappa -1)}+q^{\kappa -1}-1}
\end{equation}
which is also shown in paper~\cite{angelini2015spectral}. 

\begin{figure}
    \centering
    \includegraphics[width=15cm]{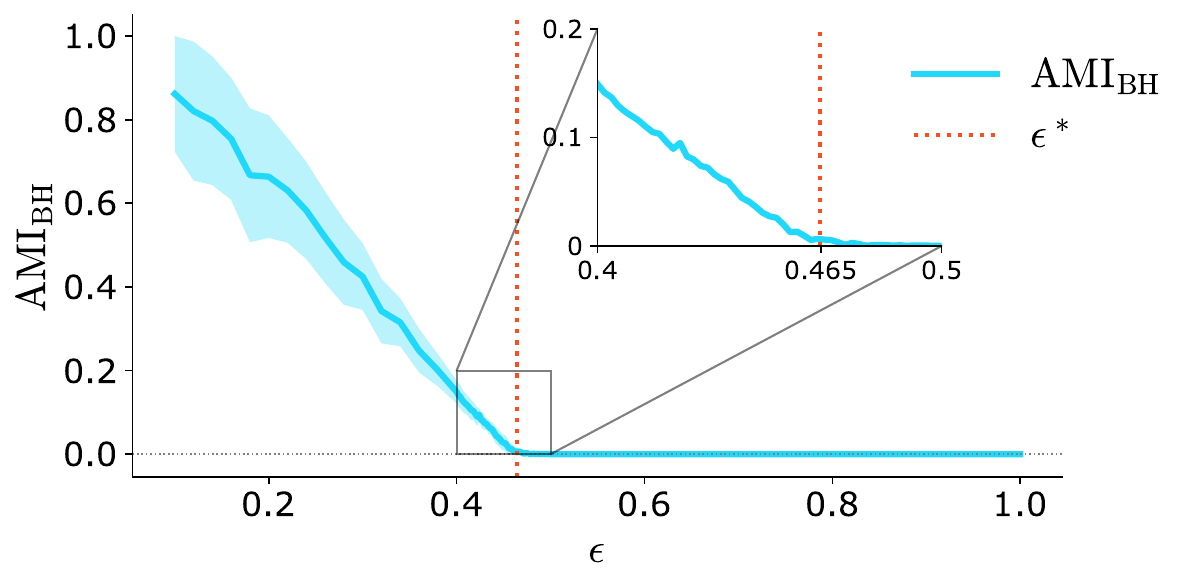}\\
	\caption{\small\it The experiment run on $3$-uniform hyper graph. The blue line is the $\rm AMI_{BH}$ at different $\epsilon$. For each point, we run experiment for 50 times; solid lines show means and shaded bands show 95\% confidence intervals. The orange vertical dash line is the critical $\epsilon^*$ where the detectability phase transition happen. \label{figAMIepsilonUniform}}
\end{figure}

We fix the $\kappa $-uniform hyper graph size $n=30000$, number of communities $q=2$, order $\kappa =3$ and average degree $d=10$. For different $\epsilon$, we calculate the $c_{\rm in}$ and $c_{\rm out}$ with equation~(\ref{cincoutby_d_epsilon}) and generate a hyper graph with $\kappa $-uniform symmetric HSBM.
Then we use Bethe Hessian method to detect communities of the hyper graph.  As the result shown in Figure~\ref{figAMIepsilonUniform}, the critical $\epsilon^*$ is close to the point that $\rm AMI_{BH}$ start to be 0. In this experiment, the $\epsilon^*=0.465$.

\subsection{Experiment on nonuniform hyper graph}\label{expfornonuniform}
We fix the nonuniform hyper graph size $n=30000$, number of communities $q=2$, $\calK=\left\{2, 3\right\}$ and average degree $d=10$. For different $\epsilon$, we calculate the $c_{\rm in}$ and $c_{\rm out}$ with equation~(\ref{cincoutby_d_epsilon}) and generate a hyper graph with nonuniform symmetric HSBM. 
Then we use Bethe Hessian spectral clustering and belief propagation to detect communities of the hyper graph, the result is recorded as $\rm AMI_{BH}$ and $\rm AMI_{BP}$. 

\begin{figure}
    \centering
    \includegraphics[width=15cm]{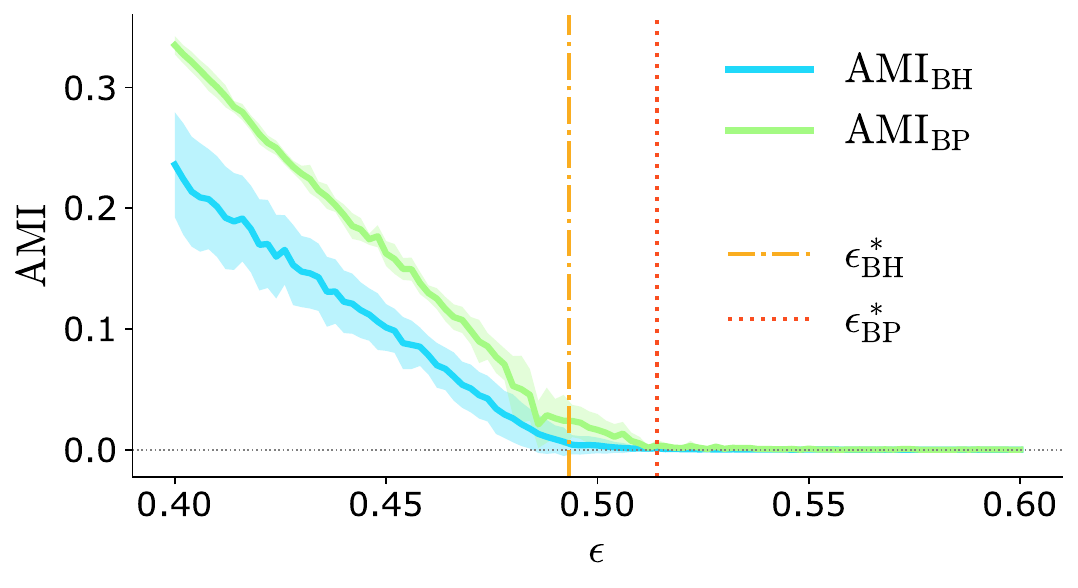}\\
	\caption{\small\it The experiment run on nonuniform hyper graph with 2 communities. For each $\epsilon$, we run bethe hessian method 100 times and belief propagation for 10 times; solid lines show means and shaded bands show 95\% confidence intervals. 
    \label{figAMIepsilonNonUniformk=2}}
\end{figure}

We define the critical $\epsilon_{\rm BH}^*$ and $\epsilon_{\rm BP}^*$ corresponding with $\snr_{\rm BH}=1$~\eqref{phi_by_dindout} and $\snr_{\rm BP}=1$~\eqref{detectability_bp}.
As the result shown in Figure~\ref{figAMIepsilonNonUniformk=2}, the $\epsilon_{\rm BH}^*$ is close to the point that $\rm AMI_{BH}$ start to be 0, the $\epsilon_{\rm BP}^*$ is close to the point that $\rm AMI_{BP}$ start to be 0. 
These results verify that the detectability $\snr_{\rm BH}$~\eqref{phi_by_dindout} of spectral clustering with Bethe Hessian is same as the non-backtracking matrix. And $\snr_{\rm BH}$ is weaker than detectability $\snr_{\rm BP}$~\eqref{detectability_bp} of belief propagation, which we believe is the theoretical detectability limit.

\subsection{Comparison with more baseline methods}
To evaluate the performance of our proposed method, we compare it against two established baselines: the Laplacian-based spectral clustering framework~\cite{zhou2006learning} and the h-Louvain algorithm, which is based on hypergraph modularity optimization~\cite{kaminski2024modularity}.
For the Laplacian, we give the true number of communities $q$ to the spectral clustering, which means we select the smallest $q$ eigenvectors of Laplacian matrix for clustering. The h-Louvain algorithm have its own model selection approach, so we can not restrict it to detect $q$ communities.

\begin{figure}
    \centering
    \includegraphics[width=15cm]{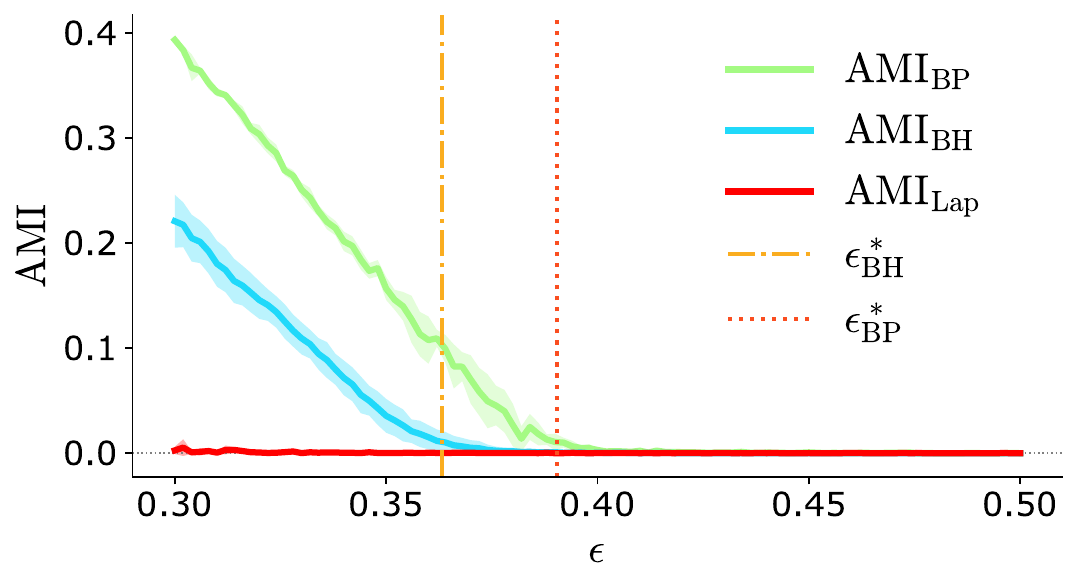}\\
	\caption{\small\it An comparison experiment with nonuniform hypergraphs with $n=30000$ nodes, $q=3$ communities, $\calK=\{2, 3\}$ hyperedge orders and $d=10$ mean node degree. For each $\epsilon$, we run Laplacian spectral clustering 10 times; solid lines show means and shaded bands show 95\% confidence intervals.\label{figAMIepsilonNonUniformLapk=3}}
\end{figure}

Under the large-scale parameter configuration used in Figure~\ref{fig_detectability_exp} ($n=30000, q=3$), the h-Louvain algorithm is computationally prohibitive due to its excessive execution time. Consequently, for this setting, we restrict our comparison to Laplacian spectral clustering, as illustrated in Figure~\ref{figAMIepsilonNonUniformLapk=3}.

\begin{figure}
    \centering
    \includegraphics[width=15cm]{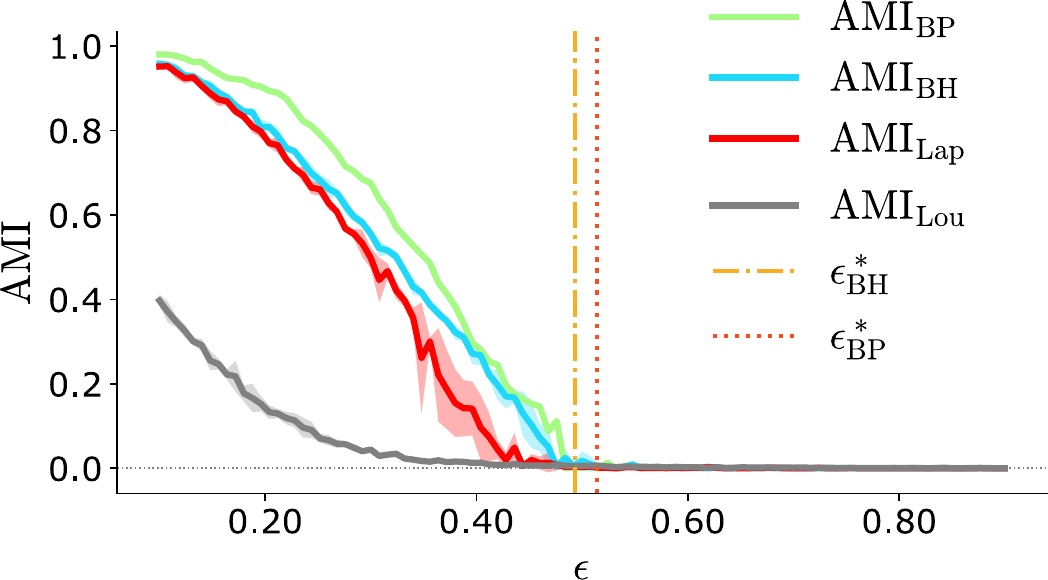}\\
	\caption{\small\it An comparison experiment with nonuniform hypergraphs with $n=3000$ nodes, $q=2$ communities, $\calK=\{2, 3\}$ hyperedge orders and $d=10$ mean node degree. For each $\epsilon$, we run each method 10 times; solid lines show means and shaded bands show 95\% confidence intervals.\label{figAMIepsilonNonUniformLapLouk=3}}
\end{figure}

\begin{figure}
    \centering
    \includegraphics[width=15cm]{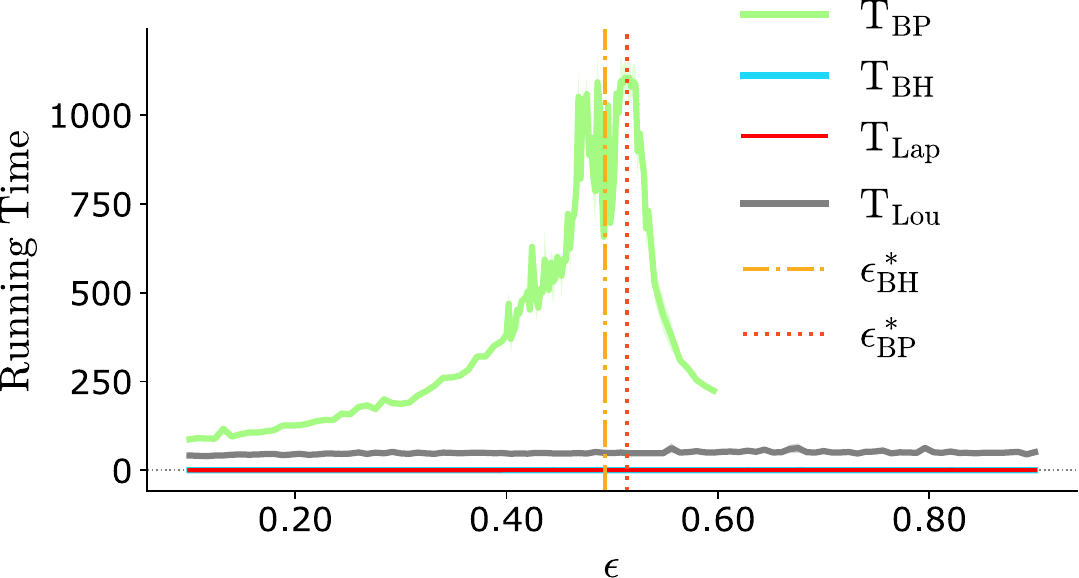}\\
	\caption{\small\it An running time comparison experiment with nonuniform hypergraphs with $n=3000$ nodes, $q=2$ communities, $\calK=\{2, 3\}$ hyperedge orders and $d=10$ mean node degree. For each $\epsilon$, we run each method 10 times; solid lines show means and shaded bands show 95\% confidence intervals.\label{figRunTimeepsilonNonUniformLapLouk=3}}
\end{figure}

To facilitate a comprehensive comparison that includes h-Louvain, we conducted additional experiments on a reduced scale with $n=3000$. These results are presented in Figure~\ref{figAMIepsilonNonUniformLapLouk=3}, where the performance of all four methods is evaluated. We also compare their running time in Figure~\ref{figRunTimeepsilonNonUniformLapLouk=3}.

The Figure~\ref{figAMIepsilonNonUniformLapk=3},  Figure~\ref{figAMIepsilonNonUniformLapLouk=3} illustrate that our proposed Bethe Hessian method has a better performance than Hypergraph Laplacian and Hypergraph Louvain and approaches the detectability limit more closely. Figure~\ref{figRunTimeepsilonNonUniformLapLouk=3} shows that BH is more efficient in general, particularly comparing with BP when it is close to the detectability limit.

\section{Implementation details of Experiment}
\subsection{Efficient approach to generate hyper graph with HSBM}\label{efficientgeneration}
The main idea for efficiently generating hyper graph with HSBM is to sample $\kappa$-order hyperedges for each possible combination of community assignment of the $\kappa$ nodes and $\kappa\in \calK$. We first need to find all possible ordered combination of $q$ communities that each community can exist multiple times. For example, if $q=2$, then the combination of community assignment for a 3-order hyperedge could be $\left\{ 
0, 0, 0 \right\}$, $\left\{ 
0, 0, 1 \right\}$, $\left\{ 
0, 1, 1 \right\}$, $\left\{ 
1, 1, 1 \right\}$.

Then, for each of community assignment $\overrightarrow{\psi}$, we calculate the expected number of generated hyperedge as
\begin{equation}
    \begin{aligned}
        m_{\overrightarrow{\psi}} = \sum_{\psi\in [q]}\binom{n\times n_\psi}{\#\left\{\text{$\psi$ in $\overrightarrow{\psi}$}\right\}}\bm{\Omega}_{\overrightarrow{\psi}}
    \end{aligned}\enspace ,
\end{equation}
where $\#\left\{\text{$\psi$ in $\overrightarrow{\psi}$}\right\}$ means the occurrence count of $\psi$ in the $\overrightarrow{\psi}$. 

Finally, for each of community assignment $\overrightarrow{\psi}$, we sample hyperedges $m_{\overrightarrow{\psi}}$ times. Each sampling we random select $\#\left\{\text{$\psi$ in $\overrightarrow{\psi}$}\right\}$ nodes in community $\psi$ for each $\psi \in \overrightarrow{\psi}$. With this approach, we reduced the time wasted on hyperedges that will not be generated.

\subsection{Dynamic programming for updating hyperedge to node message}\label{dp_hyperedgetonode}
The update of hyperedge to node message can be represented as
\begin{equation}
    \begin{aligned}
        \log(\hat{b}^{e\to i}_{\psi_i})=-\log Z^{e\to i}+\underbracket{\log(\sum_{\overrightarrow{\psi_{e/i}}\in[q]^{|e|-1}}\bm{C}_{\overrightarrow{\psi_e}}\prod_{j\in e/i}b^{j\to e}_{\psi_j})}_{\eta(e, i, \psi_i)}
    \end{aligned}\enspace ,
\end{equation}
where $Z^{e\to i}$ is the normalization term. For simplicity, we assume $e=\left\{1, 2, ..., \kappa\right\}$, $i=1$, then another term $\eta(e, i, \psi_i)$ can be expand as
\begin{equation}\label{eta}
    \begin{aligned}
    \mathrm{exp}({\eta(e, 1, \psi_1)})=\sum_{\psi_2\in[q]}\sum_{\psi_3\in[q]}...\sum_{\psi_\kappa\in[q]}\bm{C}_{\overrightarrow{\psi_e}}b^{2\to e}_{\psi_2}b^{3\to e}_{\psi_3}...b^{\kappa\to e}_{\psi_\kappa}
    \end{aligned}\enspace .
\end{equation}
There are $(\kappa-1)q^{\kappa-1}$ times of multiplication if we directly calculate $\eta(e, 1, \psi_1)$, which is unacceptable for high order $\kappa$. However, based on the specific $\bm{C}$~\eqref{nonuniform_cincout} in the symmetric HSBM, we can calculate $\eta(e, i, \psi_i)$ with dynamic programming
\begin{equation}
    \begin{aligned}
    \mathrm{exp}({\eta(e, 1, \psi_1)})
    =&\sum_{\psi_\kappa\in[q]}b^{\kappa\to e}_{\psi_\kappa}\sum_{\psi_2\in[q]}\sum_{\psi_3\in[q]}...\sum_{\psi_{\kappa-1}\in[q]}\bm{C}_{\overrightarrow{\psi_e}}b^{2\to e}_{\psi_2}b^{3\to e}_{\psi_3}...b^{\kappa-1\to e}_{\psi_{\kappa-1}}\\
    =&\sum_{\psi_\kappa\in[q]/\psi_1}b^{\kappa\to e}_{\psi_\kappa}\underbracket{\sum_{\psi_2\in[q]}\sum_{\psi_3\in[q]}...\sum_{\psi_{\kappa-1}\in[q]}c_{\rm out}b^{2\to e}_{\psi_2}b^{3\to e}_{\psi_3}...b^{\kappa-1\to e}_{\psi_{\kappa-1}}}_{\mathrm{exp}({\eta(e/\kappa, 1, \psi_1)})+(c_{\rm out}-c_{\rm in})b^{2\to e}_{\psi_1}b^{3\to e}_{\psi_1}...b^{\kappa-1\to e}_{\psi_{1}}}\\
    &+b^{\kappa\to e}_{\psi_1}\underbracket{\sum_{\psi_2\in[q]}\sum_{\psi_3\in[q]}...\sum_{\psi_{\kappa-1}\in[q]}\bm{C}_{\overrightarrow{\psi_e}}b^{2\to e}_{\psi_2}b^{3\to e}_{\psi_3}...b^{\kappa-1\to e}_{\psi_{\kappa-1}}}_{\text{$\mathrm{exp}({\eta(e/\kappa, 1, \psi_1)})$}}\\
    =&\sum_{\psi_\kappa\in[q]}b^{\kappa\to e}_{\psi_\kappa}\mathrm{exp}({\eta(e/\kappa, 1, \psi_1)})+(c_{\rm out}-c_{\rm in})\sum_{\psi_\kappa\in[q]/\psi_1}b^{\kappa\to e}_{\psi_\kappa}b^{2\to e}_{\psi_1}b^{3\to e}_{\psi_1}...b^{\kappa-1\to e}_{\psi_{1}}\\
    =&\mathrm{exp}({\eta(e/\kappa, 1, \psi_1)})+(c_{\rm out}-c_{\rm in})\sum_{\psi_\kappa\in[q]/\psi_1}b^{\kappa\to e}_{\psi_\kappa}b^{2\to e}_{\psi_1}b^{3\to e}_{\psi_1}...b^{\kappa-1\to e}_{\psi_{1}}
    \end{aligned}\enspace .
\end{equation}
By removing nodes of $e$ one by one from $\kappa$, finally we can get
\begin{equation}\label{dp_eta}
    \begin{aligned}
        \mathrm{exp}({\eta(\left\{1,2\right\}, 1, \psi_1)})
        =&c_{\rm in}b^{2\to \left\{1,2\right\}}_{\psi_1}+c_{\rm out}\sum_{\psi_2\in[q]/\psi_1}b^{2\to \left\{1,2\right\}}_{\psi_2}\\
        =&c_{\rm in}+(c_{\rm out}-c_{\rm in})\sum_{\psi_2\in[q]/\psi_1}b^{2\to \left\{1,2\right\}}_{\psi_2}
    \end{aligned} \enspace .
\end{equation}
The number of multiplication of~\eqref{dp_eta} is $\kappa+\kappa-1+..+2=O(\kappa^2)$ which is smaller than directly calculation of~\eqref{eta} for high order $\kappa$.

\subsection{The calculation of free energy}
According to~\cite{ruggeri2024message}, the free energy of the belief propagation on hypergraph can be estimated as
\begin{equation}\label{free-energy}
\begin{aligned}
\mathcal{F}&\approx-\sum_{i\in[n]}f_i+\sum_{e\in\boldsymbol{E}}(|e|-1)f_e\\
\end{aligned} \enspace ,
\end{equation}
where, based on our HSBM, the $f_i$ and $f_e$ is
\begin{equation}
    \begin{aligned}
        f_i&=\log\left(\sum_{\psi_i\in[q]}n_{\psi_i}\prod_{e\in\partial i \cap \boldsymbol{E}}\sum_{\overrightarrow{\psi_{e/i}}\in[q]^{|e|-1}}\left(\bm{\Omega}_{\overrightarrow{\psi_e}}\right)^{A_e}\left(1-\bm{\Omega}_{\overrightarrow{\psi_e}}\right)^{1-A_e}\prod_{j\in\partial e/i}b^{j\to e}_{\psi_j}\right)\\
f_e&=\log\left(\sum_{\overrightarrow{\psi_{e}}\in[q]^{|e|}}\left(\bm{\Omega}_{\overrightarrow{\psi_e}}\right)^{A_e}\left(1-\bm{\Omega}_{\overrightarrow{\psi_e}}\right)^{1-A_e}\prod_{j\in\partial e}b^{j\to e}_{\psi_j}\right)\\
    \end{aligned}
\end{equation}
Firstly, let's see $f_i$. Based on the discussion of hyperedge to node belief $\hat{b}^{e\to i}_{\psi_i}$ in Appendix~\ref{appendix-discuss_bp_h2n}, we have
\begin{equation}
\begin{aligned}
f_i&=\log\left(\sum_{\psi_i\in[q]}n_{\psi_i}
\prod_{e\in\partial i \cap \calE}\left(\sum_{\overrightarrow{\psi_{e/i}}\in[q]^{|e|-1}}\left(\bm{\Omega}_{\overrightarrow{\psi_e}}\right)\prod_{j\in\partial e/i}b^{j\to e}_{\psi_j}\right)
\prod_{e\in\partial i \cap \boldsymbol{E}/\calE}\left(\sum_{\overrightarrow{\psi_{e/i}}\in[q]^{|e|-1}}\left(1-\bm{\Omega}_{\overrightarrow{\psi_e}}\right)\prod_{j\in\partial e/i}b^{j\to e}_{\psi_j}\right)\right) \\
&=\log\left(\sum_{\psi_i\in[q]}n_{\psi_i}
\prod_{e\in\partial i \cap \calE}\left(\frac{1}{n^{|e|-1}}\sum_{\overrightarrow{\psi_{e/i}}\in[q]^{|e|-1}}\bm{C}_{\overrightarrow{\psi_e}}\prod_{j\in\partial e/i}b^{j\to e}_{\psi_j}\right)
\prod_{e\in\partial i \cap \boldsymbol{E}/\calE}e^{-\frac{1}{n^{|e|-1}}\sum_{\overrightarrow{\psi_{e/i}}\in[q]^{|e|-1}}\bm{C}_{\overrightarrow{\psi_e}}\prod_{j\in e/i}b^{j}_{\psi_j}}\right)\\
&=\log\left(\sum_{\psi_i\in[q]}n_{\psi_i}e^{-h^i_{\psi_i}}\prod_{e\in\partial i \cap \calE}\frac{1}{n^{|e|-1}}\sum_{\overrightarrow{\psi_{e/i}}\in[q]^{|e|-1}}\bm{C}_{\overrightarrow{\psi_e}}\prod_{j\in\partial e/i}b^{j\to e}_{\psi_j} \right)\\
&=\log\left(\sum_{\psi_i\in[q]}n_{\psi_i}e^{-h_{\psi_i}}\prod_{e\in\partial i \cap \calE}\frac{1}{n^{|e|-1}}e^{\eta(e, i, \psi_i)}\right)
\end{aligned} \enspace ,
\end{equation}
where external field $h_{\psi_i}$ is defined in Appendix~\ref{section_externalfield} and $\eta(e, i, \psi_i)$ is defined in Appendix~\ref{dp_hyperedgetonode}.

About the second summation of $\cal F$~\eqref{free-energy}, we have:
\begin{equation}\label{fe_2ndsum}
\begin{aligned}
\sum_{e\in \boldsymbol{E}}(|e|-1)f_e&=\sum_{e\in \calE}\log\left(\sum_{\overrightarrow{\psi_{e}}\in[q]^{|e|}}\bm{\Omega}_{\overrightarrow{\psi_e}}\prod_{j\in\partial e}b^{j\to e}_{\psi_j}\right)^{|e|-1}+\sum_{e\in \boldsymbol{E}/\calE}\log\left(\sum_{\overrightarrow{\psi_{e}}\in[q]^{|e|}}\left(1-\bm{\Omega}_{\overrightarrow{\psi_e}}\right)\prod_{j\in\partial e}b^{j\to e}_{\psi_j}\right)^{|e|-1}\\
\end{aligned}
\end{equation}
When $e\in \calE$, the first summation of~\eqref{fe_2ndsum} is
\begin{equation}
\begin{aligned}
\log\left(\sum_{\overrightarrow{\psi_{e}}\in[q]^{|e|}}\bm{\Omega}_{\overrightarrow{\psi_e}}\prod_{j\in\partial e}b^{j\to e}_{\psi_j}\right)^{|e|-1}&=(|e|-1)\log\left(\frac{1}{n^{|e|-1}}\underbracket{\sum_{\overrightarrow{\psi_{e}}\in[q]^{|e|}}\bm{C}_{\overrightarrow{\psi_e}}\prod_{j\in\partial e}b^{j\to e}_{\psi_j}}_{\tilde{\eta}(e)}\right)
\end{aligned}
\end{equation}
Similar with Appendix~\ref{dp_hyperedgetonode}, we also use dynamic programming to calculate $\tilde{\eta}(e)$. Assume $e=\left\{1, 2, ..., m\right\}$, set $\tilde{\eta}(e)=\tilde{\eta}(e, m)$, expand $\tilde{\eta}(e, m)$ we have
\begin{equation}\label{tildeeta}
\begin{aligned}
\tilde{\eta}(e, m)&=\sum_{\psi_m\in [q]}b_{\psi_m}^{m\to e}\sum_{\psi_1\in [q]}\sum_{\psi_2\in [q]}...\sum_{\psi_{m-1}\in [q]}\bm{C}_{\overrightarrow{\psi_e}}\prod_{j\in\left\{1, 2,..., m-1\right\}}b_{\psi_j}^{j\to e}\\
&=\sum_{\psi_m\in [q]}b_{\psi_m}^{m\to e}\left(\tilde{\eta}(e, m-1)+(\cout-\cin)\sum_{\psi\in[q]/\psi_m}\prod_{j\in\left\{1, 2,..., m-1\right\}}b_\psi^{j\to e}\right)\\
&=\tilde{\eta}(e, m-1)+(\cout-\cin)\sum_{\psi_m\in [q]}b_{\psi_m}^{m\to e}\sum_{\psi\in[q]/\psi_m}\prod_{j\in\left\{1, 2,..., m-1\right\}}b_\psi^{j\to e}
\end{aligned}
\end{equation}
Through dynamic programming shown in~\eqref{tildeeta}, we can reduce the number of multiplication from $q^m(m-1)$ to $\sum_{m=m}^{2}q(q(m-2)+1)+1=O(q^2m^2)$. 
The terminate condition for above dynamic programming $\tilde{\eta}(e, 2)$ is
\begin{equation}
\begin{aligned}
\tilde{\eta}(e, 2)&=\sum_{\psi_1\in[q]}\left(\cin b^{1\to e}_{\psi_1}b^{2\to e}_{\psi_1}+\cout b^{1\to e}_{\psi_1}\sum_{\psi_2\in[q]/\psi_1}b^{2\to e}_{\psi_2}\right)\\
&=\sum_{\psi_1\in[q]}\left((\cin-\cout)b^{1\to e}_{\psi_1}b^{2\to e}_{\psi_1}+\cout b^{1\to e}_{\psi_1}\sum_{\psi_2\in[q]}b^{2\to e}_{\psi_2}\right)\\
&=\cout+(\cin-\cout)\sum_{\psi_1\in[q]}b^{1\to e}_{\psi_1}b^{2\to e}_{\psi_1}
\end{aligned}
\end{equation}
When $e\notin \calE$, the second summation is
\begin{equation}
\begin{aligned}
&\sum_{e\in \boldsymbol{E}/\calE}\log\left(\sum_{\overrightarrow{\psi_{e}}\in[q]^{|e|}}\left(1-\bm{\Omega}_{\overrightarrow{\psi_e}}\right)\prod_{j\in\partial e}b^{j\to e}_{\psi_j}\right)^{|e|-1}\\
=& \log\left(\prod_{e\in \boldsymbol{E}/\calE}\left[\sum_{\overrightarrow{\psi_{e}}\in[q]^{|e|}}\left(1-\bm{\Omega}_{\overrightarrow{\psi_e}}\right)\prod_{j\in\partial e}b^{j\to e}_{\psi_j}\right]^{|e|-1}\right)\\
=& \log\left(\prod_{e\in \boldsymbol{E}/\calE}\left[1-\frac{1}{n^{|e|-1}}\sum_{\overrightarrow{\psi_{e}}\in[q]^{|e|}}\bm{C}_{\overrightarrow{\psi_e}}\prod_{j\in\partial e}b^{j\to e}_{\psi_j}\right]^{|e|-1}\right)\\
\approx& \log\left(\prod_{e\in \boldsymbol{E}}e^{-\frac{|e|-1}{n^{|e|-1}}\sum_{\overrightarrow{\psi_{e}}\in[q]^{|e|}}\bm{C}_{\overrightarrow{\psi_e}}\prod_{j\in\partial e}b^{j\to e}_{\psi_j}}\right)\\
\approx& \sum_{e\in \boldsymbol{E}}\frac{1-|e|}{n^{|e|-1}}\sum_{\overrightarrow{\psi_{e}}\in[q]^{|e|}}\bm{C}_{\overrightarrow{\psi_e}}\prod_{j\in\partial e}b^{j}_{\psi_j}\\
=&\sum_{\kappa\in \calK}\frac{1-\kappa}{n^{\kappa-1}}\sum_{f\in\boldsymbol{E}^{\kappa}}\left[(\cin-\cout)\sum_{\psi\in[q]}\prod_{j\in\partial f}b^j_{\psi}+\cout\sum_{\overrightarrow{\psi_{e}}\in[q]^{|e|}}\prod_{j\in\partial f}b^{j}_{\psi_j}\right]\\
=&\sum_{\kappa\in \calK}\frac{1-\kappa}{n^{\kappa-1}}\left[\binom{n}{\kappa}\cout+(\cin-\cout)\sum_{\psi\in [q]}\frac{1}{\kappa!}\left(\sum_{j\in [n]}b^j_\psi\right)^\kappa\right]\\
\approx&\sum_{\kappa\in \calK}\left[(1-\kappa)\cout+\frac{(1-\kappa)(\cin-\cout)}{n^{\kappa-1}\kappa!}\sum_{\psi\in [q]}\left(\sum_{j\in [n]}b^j_\psi\right)^\kappa\right]
\end{aligned}\enspace .
\end{equation}
With above equations, we can calculate free energy~\eqref{free-energy} efficiently.

\section{Supplementary analysis: community structure dominated by hyperedge order and shape.}
\begin{figure}
    \centering
    \includegraphics[width=10cm]{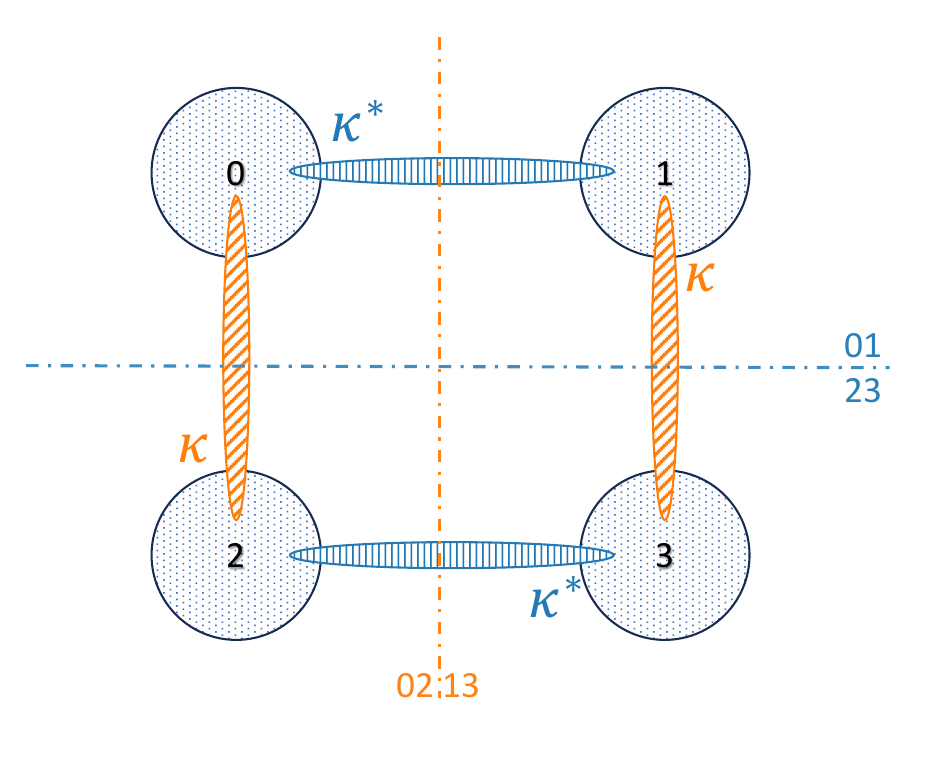}\\
	\caption{\small\it Illustration of community structure of hypergraph used in experiment for exploring the effect of order to detectability. \label{fig.append.ordereffectillus}}
\end{figure}

We designed a supplementary experiment to investigate how the dominance of different hyperedge orders influences the detected community structure using Bethe Hessian-based Spectral Clustering.
We consider a hypergraph of $n$ nodes equally partitioned into $q=4$ communities, $\left\{0, 1, 2, 3\right\}$.
The structure is designed to favor two competing community patterns:
all the $\kappa$-hyperedges are placed between nodes in communities $\left\{0,2\right\}$ or $\left\{1,3\right\}$, all the $\kappa^*$-hyperedges are placed between communities $\left\{0,1\right\}$ or $\left\{2,3\right\}$, as illustrated in Figure~\ref{fig.append.ordereffectillus}.

If we apply Bethe Hessian-based spectral clustering algorithm~\ref{scwithbh} to detect 2 communities in this hypergraph considering only the assortative case, two distinct, coarse-grained community structures could be detected:
\begin{itemize}
    \item $\kappa^*$-dominated structure with communities $\left\{0,1\right\}$ are merged into one community, $\left\{2,3\right\}$ are merged into the second.
    \item $\kappa$-dominated structure with communities $\left\{0,2\right\}$ are merged into one community, $\left\{1, 3\right\}$ are merged into the second.
\end{itemize}
Intuitively, if the number of $\kappa$-order hyperedges $|\mathcal{E}^{(\kappa)}|$ significantly outweighs $|\mathcal{E}^{(\kappa^*)}|$, the $\kappa$-dominated community structure should prevail.
Our objective is to determine the critical switching ratio of $|\mathcal{E}^{(\kappa)}|$ and $|\mathcal{E}^{(\kappa^*)}|$, at which the detected community structure transitions from the $\kappa^*$-dominated structure to the $\kappa$-dominated one.

\subsection{Simple case with $\kappa=2$, $\kappa^*=3$}
We first consider a simplified scenario with $\kappa=2$, $\kappa^*=3$. To generate the hypergraph shown in Figure~\ref{fig.append.ordereffectillus} using the HSBM, we define the affinity tensors as follows:
\begin{equation}
    \begin{aligned}
        \bmC^{(2)}_{\bm{z}}
        &=\left\{\begin{matrix}
            a,& \bm{z}\in\left\{(0, 2), (2, 0), (1, 3), (3, 1)\right\}\\
            0,& otherwise
        \end{matrix}\right. \\
        \bmC^{(3)}_{\bm{z}}
        &=\left\{\begin{matrix}
            a^*,& \bm{z}\in\begin{matrix}
            \left\{(0, 0,1),(0,1,0), (1,0,1),\right.&\\
            (0,1,1), (1,0,1), (1,1,0),&\\
            (2, 2,3), (2, 3,2),(3,2,2),&\\
            \left.(2,3,3),(3,2,3), (3,3,2)\right\}
            \end{matrix}\\
            0,& otherwise
        \end{matrix}\right.
    \end{aligned} \enspace ,
\end{equation}
where $a$ and $a^*$ are adjusted parameters.
By observing the definition of $\widehat{c}^{(\kappa)}_{ab}$ in Appendix~\ref{append:detectability_literature}, we can construct the matrix $\widehat{\bmC}^{(\kappa)}$ by fixing two dimensions and aggregating the remaining $\kappa-2$ dimension of scaled affinity tensor $\frac{1}{q^{\kappa-2}(\kappa-2)!}\bmC^{(\kappa)}$.
In this scenario, we calculate the corresponding $4\times 4$ matrices $\widehat{\bmC}^{(2)}$ and $\widehat{\bmC}^{(3)}$:
\begin{equation}\label{hatC_23}
    \begin{aligned}
        \widehat{\bmC}^{(2)}&=\frac{1}{4^{2-2}(2-2)!}\left[\begin{matrix}
                            0,& 0,&a,&0\\
                            0,& 0,&0,&a\\
                            a,& 0,&0,&0\\
                            0,& a,&0,&0\\
                            \end{matrix}\right] \\
        \widehat{\bmC}^{(3)}&=\frac{1}{4^{3-2}(3-2)!}\left[\begin{matrix}
                            a^*,& 2a^*,&0,&0\\
                            2a^*,& a^*,&0,&0\\
                            0,& 0,&a^*,&2a^*\\
                            0,& 0,&2a^*,&a^*\\
                            \end{matrix}\right]
    \end{aligned} \enspace .
\end{equation}
We utilize the $\snr_{\rm BH}$ based on $\widehat{\bmC}^{(\kappa)}$ as shown in~\eqref{append:detectabilityforanyq}, as
\begin{equation}\label{phi_hatC}
    \snr_{\rm BH}:=\frac{\left(\sum_{\kappa \in \calK}\widehat{c}_{\rm in}^{(\kappa )}-\widehat{c}_{\rm out}^{(\kappa )}\right)^2}{q^2\sum_{\kappa \in \calK}(\kappa -1)d^{(\kappa )}} \enspace ,
\end{equation}
where $d^{(\kappa)}=\frac{\widehat{c}_{\rm in}^{(\kappa )}+(q-1)\widehat{c}_{\rm out}^{(\kappa )}}{q(\kappa -1)}$. This formulation allows us to leverage techniques from the detectability analysis of hierarchical community structures on network~\cite{peel2024detectability}.
The finest 4 communities $\left\{0,1,2,3\right\}$ and two coarse-grained community structure, the $\kappa$-dominated $\left\{\left\{0, 2\right\}, \left\{1, 3\right\}\right\}$, the $\kappa^*$-dominated $\left\{\left\{0, 1\right\}, \left\{2, 3\right\}\right\}$, form two hierarchical community structure. 
We can aggregate the matrix $\widehat{\bmC}^{(2)}$ and $\widehat{\bmC}^{(3)}$ by different coarser communities to form $2\times 2$ matrices, and then derive their corresponding $\snr_{\mathrm{BH} \; 02;13}$ and $\snr_{\mathrm{BH} \; 01;23}$ based on~\eqref{phi_hatC}.
Although $\snr_{\mathrm{BH} \; 02;13}=1$ or $\snr_{\mathrm{BH} \; 01;23}=1$ can not rigorously be the detectability limit for $\kappa$ or $\kappa^*$ dominated community structure, the switching point between the these two competing community structure can be approximated by solving for the equality $\snr_{\mathrm{BH} \; 02;13}=\snr_{\mathrm{BH} \; 01;23}$.

For the $\kappa^*$ dominated community structure, the aggregation process of $\widehat{\bmC}$ is by averaging each block defined by the community partition $\left\{\left\{0, 1\right\}, \left\{2, 3\right\}\right\}$.
Using the matrices in~\eqref{hatC_23}, the aggregated matrices $\widehat{\bmC}_{(\kappa^*)}$ are thus calculated as:
\begin{equation}
    \begin{aligned}
        \widehat{\bmC}^{(2)}_{(\kappa^*)}&=\left[\begin{matrix}
                            0,& a/2\\
                            a/2,& 0
                            \end{matrix}\right] \\
        \widehat{\bmC}^{(3)}_{(\kappa^*)}&=\frac{1}{4}\left[\begin{matrix}
                            \frac{3}{2}a^*,&0\\
                            0,&\frac{3}{2}a^*
                            \end{matrix}\right]
    \end{aligned} \enspace .
\end{equation}
The average $\kappa$-degree $d^{(\kappa)}$ can be calculated by the average of one row of $\frac{1}{\kappa-1}\widehat{\bmC}^{(\kappa)}$, which remains constant regardless of the aggregation process above:
\begin{equation}
    \begin{aligned}
        d^{(2)}&=\frac{a}{4}\\
        d^{(3)}&=\frac{3a^*}{32}
    \end{aligned} \enspace .
\end{equation}
Then based on~\eqref{phi_hatC}, we have
\begin{equation}
    \snr_{\mathrm{BH} \; 01;23}=\frac{(-\frac{a}{2}+\frac{3}{8}a^*)^2}{2^2[(2-1)d^{(2)}+(3-1)d^{(3)}]} \enspace .
\end{equation}

Similarly, for the $\kappa$-dominated community structure $\left\{\left\{0, 2\right\}, \left\{1, 3\right\}\right\}$, the aggregated matrices are:
\begin{equation}
    \begin{aligned}
        \widehat{\bmC}^{(2)}_{(\kappa)}&=\left[\begin{matrix}
                            a/2,& 0\\
                            0,& a/2
                            \end{matrix}\right] \\
        \widehat{\bmC}^{(3)}_{(\kappa)}&=\frac{1}{4}\left[\begin{matrix}
                            \frac{1}{2}a^*,&a^*\\
                            a^*,&\frac{1}{2}a^*
                            \end{matrix}\right]
    \end{aligned} \enspace ,
\end{equation}
leading to 
\begin{equation}
    \snr_{\mathrm{BH} \; 02;13}=\frac{(\frac{a}{2}+\frac{1}{8}a^*-\frac{1}{4}a^*)^2}{2^2[(2-1)d^{(2)}+(3-1)d^{(3)}]} \enspace .
\end{equation}
Solving for the switching point $\snr_{\mathrm{BH} \; 01;23}=\snr_{\mathrm{BH} \; 02;13}$:
\begin{equation}
    \begin{aligned}
        (-\frac{a}{2}+\frac{3a^*}{8})^2&=(\frac{a}{2}+\frac{a^*}{8}-\frac{a^*}{4})^2\\
        (\frac{3a^*}{8})^2-\frac{3}{8}aa^*&=(\frac{a^*}{8})^2-\frac{1}{8}aa^*\\
        \frac{1}{8}(a^*)^2-\frac{1}{4}aa^*&=0\\
        \frac{a}{a^*}=\frac{1}{2}
    \end{aligned} \enspace .
\end{equation}
Thus, the theoretical switching condition for the affinity tensor parameters is $\frac{a}{a^*}=\frac{1}{2}$.

To determine the corresponding hyperedge ratio $\rho=|\mathcal{E}^{(2)}|/|\mathcal{E}^{(3)}|$, we estimate $|\mathcal{E}^{(\kappa)}|$ using the approximate formula:
\begin{equation}\label{append:estimate_Ek}
    \begin{aligned}
        |\mathcal{E}^{(\kappa)}|&=2\left[\binom{2n/q}{\kappa}-2\binom{n/q}{\kappa}\right]\frac{a}{n^{\kappa-1}}\\
        &\approx\frac{2a}{n^{\kappa-1}\kappa!}\left[(2n/q)^\kappa-2(n/q)^\kappa\right]\\
        &=\frac{2an}{q^\kappa\kappa!}(2^\kappa-2)
    \end{aligned} \enspace .
\end{equation}
Substituting $\kappa=2$, $\kappa^*=3$ and $q=4$, we have
\begin{equation}
    \begin{aligned}
        \rho^*_{theo}=\frac{|\mathcal{E}^{(2)}|}{|\mathcal{E}^{(3)}|}=\frac{2a/(4^22!)}{6a^*/(4^33!)}=4\frac{a}{a^*}=2
    \end{aligned} \enspace .
\end{equation}
The theoretical switching point in terms of the hyperedge ratio is $\rho^*_{theo}=2$.

\begin{figure}
    \centering
    \includegraphics[width=\linewidth]{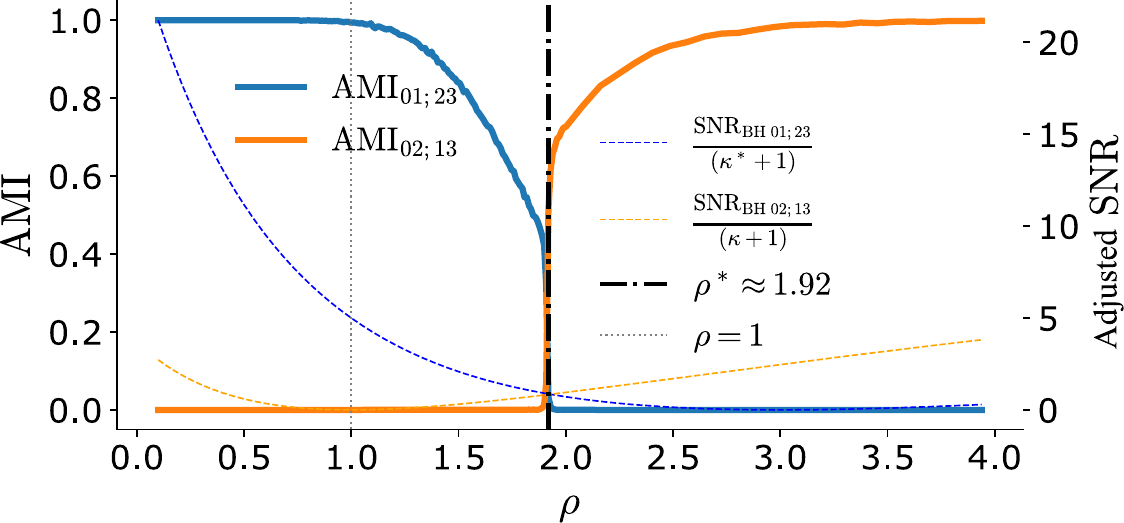}\\
	\caption{\small\it Experimental results with $n=8000$, $d=50$, $\kappa=2$ and $\kappa^*=3$. We show the $\mathrm{AMI}_{01;23}$ and $\mathrm{AMI}_{02;13}$ as the $\rm AMI$ of detected communities by Bethe Hessian based-spectral clustering to the $\kappa^*$ and $\kappa$ dominated community structure.  \label{fig.append.ordereffect.n8000k2kstar3}}
\end{figure}

To experimentally observe the transition, we conducted a simulation using Bethe Hessian based-spectral clustering to detect 2 communities on hypergraphs generated by fixing the total average degree $d$ and varying the ratio $\rho$. By definition:
\begin{equation}\label{append:define_d_epsilon}
    \begin{aligned}
        d&=\frac{\kappa |\mathcal{E}^{(\kappa)}|+\kappa^* |\mathcal{E}^{(\kappa^*)}|}{n}\\
        \rho&=\frac{|\mathcal{E}^{(\kappa)}|}{|\mathcal{E}^{(\kappa^*)}|}
    \end{aligned}\enspace .
\end{equation}
Solving these relations for the hyperedge counts, we get:
\begin{equation}
    \begin{aligned}
        |\mathcal{E}^{(\kappa^*)}|&=\frac{nd}{\kappa\rho+\kappa^*}\\
        |\mathcal{E}^{(\kappa)}|&=\frac{nd\rho}{\kappa\rho+\kappa^*}
    \end{aligned} \enspace .
\end{equation}
Using the estimation of $|\mathcal{E}^{(\kappa)}|$ in~\eqref{append:estimate_Ek}, we derive the affinity tensor parameters $a$ and $a^*$:
\begin{equation}
    \begin{aligned}
        a&=\frac{4^\kappa\kappa!d\rho}{2(2^\kappa-2)(\kappa\rho+\kappa^*)}\\
        a^*&=\frac{4^{\kappa^*}\kappa^*!d}{2(2^{\kappa^*}-2)(\kappa\rho+\kappa^*)}
    \end{aligned} \enspace .
\end{equation}
Based on these derived values for $a$ and $a^*$, we set up the affinity tensor and generated the hypergraph for community detection. 
The results are shown in Figure~\ref{fig.append.ordereffect.n8000k2kstar3}.

We observed that the derived $\rho^*_{theo}$ has a little bias to the sharp transition point.
Instead, the transition point $\rho^*$ can be better aligned by the cross point of two manually modified $\snr_{\mathrm{BH} \; 01;23}$ and $\snr_{\mathrm{BH} \; 02;13}$:
\begin{equation}
    \frac{\snr_{\mathrm{BH} \; 01;23}}{\kappa^*+1}=\frac{\snr_{\mathrm{BH} \; 02;13}}{\kappa+1} \enspace .
\end{equation}
Next, we proceed to consider a more general case with arbitrary $\kappa$ and $\kappa^*$.

\subsection{General case with arbitrary $\kappa$, $\kappa^*$}
Similar as last section, we first define the affinity tensor under scenario of arbitrary $\kappa$, $\kappa^*$:
\begin{equation}
    \begin{aligned}
        \bmC^{(\kappa)}_{\bm{z}}
        &=\left\{\begin{matrix}
            a,& \bm{z}\in\left\{\left\{(0, 2)^\kappa/(0)^\kappa,(2)^{\kappa}\right\}, \left\{(1, 3)^\kappa/(1)^\kappa,(3)^{\kappa}\right\}\right\}\\
            0,& otherwise
        \end{matrix}\right. \\
        \\
        \bmC^{(\kappa^*)}_{\bm{z}}
        &=\left\{\begin{matrix}
            a^*,& \bm{z}\in\left\{\left\{(0, 1)^{\kappa^*}/(0)^{\kappa^*},(1)^{\kappa^*}\right\}, \left\{(2, 3)^{\kappa^*}/(2)^{\kappa^*},(3)^{\kappa^*}\right\}\right\} \\
            0,& otherwise
        \end{matrix}\right.
    \end{aligned} \enspace ,
\end{equation}
where $(...)^\kappa$ means all $\kappa$-length permutations of index, such as $(0,2)^\kappa$ is all $\kappa$-length index which exclusively include 0 or 2, $(0)^\kappa$ is $\left\{\underbrace{(0, 0, .., 0)}_{\kappa}\right\}$. Then we aggregate all other $\kappa-2$ dimension of tensor $\frac{1}{q^{\kappa-2}(\kappa-2)!}\bmC^{(\kappa)}$ and $\frac{1}{q^{\kappa^*-2}(\kappa^*-2)!}\bmC^{(\kappa^*)}$, derive the matrices $\widehat{\bmC}^{(\kappa)}$ and $\widehat{\bmC}^{(\kappa^*)}$:
\begin{equation}
    \begin{aligned}
        \widehat{\bmC}^{(\kappa)}&=
            \frac{1}{4^{\kappa-2}(\kappa-2)!}\left[\begin{matrix}
            2^{\kappa-2}-1,& 0,&2^{\kappa-2},&0\\
            0,& 2^{\kappa-2}-1,&0,&2^{\kappa-2}\\
            2^{\kappa-2},& 0,&2^{\kappa-2}-1,&0\\
            0,& 2^{\kappa-2},&0,&2^{\kappa-2}-1\\
            \end{matrix}\right]a \\
        \\
        \widehat{\bmC}^{(\kappa^*)}&=
            \frac{1}{4^{\kappa^*-2}(\kappa^*-2)!}\left[\begin{matrix}
            2^{\kappa^*-2}-1,& 2^{\kappa^*-2},&0,&0\\
            2^{\kappa^*-2},& 2^{\kappa^*-2}-1,&0,&0\\
            0,& 0,&2^{\kappa^*-2}-1,&2^{\kappa^*-2}\\
            0,& 0,&2^{\kappa^*-2},&2^{\kappa^*-2}-1\\
            \end{matrix}\right]a^*
    \end{aligned} \enspace .
\end{equation}

Before aggregation with different coarse-grained community structure, we first derive $d^{(\kappa)}$ and $d^{\kappa^*}$ since they remain constant after the aggregation:
\begin{equation}
    \begin{aligned}
    d^{(\kappa)}&=\frac{1}{\kappa-1}\frac{a}{4^{\kappa-2}(\kappa-2)!}\frac{2*2^{\kappa-2}-1}{4}\\
    &=\frac{a}{4^{\kappa-1}(\kappa-1)!}(2^{\kappa-1}-1)\\
    d^{(\kappa^*)}&=\frac{a^*}{4^{\kappa^*-1}(\kappa^*-1)!}(2^{\kappa^*-1}-1)
    \end{aligned} \enspace .
\end{equation}

For $\kappa^*$-dominated community structure, $\left\{\left\{0, 1\right\}, \left\{2, 3\right\}\right\}$, the aggregated matrices $\widehat{\bmC}_{(\kappa^*)}$ are:
\begin{equation}
    \begin{aligned}
    	\widehat{\bmC}_{(\kappa^*)}^{(\kappa)}&=\frac{a}{4^{\kappa-2}(\kappa-2)!}\left[\begin{matrix}
    		(2^{\kappa-2}-1)/2,& (2^{\kappa-2})/2\\
    		(2^{\kappa-2})/2,& (2^{\kappa-2}-1)/2\\
    	\end{matrix}\right] \\
        \\
    	\widehat{\bmC}_{(\kappa^*)}^{(\kappa^*)}&=\frac{a^*}{4^{\kappa^*-2}(\kappa^*-2)!}\left[\begin{matrix}
    		2^{\kappa^*-2}-\frac{1}{2},& 0\\
    		0,& 2^{\kappa^*-2}-\frac{1}{2}\\
    	\end{matrix}\right] 
	\end{aligned} \enspace ,
\end{equation}
leading to
\begin{equation}
    \snr_{\mathrm{BH} \; 01;23}=\frac{\left[-\frac{a}{4^{\kappa-2}(\kappa-2)!}\frac{1}{2}+\frac{a^*}{4^{\kappa^*-2}(\kappa^*-2)!}(2^{\kappa^*-2}-\frac{1}{2})\right]^2}{2^2[(\kappa-1)d^{(\kappa)}+(\kappa^*-1)d^{(\kappa^*)}]} \enspace .
\end{equation}
For $\kappa$-dominated community structure, $\left\{\left\{0, 2\right\}, \left\{1, 3\right\}\right\}$, the aggregated matrices $\widehat{\bmC}_{(\kappa)}$ are:
\begin{equation}
    \begin{aligned}
        \widehat{\bmC}_{(\kappa)}^{(\kappa)}&=\frac{a}{4^{\kappa-2}(\kappa-2)!}\left[\begin{matrix}
    		2^{\kappa-2}-\frac{1}{2},& 0\\
    		0,& 2^{\kappa-2}-\frac{1}{2}\\
    	\end{matrix}\right] \\
        \\
    	\widehat{\bmC}_{(\kappa)}^{(\kappa^*)}&=\frac{a^*}{4^{\kappa^*-2}(\kappa^*-2)!}\left[\begin{matrix}
    		(2^{\kappa^*-2}-1)/2,& (2^{\kappa^*-2})/2\\
    		(2^{\kappa^*-2})/2,& (2^{\kappa^*-2}-1)/2\\
    	\end{matrix}\right] 
	\end{aligned} \enspace ,
\end{equation}
leading to
\begin{equation}
    \snr_{\mathrm{BH} \; 02;13}=\frac{\left[\frac{a}{4^{\kappa-2}(\kappa-2)!}(2^{\kappa-2}-\frac{1}{2})-\frac{a^*}{4^{\kappa^*-2}(\kappa^*-2)!}\frac{1}{2}\right]^2}{2^2[(\kappa-1)d^{(\kappa)}+(\kappa^*-1)d^{(\kappa^*)}]} \enspace .
\end{equation}

Next, we solve for the switching point $\snr_{\mathrm{BH} \; 01;23}=\snr_{\mathrm{BH} \; 02;13}$. For convenience, we define
\begin{equation}
    \begin{aligned}
        b&=\frac{a}{4^{\kappa-2}(\kappa-2)!}\\
        b^*&=\frac{a^*}{4^{\kappa^*-2}(\kappa^*-2)!}
    \end{aligned} \enspace ,
\end{equation}
then
\begin{equation}
    \begin{aligned}
    \snr_{\mathrm{BH} \; 01;23}&=\snr_{\mathrm{BH} \; 02;13}\\
    (b^*2^{\kappa^*-2}-\frac{b}{2}-\frac{b^*}{2})^2&=(b2^{\kappa-2}-\frac{b}{2}-\frac{b^*}{2})^2\\
    (b^*2^{\kappa^*-2})^2-2^{\kappa^*-2}bb^*-2^{\kappa^*-2}(b^*)^2&=(b2^{\kappa-2})^2-2^{\kappa-2}bb^*-2^{\kappa-2}(b)^2\\
    2^{\kappa^*-2}(2^{\kappa^*-2}-1)(b^*)^2-(2^{\kappa^*-2}-2^{\kappa-2})bb^*-2^{\kappa-2}(2^{\kappa-2}-1)b^2&=0\\
    (2^{\kappa^*-2}b^*-2^{\kappa-2}b)\underbrace{[(2^{\kappa^*-2}-1)b^*+(2^{\kappa-2}-1)b]}_{+}&=0\\
    \frac{b}{b^*}&=\frac{2^{\kappa^*-2}}{2^{\kappa-2}}\\
    \frac{a}{a^*}\frac{4^{\kappa^*-2}(\kappa^*-2)!}{4^{\kappa-2}(\kappa-2)!}&=\frac{2^{\kappa^*-2}}{2^{\kappa-2}}\\
    \frac{a}{a^*}=\frac{2^{\kappa-2}(\kappa-2)!}{2^{\kappa^*-2}(\kappa^*-2)!}
    \end{aligned} \enspace .
\end{equation}
Based on the formula for $|\mathcal{E}^{(\kappa)}|$ in~\eqref{append:estimate_Ek}, the hyperedge ratio at the switching point is
\begin{equation}
    \begin{aligned}
        \rho^*_{theo}=\frac{|\mathcal{E}^{(\kappa)}|}{|\mathcal{E}^{(\kappa^*)}|}
        &=\frac{\frac{2an}{q^\kappa\kappa!}(2^\kappa-2)}{\frac{2a^*n}{q^{\kappa^*}\kappa^*!}(2^{\kappa^*}-2)}\\
        &=\frac{(2^\kappa-2)/(4^\kappa\kappa!)}{(2^{\kappa^*}-2)/(4^{\kappa^*}\kappa^*!)}\frac{a}{a^*}\\
        &=\frac{2^{\kappa-2}(\kappa-2)!(2^\kappa-2)/(4^\kappa\kappa!)}{2^{\kappa^*-2}(\kappa^*-2)!(2^{\kappa^*}-2)/(4^{\kappa^*}\kappa^*!)}\\
        &=\frac{(2^\kappa-2)/(2^{\kappa+2}\kappa(\kappa-1))}{(2^{\kappa^*}-2)/(2^{\kappa^*+2}\kappa^*(\kappa^*-1))}\\
        &=2^{\kappa^*-\kappa}\frac{2^{\kappa-1}-1}{2^{\kappa^*-1}-1}\frac{\binom{\kappa^*}{2}}{\binom{\kappa}{2}}
    \end{aligned} \enspace .
\end{equation}

\begin{figure}
    \centering
    \includegraphics[width=\linewidth]{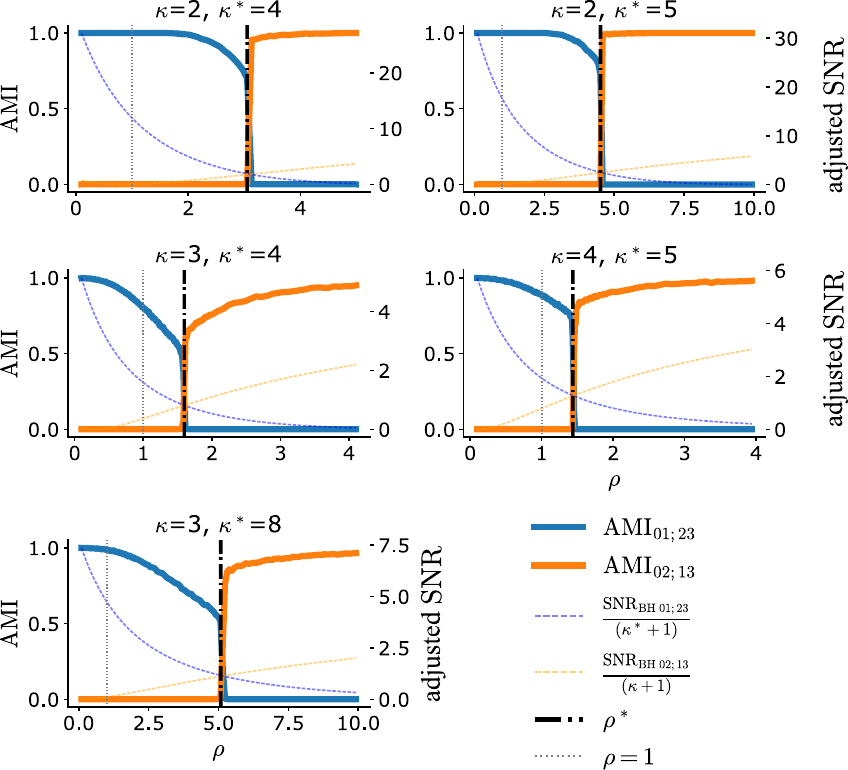}\\
	\caption{\small\it Experimental results with general $\kappa$ and $\kappa^*$. All experiments are proceed on $n=8000$ hypergraphs. For first row, the average degree $d=50$ for better observing the sharp transition. For others $d=10$ is enough. We show the $\mathrm{AMI}_{01;23}$ and $\mathrm{AMI}_{02;13}$ as the $\rm AMI$ of detected communities by Bethe Hessian based-spectral clustering to the $\kappa^*$ and $\kappa$ dominated community structure.  \label{fig.append.ordereffect.more}}
\end{figure}

However, we still find this switching point has a little bias to the actual transition point, which can be better aligned by:
\begin{equation}
    \frac{\snr_{\mathrm{BH} \; 01;23}}{\kappa^*+1}=\frac{\snr_{\mathrm{BH} \; 02;13}}{\kappa+1} \enspace ,
\end{equation}
as shown in Figure~\ref{fig.append.ordereffect.more}.

\subsection{Order trade-offs for Other Method}
\begin{figure}
    \centering
    \includegraphics[width=\linewidth]{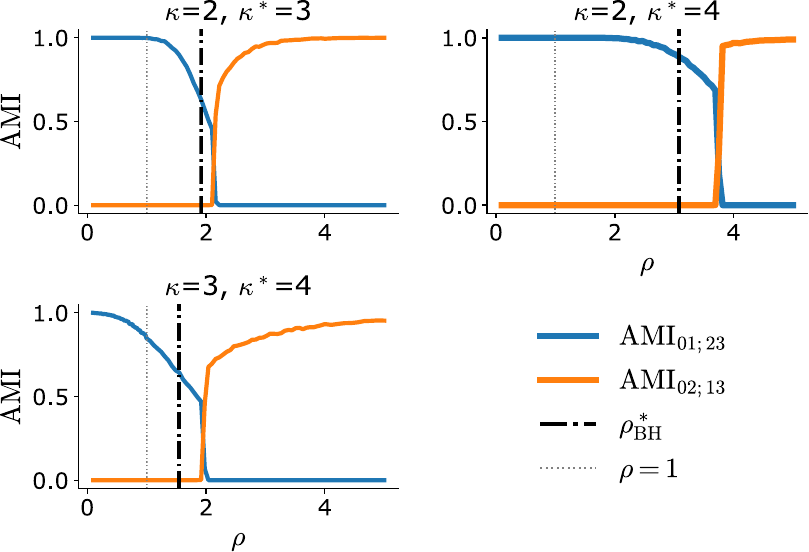}\\
	\caption{\small\it Experimental results with Laplacian based spectral clustering. All experiments are proceed on hypergraphs with $n=8000$. For first row, the average degree $d=50$ for better observing the sharp transition. For second row, $d=10$ is enough. We show the $\mathrm{AMI}_{01;23}$ and $\mathrm{AMI}_{02;13}$ as the $\rm AMI$ of detected communities by Laplacian based-spectral clustering to the $\kappa^*$ and $\kappa$ dominated community structure. We also show the switching point of Bethe Hessian method $\rho^*_{\rm BH}$ for comparison.\label{fig.append.ordereffect.lap}}
\end{figure}

Although we primarily examine the order trade-offs of the Bethe Hessian, our intent is to show that such trade-offs are intrinsic to hypergraph community detection in general. We demonstrated this by the Hypergraph Laplacian~\cite{zhou2006learning} in Figure~\ref{fig.append.ordereffect.lap}. Clearly, the Laplacian method also possesses a bias toward keeping higher-order hyperedges within communities, albeit with a different switching point compared to our proposed Bethe Hessian method.

\subsection{Community structure dominated by hyperedge shape}
\begin{figure}
    \centering
    \includegraphics[width=15cm]{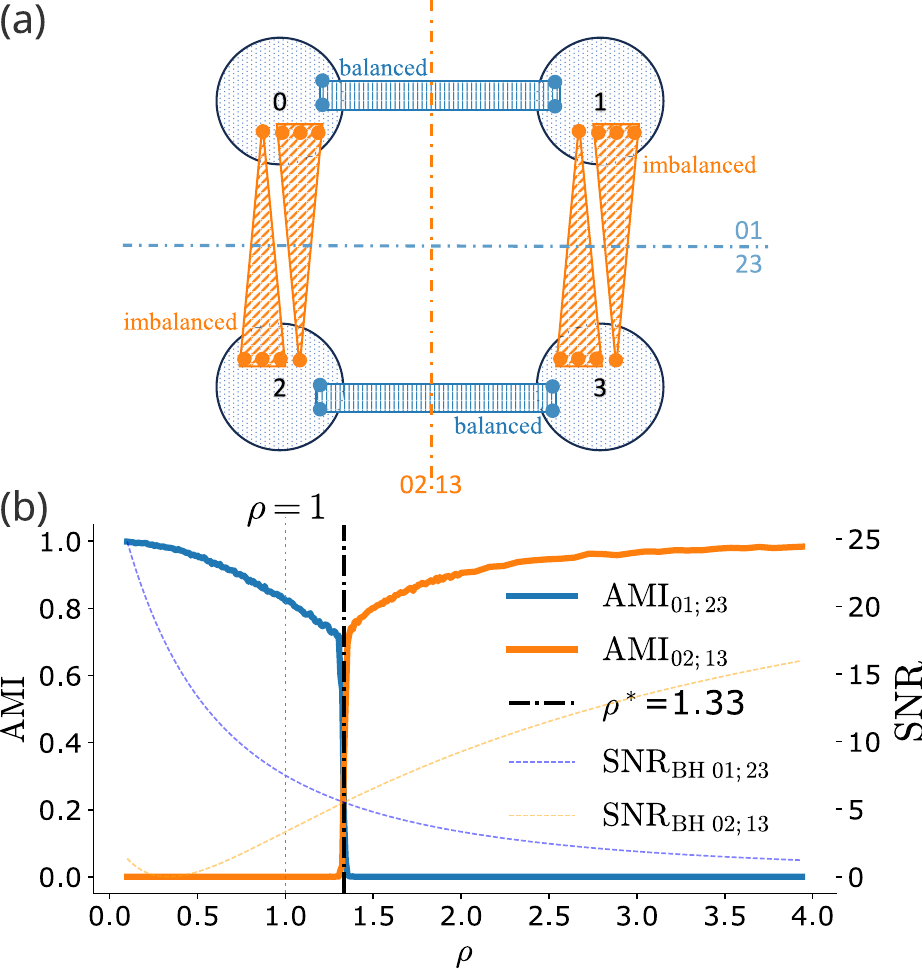}\\
	\caption{\small\it (a) Illustration of community structure of hypergraph used in experiment for exploring the effect of shape to detectability. All the hyperedges are 4-order, only the different shape. The hyperedges shaped by rectangle between communities $\left\{0, 1\right\}$ or $\left\{2, 3\right\}$ are balanced, the hyperedges shaped by triangle between communities $\left\{0, 2\right\}$ or $\left\{1, 3\right\}$ are imbalanced. (b) Experimental results with $n=8000$, $d=10$, $\kappa=4$ and $\kappa^*=4$. We show the $\mathrm{AMI}_{\kappa^*}$ and $\mathrm{AMI}_{\kappa}$ as the $\rm AMI$ of detected communities by Bethe Hessian based-spectral clustering to community structure dominated by balanced and imbalanced 4-order hyperedges.  \label{fig.append.shapeeffectillus}}
\end{figure}

Beyond the order of hyperedges, their "shape"—the way nodes are distributed across communities—also significantly influences the detection of coarse-grained community structures.
To investigate this, we designed an experiment using a hypergraph with a finest community structure of four communities $\left\{0, 1, 2, 3\right\}$, where all hyperedges are of order 4.
We define two distinct hyperedge shapes based on their node distribution:
\begin{itemize}
    \item Balanced hyperedges: These 4-order hyperedges span two communities by connecting two nodes from one community and two from another. In our model, these only exist between the community pairs $\left\{0, 1\right\}$ or $\left\{2, 3\right\}$.
    \item Imbalanced hyperedges: These 4-order hyperedges span two communities by connecting one node from one community and three from another. In our model, these only exist between the community pairs $\left\{0, 2\right\}$ or $\left\{1, 3\right\}$.
\end{itemize}
This setup, illustrated in Figure~\ref{fig.append.shapeeffectillus} (a), creates two competing coarse-grained community structures: one based on balanced hyperedges $\left\{\left\{0, 1\right\}, \left\{2, 3\right\}\right\}$ and another based on imbalanced hyperedges $\left\{\left\{0, 2\right\}, \left\{1, 3\right\}\right\}$.

To generate this hypergraph using the HSBM, we define the affinity tensor $\bmC^{(4)}$ as follows. Let $a^*$ be the affinity parameter for balanced hyperedges and $a$ for imbalanced ones:
\begin{equation}
    \begin{aligned}
        \bmC^{(4)}_{\bm{z}}
        &=\left\{\begin{matrix}
            a,& \bm{z}\in\begin{matrix}
            \left\{(0, 0,0, 2),(0,0, 2,0), (0, 2,0,0), (2, 0, 0, 0)\right.&\\
            (0,2,2, 2), (2,0,2, 2), (2,2,0, 2), (2, 2, 2, 0)&\\
            (1, 1,1, 3), (1,1, 3,1),(1,3,1, 1), (3, 1, 1, 1)&\\
            \left.(3,3,3, 1),(3,3, 1,3), (3,1, 3,3), (1, 3, 3, 3)\right\}
            \end{matrix}\\
            a^*, & \bm{z}\in\begin{matrix}
            \left\{(0, 0,1, 1),(0,1,1,0), (0, 1,0,1),\right.&\\
            (1, 0,0,1), (1, 1,0,0), (1, 0,1,0),&\\
            (2, 2,3, 3), (2, 3,2, 3),(2,3, 3, 2),&\\
            \left.(3, 2,2,3),(3,2,3, 2), (3, 3,2,2)\right\}
            \end{matrix}\\
            0,& otherwise
        \end{matrix}\right.
    \end{aligned} \enspace .
\end{equation}
For consistency, we denote the expected number of balanced hyperedges as $|\mathcal{E}^{(\kappa^*)}|$ and imbalanced hyperedges as $|\mathcal{E}^{(\kappa)}|$. These can be calculated as:
\begin{equation}
    \begin{aligned}
        |\mathcal{E}^{(\kappa^*)}|&=2\binom{n/q}{2}\binom{n/q}{2}\frac{a^*}{n^3}=2\frac{(n/q)^4}{4}\frac{a^*}{n^3}=2\frac{na^*}{4*4^4}\\
        |\mathcal{E}^{(\kappa)}|&=4\binom{n/q}{3}\binom{n/q}{1}\frac{a}{n^3}=4\frac{(n/q)^4}{6}\frac{a}{n^3}=4\frac{na}{6*4^4}\\
    \end{aligned} \enspace .
\end{equation}
Using the definitions~\eqref{append:define_d_epsilon} of average degree $d$ and the hyperedge ratio $\rho=|\mathcal{E}^{(\kappa)}|/|\mathcal{E}^{(\kappa^*)}|$ from previous section, we can express the parameters $a$ and $a^*$ as:
\begin{equation}
    \begin{aligned}
    a&=\frac{6*4^4d\rho}{4(4\rho+4)}=\frac{6*4^2d\rho}{\rho+1}\\
    a^*&=\frac{4*4^4d}{2(4\rho+4)}=\frac{2*4^3d}{\rho+1}
    \end{aligned} \enspace .
\end{equation}

As shown in the experimental results in Figure~\ref{fig.append.shapeeffectillus} (b), the Bethe Hessian-based spectral clustering algorithm's preference switches from the balanced to the imbalanced community structure at a point $\rho^*_{theo}\neq 1$. This empirically confirms that the hyperedge shape directly affects community detectability.
To theoretically predict this switching point, we analyze under similar routine to previous sections.

First, we derive the matrix $\widehat{\bmC}^{(4)}$:
\begin{equation}
    \widehat{\bmC}^{(4)}=\frac{1}{4^{4-2}(4-2)!}\left[\begin{matrix}
a^*+2a,& 2a^*,&2a,&0\\
2a^*,& a^*+2a,&0,&2a\\
2a,& 0,&a^*+2a,&2a^*\\
0,& 2a,&2a^*,&a^*+2a\\
\end{matrix}\right] \enspace .
\end{equation}
Aggregating this matrix according to the community structure dominated by balanced 4-order hyperedge $\left\{\left\{0, 1\right\}, \left\{2, 3\right\}\right\}$ yields:
\begin{equation}
    \widehat{\bmC}^{(4)}_{(\kappa^*)}=\frac{1}{32}\left[\begin{matrix}
                \frac{3}{2}a^*+a,& a\\
                a,& \frac{3}{2}a^*+a\\
                \end{matrix}\right] \Rightarrow \snr_{\mathrm{BH} \; 01;23}=\frac{1}{2^2(4-1)d^{(4)}}\left(\frac{1}{32}\frac{3}{2}a^*\right)^2 \enspace .
\end{equation}
Similarly, aggregating for the community structure dominated by imbalanced 4-order hyperedge $\left\{\left\{0, 2\right\}, \left\{1, 3\right\}\right\}$ gives:
\begin{equation}
    \widehat{\bmC}^{(4)}_{(\kappa)}=\frac{1}{32}\left[\begin{matrix}
                \frac{1}{2}a^*+2a,& a^*\\
                a^*,& \frac{1}{2}a^*+2a\\
                \end{matrix}\right] \Rightarrow \snr_{\mathrm{BH} \; 02;13}=\frac{1}{2^2(4-1)d^{(4)}}\left(\frac{1}{32}(2a-\frac{1}{2}a^*)\right)^2 \enspace .
\end{equation}
To find the switching point, we solve the $\snr_{\mathrm{BH} \; 01;23}=\snr_{\mathrm{BH} \; 02;13}$:
\begin{equation}
    \begin{aligned}
        \snr_{\mathrm{BH} \; 01;23}&=\snr_{\mathrm{BH} \; 02;13}\\
        (\frac{3}{2}a^*)^2&=(2a-\frac{1}{2}a^*)^2\\
        2(a^*)^2-4a^2+2aa^*&=0\\
        (a^*-a)(a^*+2a)&=0\\
        a&=a^*
    \end{aligned} \enspace .
\end{equation}
Last, to find the corresponding ratio of hyperedge counts $\rho^*_{theo}$, we substitute $a=a^*$ into the ratio formula:
\begin{equation}
    \begin{aligned}
        \rho^*_{theo}=\frac{|\mathcal{E}^{(\kappa)}|}{|\mathcal{E}^{(\kappa^*)}|}
        &=\frac{4\binom{n/q}{3}\binom{n/q}{1}\frac{a}{n^3}}{2\binom{n/q}{2}\binom{n/q}{2}\frac{a}{n^3}}\\
        &\approx\frac{\frac{(n/q)^4}{3}}{\frac{(n/q)^4}{4}}\\
        &=\frac{4}{3}
    \end{aligned} \enspace .
\end{equation}
This theoretical value of $\rho^*_{theo}=\frac{4}{3}\approx 1.33$ perfectly aligns with the switching point observed in our experiments (Figure~\ref{fig.append.shapeeffectillus} (b)).
This means that balanced 4-order hyperedges provide a stronger community signal than imbalanced ones.

\begin{figure}
    \centering
    \includegraphics[width=15cm]{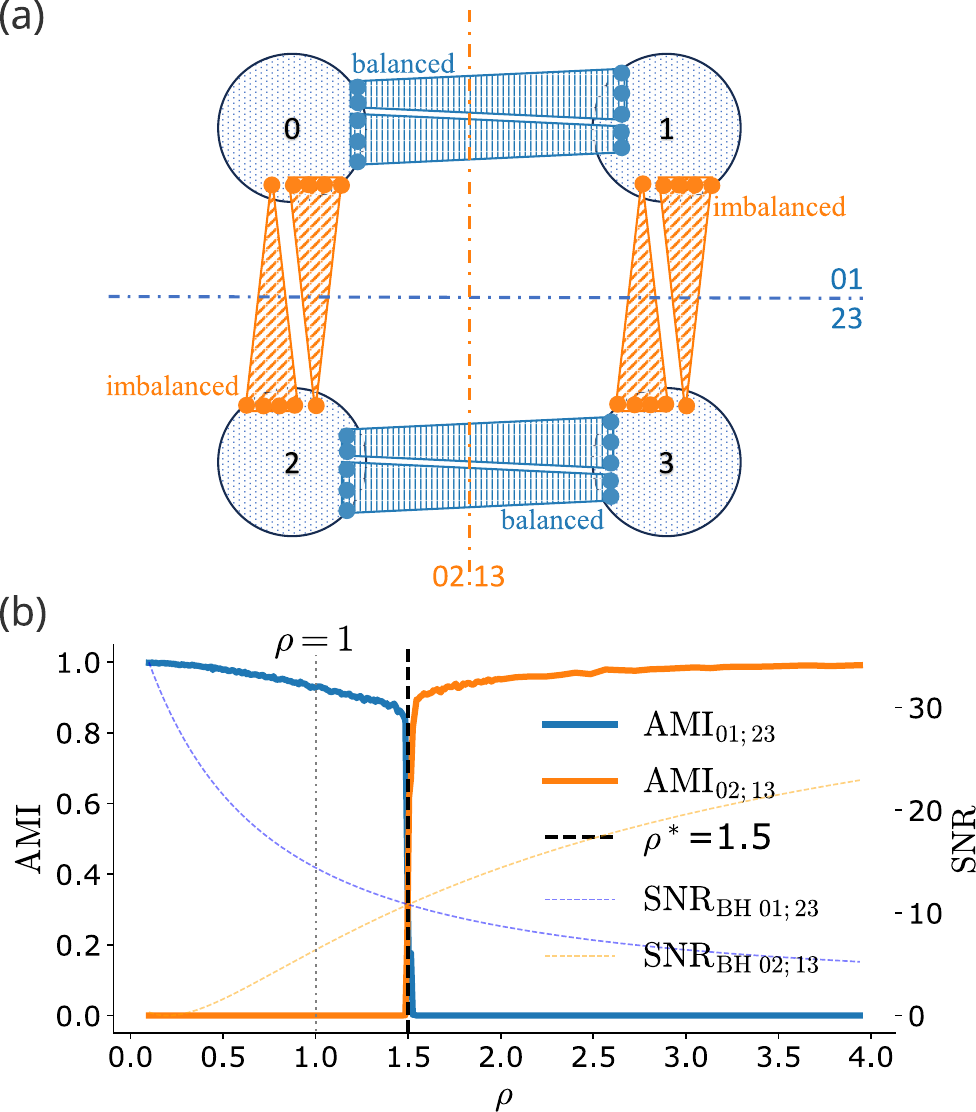}\\
	\caption{\small\it (a) Illustration of community structure of hypergraph used in experiment for exploring the effect of shape to detectability. All the hyperedges are 5-order, only the different shape. The hyperedges shaped by trapezoid between communities $\left\{0, 1\right\}$ or $\left\{2, 3\right\}$ are more balanced, composed by 2 nodes in one community, 3 nodes in another. The hyperedges shaped by triangle between communities $\left\{0, 2\right\}$ or $\left\{1, 3\right\}$ are more imbalanced, composed by 1 node in one community, 4 nodes in another. (b) Experimental results with $n=8000$, $d=10$, $\kappa=5$ and $\kappa^*=5$. We show the $\mathrm{AMI}_{\kappa^*}$ and $\mathrm{AMI}_{\kappa}$ as the $\rm AMI$ of detected communities by Bethe Hessian based-spectral clustering to community structure dominated by more balanced and more imbalanced 5-order hyperedges. \label{fig.append.shapeeffectillus5}}
\end{figure}

We also proceed similar experiment with 5-order hyperedges, as illustrated in Figure~\ref{fig.append.shapeeffectillus5} (a). We define the two distinct 5-order hyperedge shapes based on their node distribution:
\begin{itemize}
    \item More Balanced hyperedges: These 5-order hyperedges span two communities by connecting two nodes from one community and three from another. In our model, these only exist between the community pairs $\left\{0, 1\right\}$ or $\left\{2, 3\right\}$.
    \item More Imbalanced hyperedges: These 5-order hyperedges span two communities by connecting one node from one community and four from another. In our model, these only exist between the community pairs $\left\{0, 2\right\}$ or $\left\{1, 3\right\}$.
\end{itemize}

Based on the above setup, we define the affinity tensor $\bmC^{(5)}$ as follows. Let $a^*$ be the affinity parameter for more balanced hyperedges and $a$ for more imbalanced ones:
\begin{equation}
    \begin{aligned}
        \bmC^{(5)}_{\bm{z}}
        &=\left\{\begin{matrix}
            a,& \bm{z}\in
            \left\{\bm{z}|\#\left\{0\text{ in }\bm{z}\right\}=3, \#\left\{2\text{ in }\bm{z}\right\}=2\right\}\\
            a,& \bm{z}\in
            \left\{\bm{z}|\#\left\{0\text{ in }\bm{z}\right\}=2, \#\left\{2\text{ in }\bm{z}\right\}=3\right\}\\
            a,& \bm{z}\in
            \left\{\bm{z}|\#\left\{1\text{ in }\bm{z}\right\}=3, \#\left\{3\text{ in }\bm{z}\right\}=2\right\}\\
            a,& \bm{z}\in
            \left\{\bm{z}|\#\left\{1\text{ in }\bm{z}\right\}=2, \#\left\{3\text{ in }\bm{z}\right\}=3\right\}\\
            a^*,& \bm{z}\in
            \left\{\bm{z}|\#\left\{0\text{ in }\bm{z}\right\}=3, \#\left\{1\text{ in }\bm{z}\right\}=2\right\}\\
            a^*,& \bm{z}\in
            \left\{\bm{z}|\#\left\{0\text{ in }\bm{z}\right\}=2, \#\left\{1\text{ in }\bm{z}\right\}=3\right\}\\
            a^*,& \bm{z}\in
            \left\{\bm{z}|\#\left\{2\text{ in }\bm{z}\right\}=3, \#\left\{3\text{ in }\bm{z}\right\}=2\right\}\\
            a^*,& \bm{z}\in
            \left\{\bm{z}|\#\left\{2\text{ in }\bm{z}\right\}=2, \#\left\{3\text{ in }\bm{z}\right\}=3\right\}\\
            0,& otherwise
        \end{matrix}\right.
    \end{aligned} \enspace .
\end{equation}

For consistency, we denote the expected number of more balanced hyperedges as $|\mathcal{E}^{(\kappa^*)}|$ and more imbalanced hyperedges as $|\mathcal{E}^{(\kappa)}|$. These can be calculated as:
\begin{equation}
    \begin{aligned}
        |\mathcal{E}^{(\kappa^*)}|&=4\binom{n/q}{2}\binom{n/q}{3}\frac{a^*}{n^4}=4\frac{(n/q)^5}{2!3!}\frac{a^*}{n^4}=4\frac{na^*}{2!3!q^5}\\
        |\mathcal{E}^{(\kappa)}|&=4\binom{n/q}{4}\binom{n/q}{1}\frac{a}{n^4}=4\frac{(n/q)^5}{4!}\frac{a}{n^4}=4\frac{na}{4!q^5}\\
    \end{aligned} \enspace .
\end{equation}
Using the definitions~\eqref{append:define_d_epsilon} of average degree $d$ and the hyperedge ratio $\rho=|\mathcal{E}^{(\kappa)}|/|\mathcal{E}^{(\kappa^*)}|$, we can express the parameters $a$ and $a^*$ as:
\begin{equation}
    \begin{aligned}
    a&=\frac{4^44!d\rho}{5(\rho+1)}\\
    a^*&=\frac{4^42!3!d}{5(\rho+1)}
    \end{aligned} \enspace .
\end{equation}

As shown in the experimental results in Figure~\ref{fig.append.shapeeffectillus5} (b), the Bethe Hessian-based spectral clustering algorithm's preference switches from the more balanced to the more imbalanced community structure at a point $\rho^*_{theo}\neq 1$. This further empirically confirms that the hyperedge shape directly affects community detectability.
To theoretically predict this switching point, we analyze under similar routine to previous sections.

First, we derive the matrix $\widehat{\bmC}^{(4)}$:
\begin{equation}
    \widehat{\bmC}^{(5)}=\frac{1}{4^{5-2}(5-2)!}\left[\begin{matrix}
4a^*+3a,& 6a^*,&2a,&0\\
6a^*,& 4a^*+3a,&0,&2a\\
2a,& 0,&4a^*+3a,&6a^*\\
0,& 2a,&6a^*,&4a^*+3a\\
\end{matrix}\right] \enspace .
\end{equation}
Aggregating this matrix according to the community structure dominated by more balanced 5-order hyperedges $\left\{\left\{0, 1\right\}, \left\{2, 3\right\}\right\}$ yields:
\begin{equation}
    \begin{aligned}
        \widehat{\bmC}^{(5)}_{(\kappa^*)}=&\frac{1}{64*6}\left[\begin{matrix}
                5a^*+\frac{3}{2}a,& a\\
                a,& 5a^*+\frac{3}{2}a\\
                \end{matrix}\right] \\
        \Rightarrow \snr_{\mathrm{BH} \; 01;23}=&\frac{1}{2^2(5-1)d^{(5)}}\left(\frac{1}{64*6}(5a^*+\frac{3}{2}a-a)\right)^2 \enspace .
    \end{aligned}
\end{equation}
Similarly, aggregating for the community structure dominated by more imbalanced 5-order hyperedges $\left\{\left\{0, 2\right\}, \left\{1, 3\right\}\right\}$ gives:
\begin{equation}
\begin{aligned}
    \widehat{\bmC}^{(5)}_{(\kappa)}=&\frac{1}{64*6}\left[\begin{matrix}
                2a^*+\frac{5}{2}a,& 3a^*\\
                3a^*,& 2a^*+\frac{5}{2}a\\
                \end{matrix}\right] \\
    \Rightarrow \snr_{\mathrm{BH} \; 02;13}=&\frac{1}{2^2(5-1)d^{(5)}}\left(\frac{1}{64*6}(2a^*+\frac{5}{2}a-3a^*)\right)^2 \enspace .
\end{aligned}
\end{equation}
To find the switching point, we solve the $\snr_{\mathrm{BH} \; 01;23}=\snr_{\mathrm{BH} \; 02;13}$:
\begin{equation}
    \begin{aligned}
        \snr_{\mathrm{BH} \; 01;23}&=\snr_{\mathrm{BH} \; 02;13}\\
        (5a^*+\frac{1}{2}a)^2&=(\frac{5}{2}a-a^*)^2\\
        (4a^*+3a)(6a^*-2a)&=0\\
        \frac{a}{a^*}&=3
    \end{aligned} \enspace .
\end{equation}
Last, to find the corresponding ratio of hyperedge counts $\rho^*_{theo}$, we substitute $a=a^*$ into the ratio formula:
\begin{equation}
    \begin{aligned}
        \rho^*_{theo}=\frac{|\mathcal{E}^{(\kappa)}|}{|\mathcal{E}^{(\kappa^*)}|}
        &=\frac{\frac{a}{4!}}{\frac{a^*}{2!3!}}
        =\frac{3}{2}
    \end{aligned} \enspace .
\end{equation}
This theoretical value of $\rho^*_{theo}=\frac{3}{2}$ perfectly aligns with the switching point observed in our experiments (Figure~\ref{fig.append.shapeeffectillus5} (b)).
This means that more balanced 5-order hyperedges provide a stronger community signal than more imbalanced ones.

\newpage
\section{Notation}
\begin{table}[H]
    \centering
    \begin{tabular}{c|c}
         node set & $\cal V$ \\
         node  & $i, j, k, l$ \\
         number of nodes & $n$ \\
         edge(hyper) set & $\calE$ \\
         edge(hyper)  & $e, f, g$ \\
         number of edges(hyper) & $m$ \\
         order of edge(hyper) $e$ & $|e|$ \\
         order set & $\calK$ \\
         order $\kappa$ hyperedge set & $\calE^{(\kappa)}$ \\
         number of order $\kappa$ hyperedges & $m^{(\kappa)}$ \\
         average degree & $d$ \\
         diagonal degree matrix & $\bm D$ \\ 
         order $\kappa$ degree of node $i$ & $d_i^{(\kappa)}$ \\
         order $\kappa$ average degree & $d^{(\kappa)}$ \\
         order $\kappa$ degree matrix & $\bm{D}^{(\kappa)}$ \\
         number of communities & $q$ \\
         community labels & $\overrightarrow{\psi}$ \\
         adjacent matrix & $\bm A$ \\
         edge probability matrix(tensor) & $\bm{\Omega}$ \\
         rescaled edge probability matrix(tensor) & $\bm C$ \\
         incidence matrix & $\bm H$ \\
         order $\kappa$-incidence matrix & $\bm{H}^{(\kappa)}$ \\
         one-mode projection of $\bm{H}^{(\kappa)}$ & $\bm{A}^{(\kappa)}$ \\
         non-backtracking matrix & $\bmNB$ \\
         non-backtracking matrix of $\kappa$-order uniform hyper graph & $\bmNB^{(\kappa)}$ \\
         non-backtracking matrix of nonuniform hyper graph with order set $\calK$ & $\bmNB^{(\calK)}$ \\
         bethe hessian matrix & $\bm{B}$ \\
         bethe hessian matrix of $\kappa$-order uniform hyper graph & $\bm{B}^{(\kappa)}$ \\
         bethe hessian matrix of nonuniform hyper graph with order set $\calK$ & $\bm{B}^{(\calK)}$ \\
         
         \hline
    \end{tabular}
    \caption{Notations}
    \label{PQlambda}
\end{table}
\end{document}